\documentclass[twocolumn,showpacs,preprintnumbers,amsmath,amssymb]{revtex4}

\usepackage{graphicx}
\usepackage{dcolumn}
\usepackage{bm}

\newcommand{\kslash}{k\kern-1ex /}
\newcommand{\pslash}{p\kern-1ex /}
\newcommand{\qslash}{q\kern-1ex /}
\newcommand{\lslash}{l\kern-1ex /}
\newcommand{\sslash}{s\kern-1ex /}
\newcommand{\Dslash}{D\kern-1.2ex /}

\newcommand{\tr}{{\rm tr}}

\newcommand{\beqa}{\begin{eqnarray}}
\newcommand{\eeqa}{\end{eqnarray}}

\newcommand{\bd}{\begin{description}}
\newcommand{\ed}{\end{description}}
\newcommand{\la}{\langle}
\newcommand{\ra}{\rangle}
\newcommand{\ben}{\begin{eqnarray}}
\newcommand{\een}{\end{eqnarray}}
\newcommand{\nn}{\nonumber}
\def\lsim{\raise0.3ex\hbox{$<$\kern-0.75em\raise-1.1ex\hbox{$\sim$}}}
\def\gsim{\raise0.3ex\hbox{$>$\kern-0.75em\raise-1.1ex\hbox{$\sim$}}}
\def\simgt{\rlap{\lower 3.0 pt\hbox{$\mathchar \sim$}}\raise 1.5pt \hbox {$>$}}
\def\simlt{\rlap{\lower 3.0 pt\hbox{$\mathchar \sim$}}\raise 1.5pt \hbox {$<$}}

\newcommand{\msbar}{{\overline {\rm MS}}}

\newcolumntype{d}[1]{D{.}{\cdot}{#1}}
\newcolumntype{.}{D{.}{.}{-1}}
\newcolumntype{,}{D{,}{,}{-1}}

\begin{document}

\preprint{UTCCS-P-53, UTHEP-582}

\title{SU(2) and SU(3) chiral perturbation theory analyses on baryon masses\\
in 2+1 flavor lattice QCD}

\author{
 K.-I.~Ishikawa${}^{a}$,
 N.~Ishizuka${}^{b,c}$,
 T.~Izubuchi${}^{d,e}$,
 D.~Kadoh${}^{c}$\footnote{Present address: Theoretical Physics Laboratory, RIKEN, Wako 2-1, Saitama 351-0198, Japan},
 K.~Kanaya${}^{b}$,
 Y.~Kuramashi${}^{b,c}$,
 Y.~Namekawa${}^{c}$,
 M.~Okawa${}^{a}$,
 Y.~Taniguchi${}^{b,c}$,
 A.~Ukawa${}^{b,c}$,
 N.~Ukita${}^{c}$
 T.~Yoshi\'e${}^{b,c}$\\
(PACS-CS Collaboration)
}
\affiliation{
 ${}^a$Graduate School of Science, Hiroshima University, Higashi-Hiroshima, Hiroshima 739-8526, Japan\\
 ${}^b$Graduate School of Pure and Applied Sciences, University of Tsukuba, Tsukuba, Ibaraki 305-8571, Japan\\
 ${}^c$Center for Computational Sciences, University of Tsukuba, Tsukuba, Ibaraki 305-8577, Japan\\
 ${}^d$Riken BNL Research Center, Brookhaven National Laboratory, Upton,
New York 11973, USA\\
 ${}^e$Institute for Theoretical Physics, Kanazawa University, Kanazawa,
       Ishikawa 920-1192, Japan
}



\date{\today}

\begin{abstract}
We investigate the quark mass dependence of baryon
masses in 2+1 flavor lattice QCD
using SU(3) heavy baryon chiral perturbation theory up to one-loop order.
The baryon mass data used for the analyses are obtained 
for the degenerate up-down quark mass
of 3 MeV to 24 MeV and two choices of the strange quark mass
around the physical value. 
We find that the SU(3) chiral expansion fails 
to describe both the octet and the decuplet baryon data 
if phenomenological values 
are employed for the meson-baryon couplings.  
The SU(2) case is also examined for the nucleon.  We observe that higher order 
terms are controlled only around the physical point. 
We also evaluate finite size effects using SU(3) 
heavy baryon chiral perturbation theory, finding small values of order 1\% 
even at the physical point.  
\end{abstract}

\maketitle

\section{Introduction}
\label{sec:intro}

The aim of the PACS-CS project is full QCD calculations on the physical point
avoiding any contamination due to chiral extrapolations.
At the first stage of this project\cite{pacscs_nf3} we have succeeded in 
reducing the up-down quark mass from 67 MeV, the minimum value 
reached by the previous CP-PACS/JLQCD work\cite{cppacs_nf3}, 
to 3 MeV, corresponding to the decrease of pion mass from 702 MeV to 156 MeV. 
This work allowed us to make detailed chiral analyses on the pseudoscalar 
meson sector with the use of chiral perturbation theory (ChPT). 
An important finding is that the strange quark
mass is not small enough to be treated by the SU(3) ChPT up to the
next-to-leading order (NLO). For the octet and decuplet baryon masses
we simply employed a linear chiral extrapolation to the physical point
assuming isospin symmetry and analyticity of the
strange quark contribution around its physical mass.

In this article we reinvestigate the quark mass dependence of the octet and
decuplet baryon masses employing the SU(3) heavy baryon chiral
perturbation theory (HBChPT) up to
NLO\cite{jenkins1,jenkins2,walker-loud}. 
The results are compared with those of the linear chiral extrapolation 
obtained in Ref.~\cite{pacscs_nf3}. 
We also examine the convergence property of HBChPT fits to the lattice 
results. 
For the nucleon mass we examine the SU(2) HBChPT with an analytic
expansion of the low energy constants (LECs) in terms of the
strange quark mass around its physical value. 
The SU(2) BChPT analyses in 2 flavor lattice QCD were previously made  
by other collaborations and reported in Refs.~\cite{qcdsf,etm}. 
We also discuss the magnitude of 
finite size effects based on the NLO SU(3) HBChPT.

This paper is organized as follows. In Sec.~\ref{sec:linear} 
we briefly review the results obtained in Ref.~\cite{pacscs_nf3}
to make the paper self-contained. 
In Sec.~\ref{sec:hbchpt}
we apply SU(3) HBChPT analyses to the octet 
baryon masses. We present the fit results for the LECs and discuss 
the convergence behavior up to NLO.
The same analysis is repeated for the decuplet baryon masses.
Section~\ref{sec:nucleon_su2} describes
the fit results of the nucleon mass with the SU(2) HBChPT.
In Sec.~\ref{sec:fse} we discuss the magnitude 
of finite size effects for the baryon masses based on the SU(3) HBChPT. 
Our conclusions are summarized in Sec.~\ref{sec:conclusion}.    

\section{SU(2) Linear chiral fit for baryon masses}
\label{sec:linear}

We give a quick review of the chiral analyses employed
in Ref.~\cite{pacscs_nf3}. The 
extrapolation to the physical point is performed with the following fit
formula:
\ben
m&=&\alpha+\beta m_{\rm ud}^{\rm AWI}+\gamma m_{\rm s}^{\rm AWI},
\label{eq:linear_su2}
\een
where $m_{\rm ud}^{\rm AWI}$ denotes the Axial Ward Identity quark mass 
for the up-down quark and $m_{\rm s}^{\rm AWI}$ for
the strange quark. 
Since we choose $\kappa_s$ around the physical strange quark mass,
the above formula is essentially an SU(2) chiral expansion
with the strange quark contribution analytically 
expanded around its physical value. 
Actually we can rewrite the formula (\ref{eq:linear_su2}) as 
\ben
m&=&\alpha^\prime+\beta m_{\rm ud}^{\rm AWI}
+\gamma (m_{\rm s}^{\rm AWI}-m_{\rm s,ph}^{\rm AWI})
\label{eq:linear_su2_conv}
\een
where $\alpha^\prime=\alpha+\gamma m_{\rm s,ph}^{\rm AWI}$
with $m_{\rm s,ph}^{\rm AWI}$ the physical value of the strange quark
mass.
This is a linear expansion of the baryon mass around 
$(m_{\rm ud}^{\rm AWI},m_{\rm s}^{\rm AWI})=(0,m_{\rm s,ph}^{\rm AWI})$.

\begin{figure*}[h]
\vspace{13mm}
\begin{center}
\begin{tabular}{cc}
\includegraphics[width=85mm,angle=0]{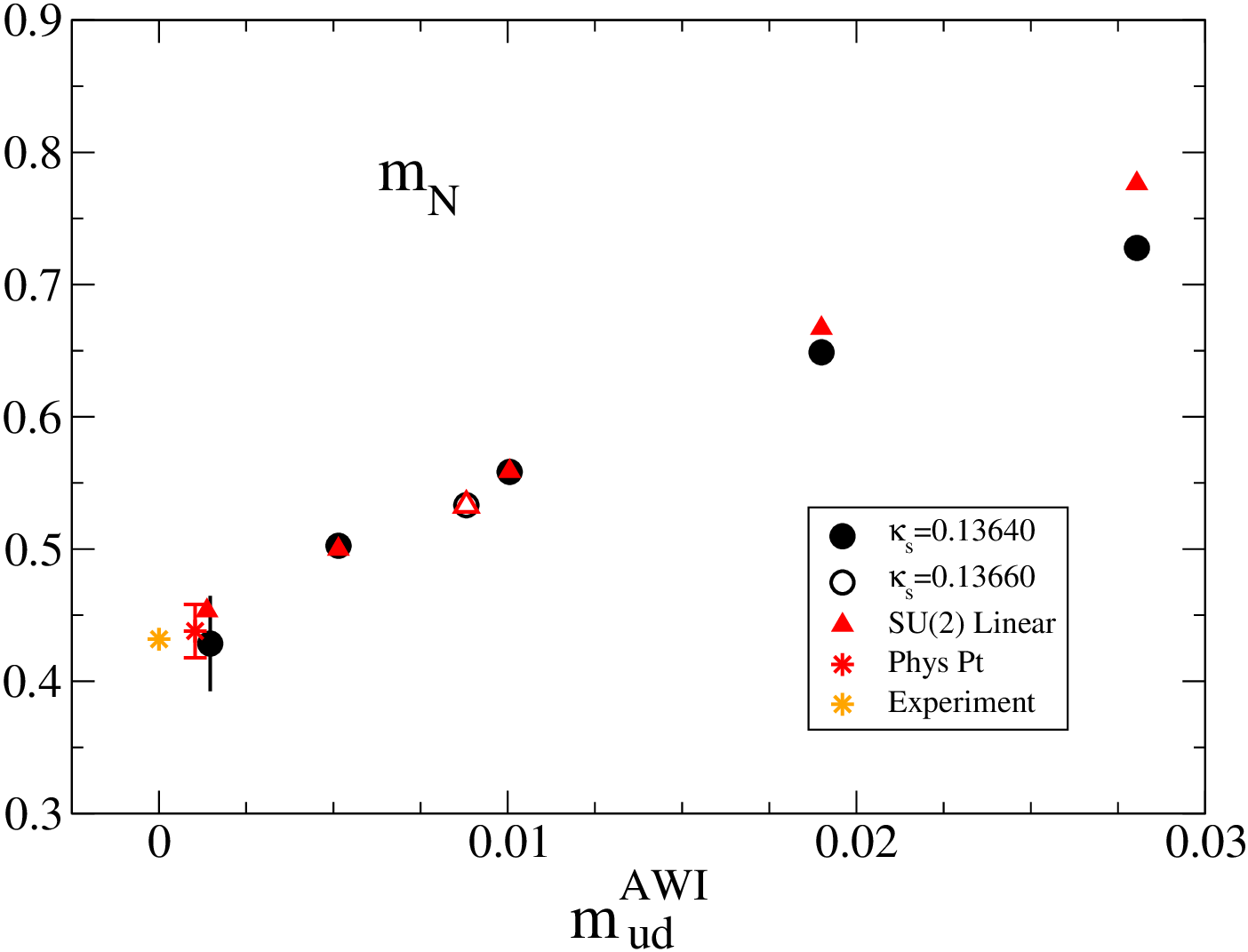}
\hspace*{2mm}&\hspace*{2mm}
\includegraphics[width=85mm,angle=0]{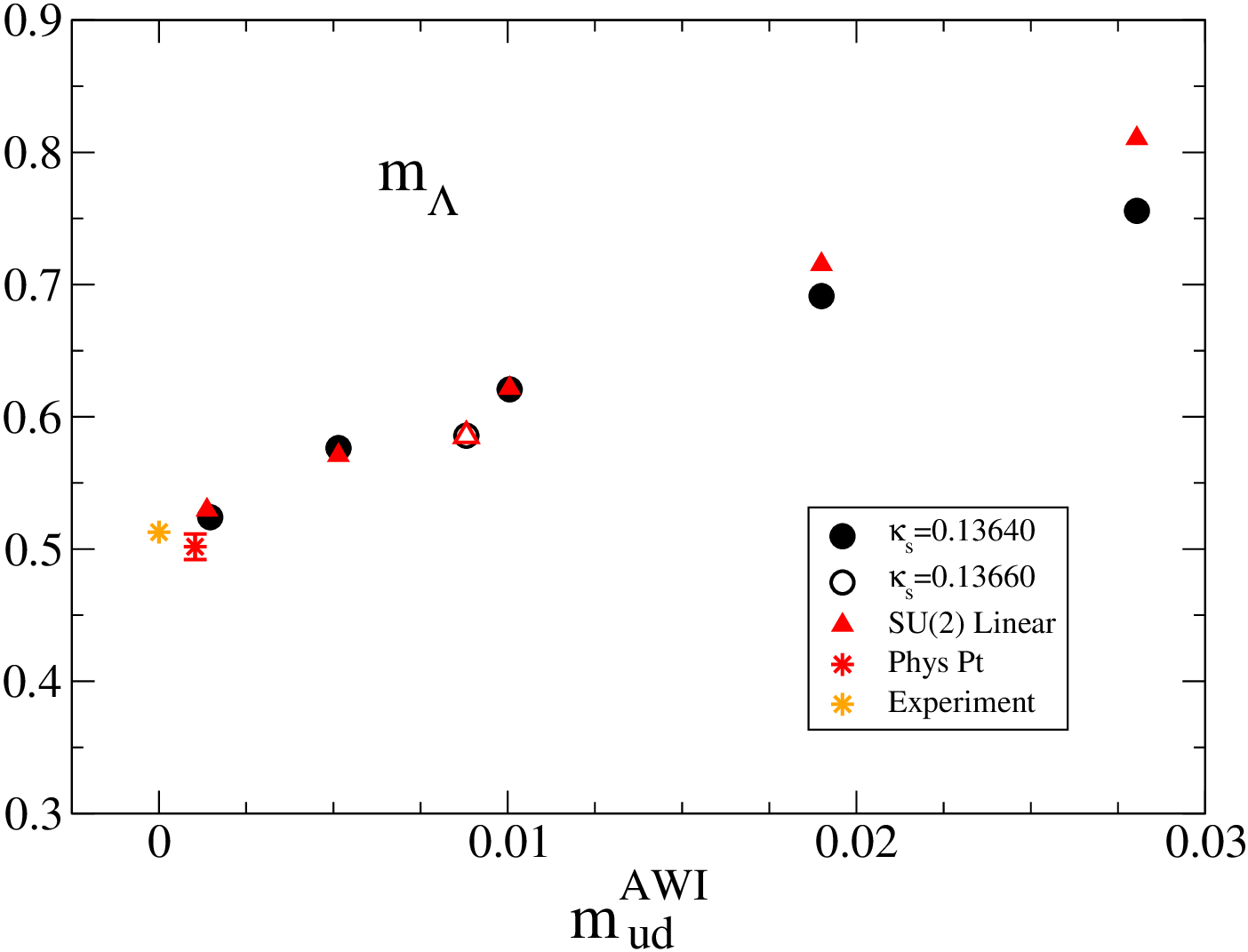}\\
\vspace*{7mm}& \\
\includegraphics[width=85mm,angle=0]{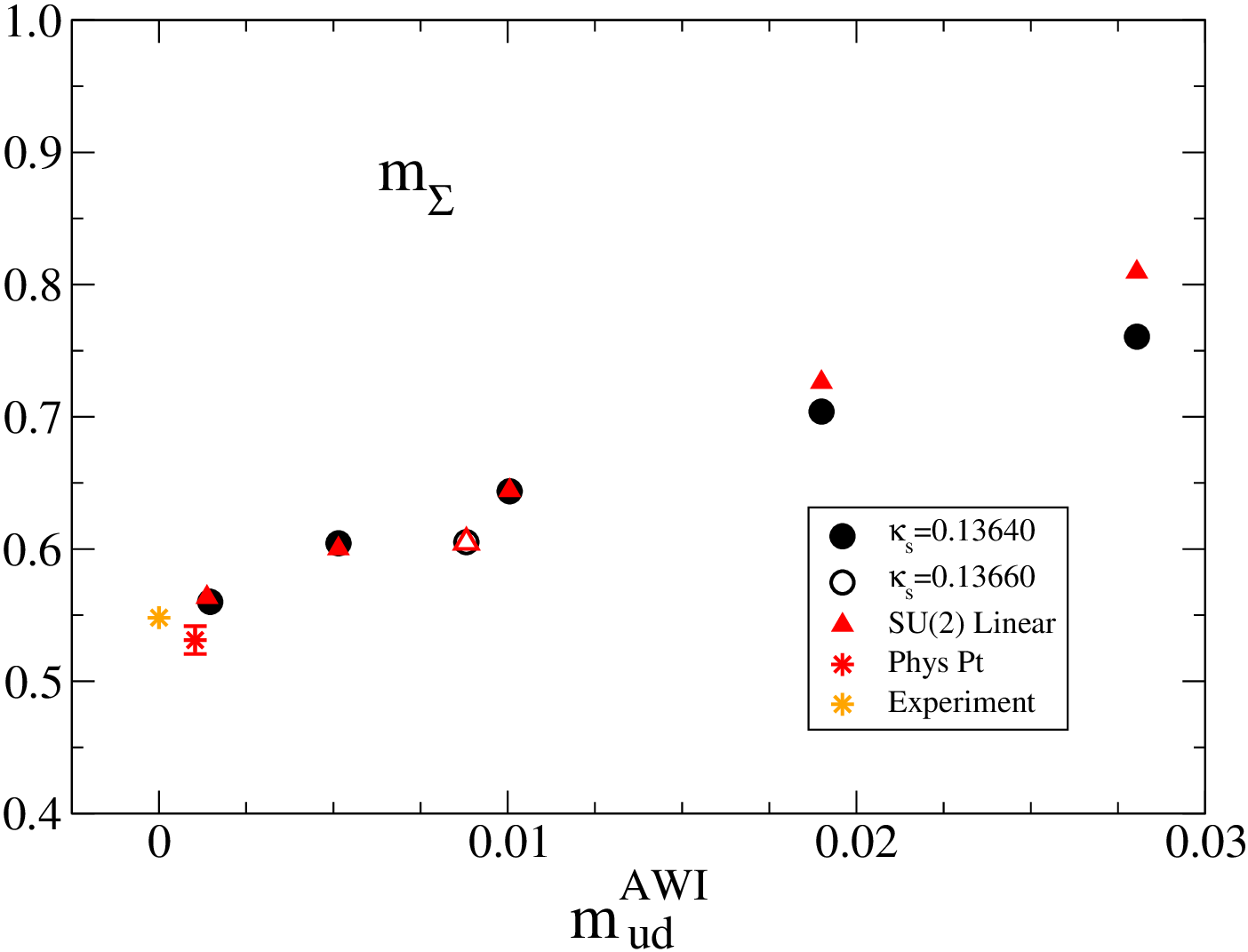}
\hspace*{2mm}&\hspace*{2mm}
\includegraphics[width=85mm,angle=0]{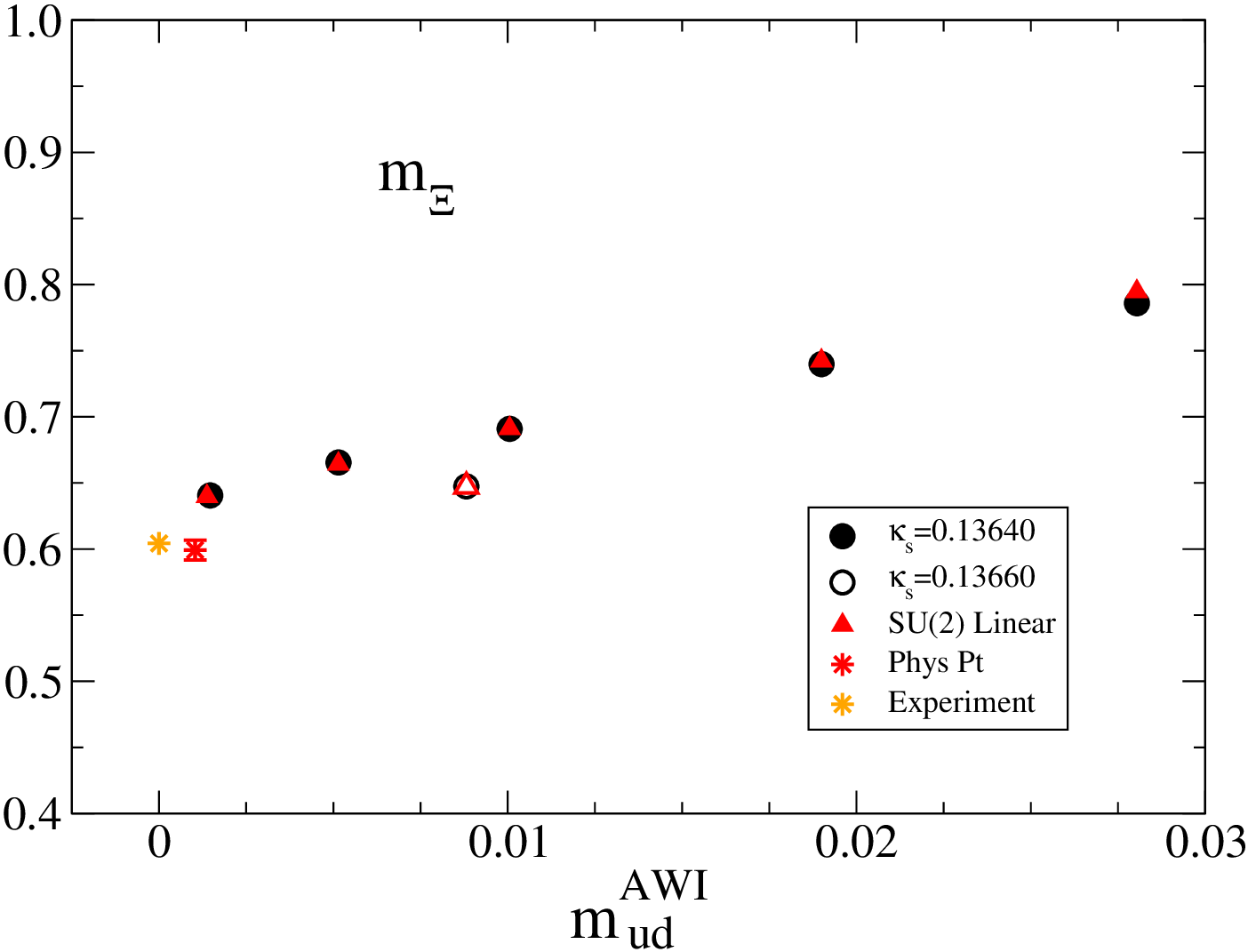}
\end{tabular}
\end{center}
\vspace{-.5cm}
\caption{\label{fig:linear_su2_o}Fit results with the linear 
formula Eq.~(\ref{eq:linear_su2}) for the octet baryon masses.
Experimental values are given in lattice units with $a^{-1}=2.176$ GeV
in Ref.~\cite{pacscs_nf3}.} 
\end{figure*}

\begin{figure*}[h]
\vspace{13mm}
\begin{center}
\begin{tabular}{cc}
\includegraphics[width=85mm,angle=0]{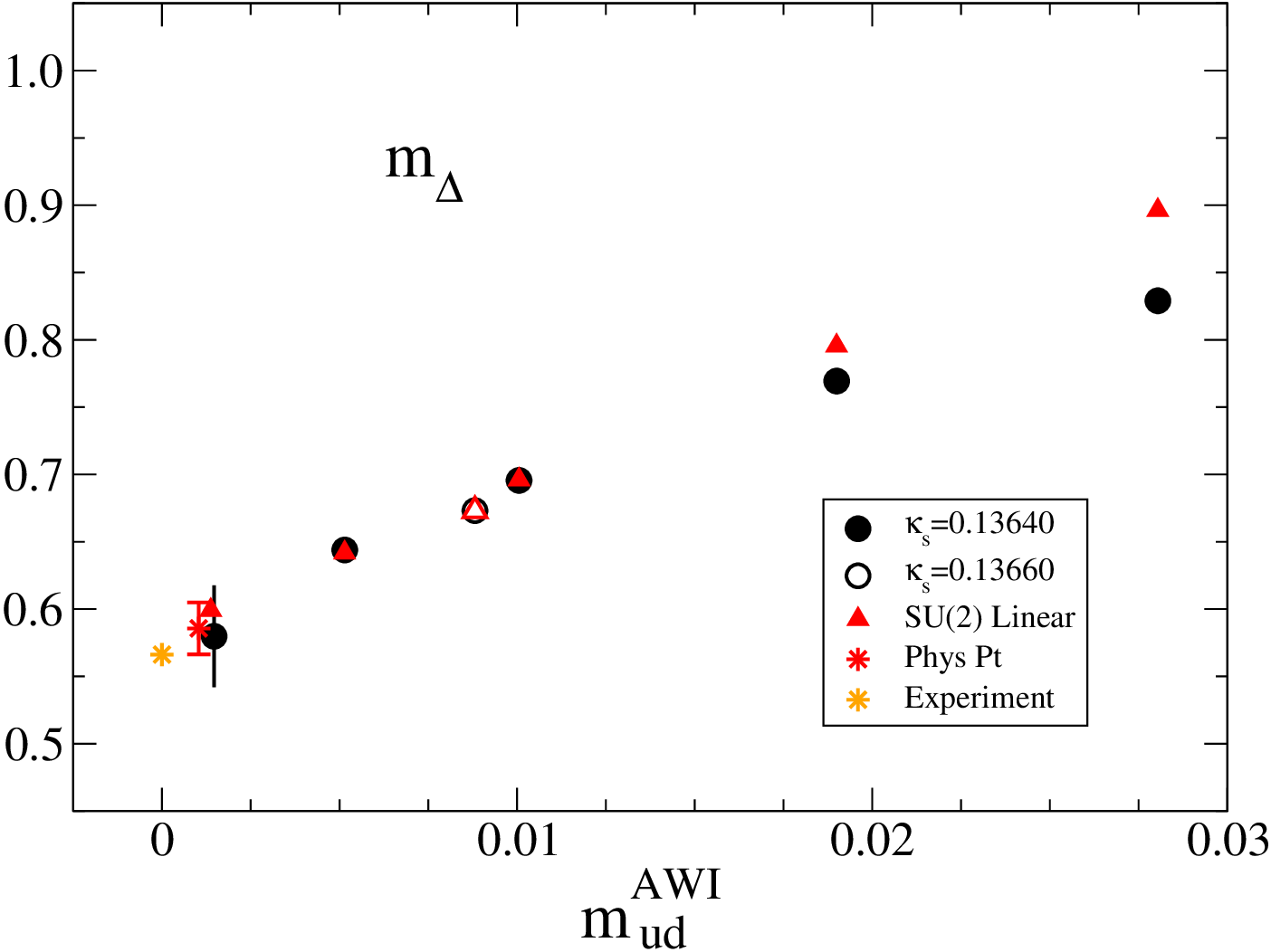}
\hspace*{2mm}&\hspace*{2mm}
\includegraphics[width=85mm,angle=0]{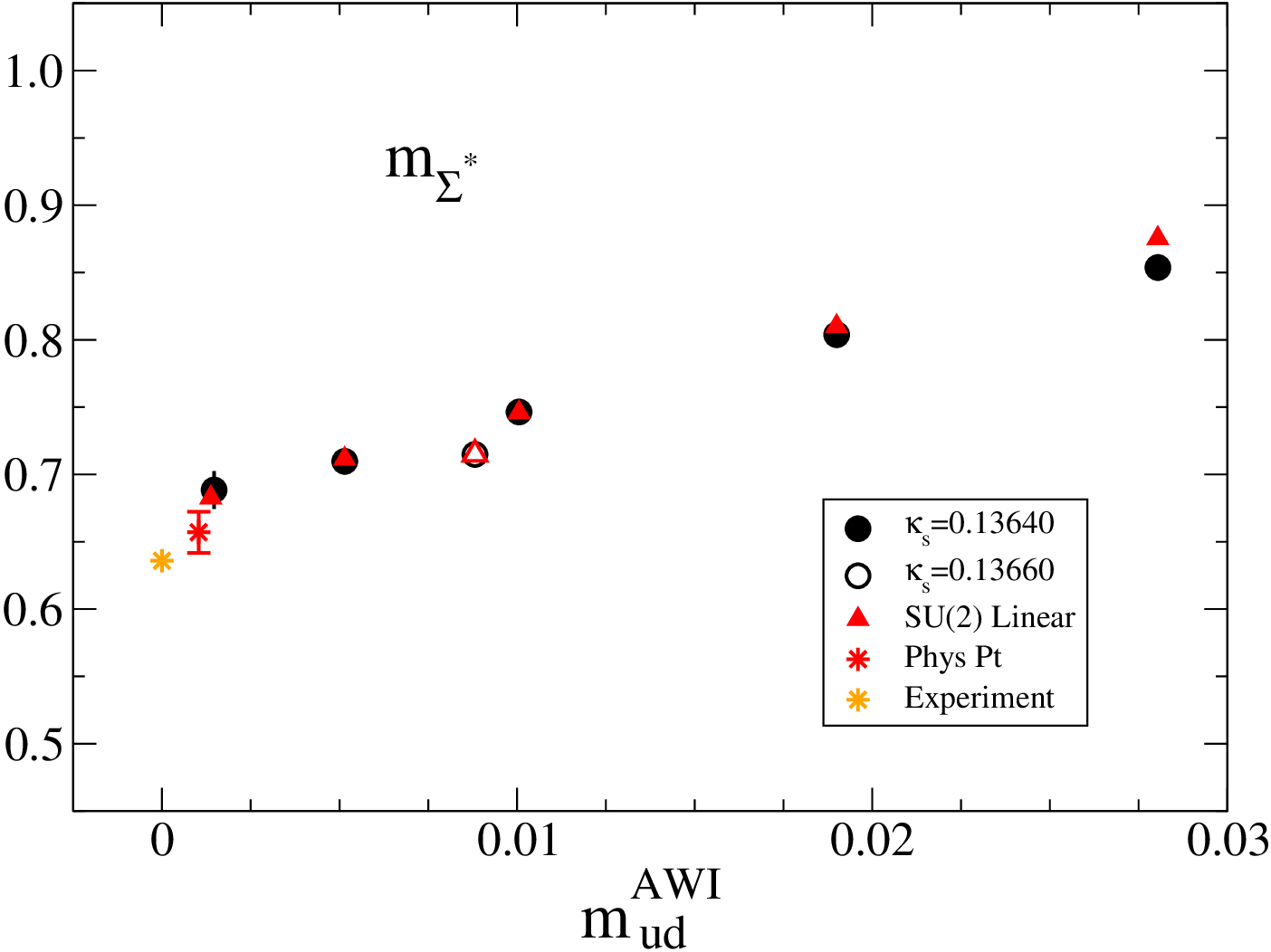}\\
\vspace*{7mm}& \\
\includegraphics[width=85mm,angle=0]{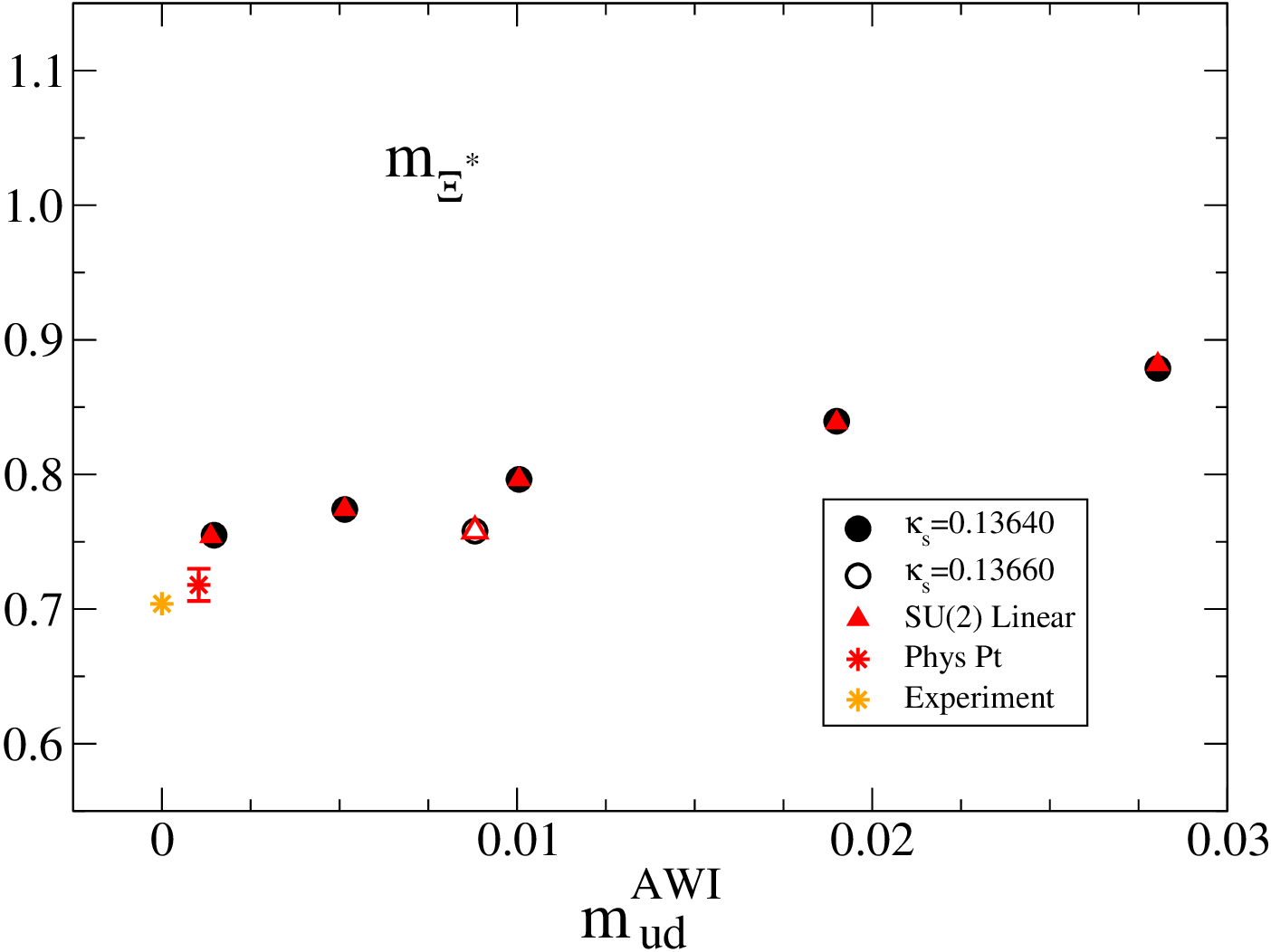}
\hspace*{2mm}&\hspace*{2mm}
\includegraphics[width=85mm,angle=0]{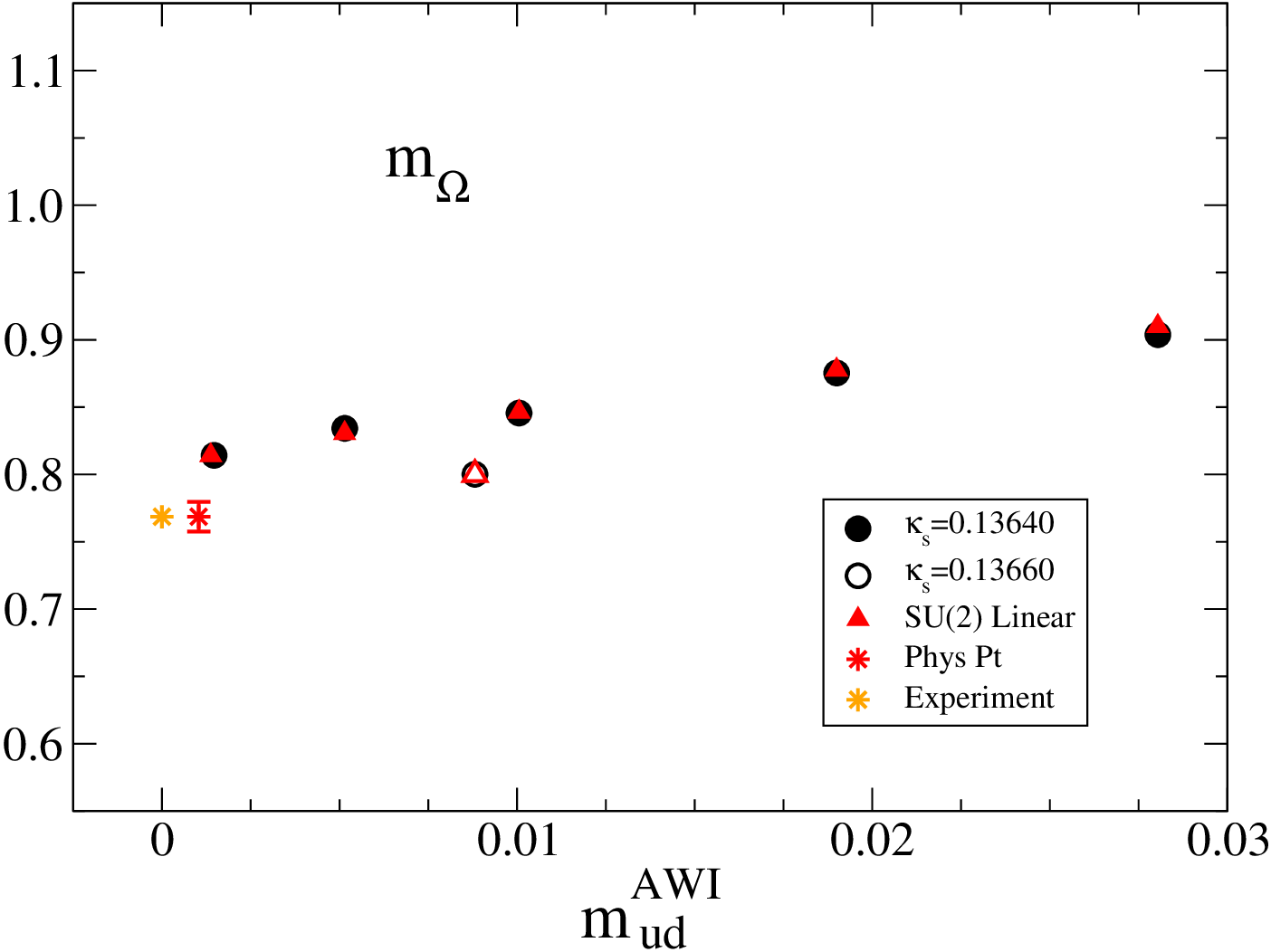}
\end{tabular}
\end{center}
\vspace{-.5cm}
\caption{\label{fig:linear_su2_d}Fit results with the linear 
formula Eq.~(\ref{eq:linear_su2}) for the decuplet baryon masses.
Experimental values are given in lattice units with $a^{-1}=2.176$ GeV
in Ref.~\cite{pacscs_nf3}.} 
\end{figure*}

\begin{figure*}[h]
\vspace{13mm}
\begin{center}
\begin{tabular}{cc}
\includegraphics[width=85mm,angle=0]{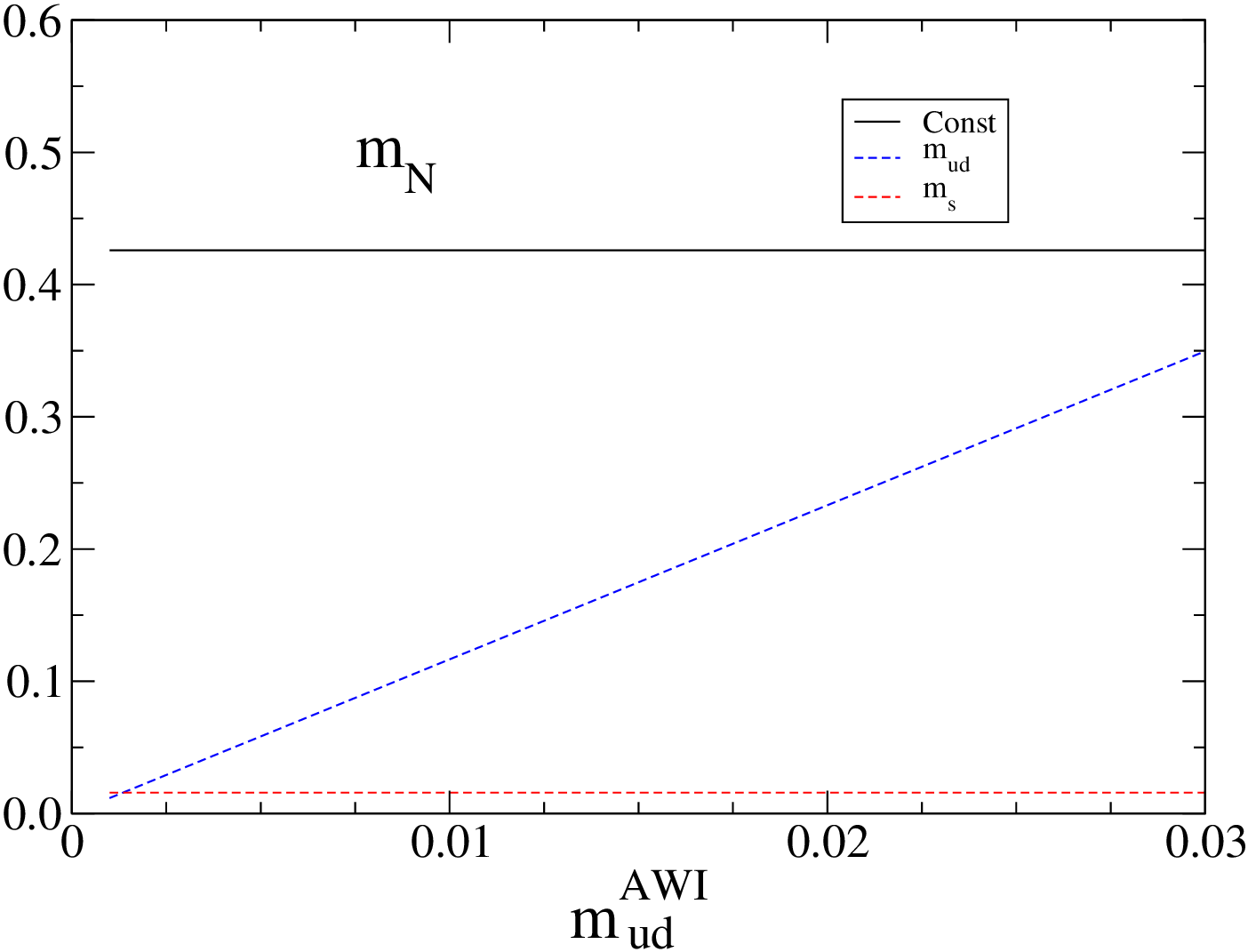}
\hspace*{2mm}&\hspace*{2mm}
\includegraphics[width=85mm,angle=0]{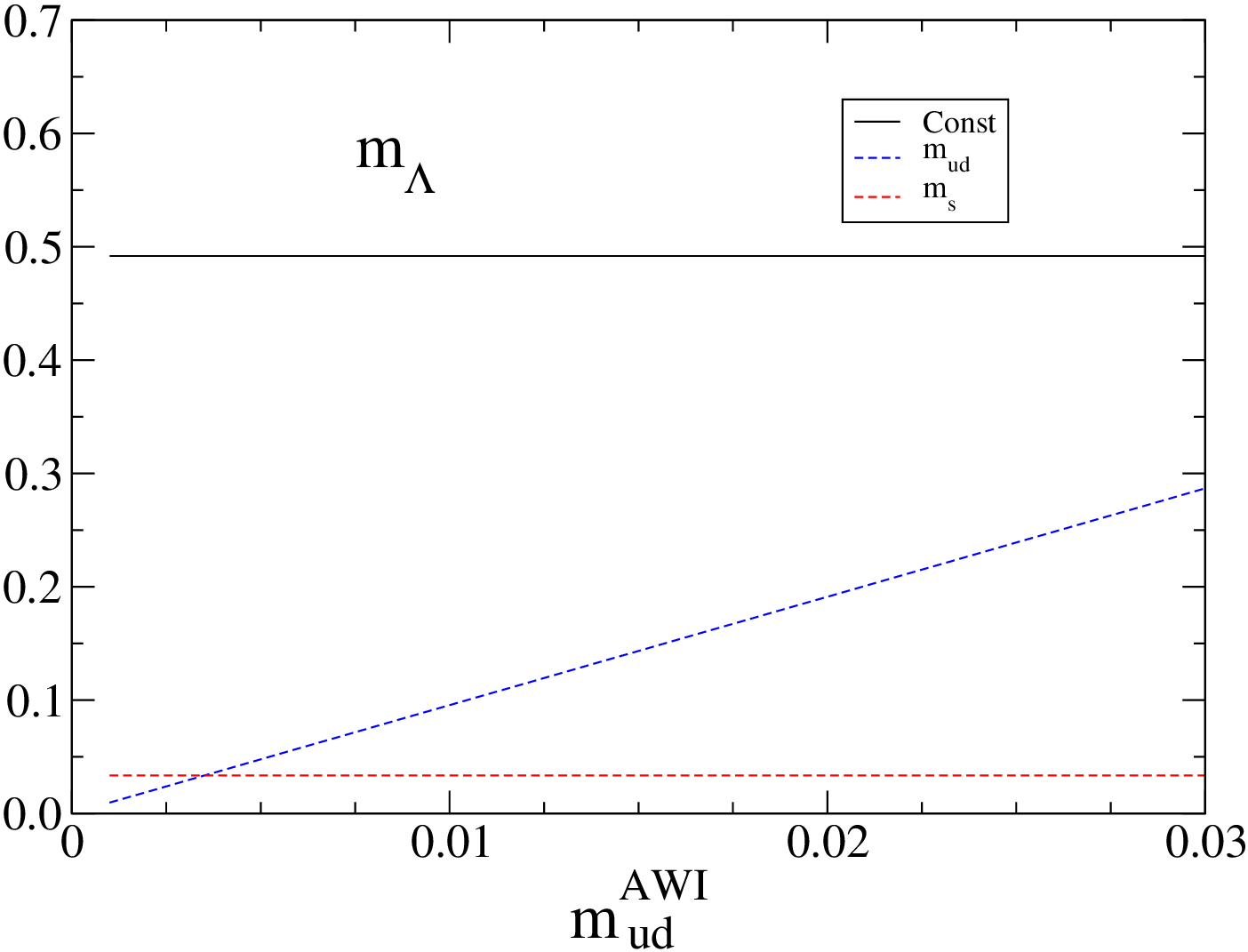}\\
\vspace*{7mm}& \\
\includegraphics[width=85mm,angle=0]{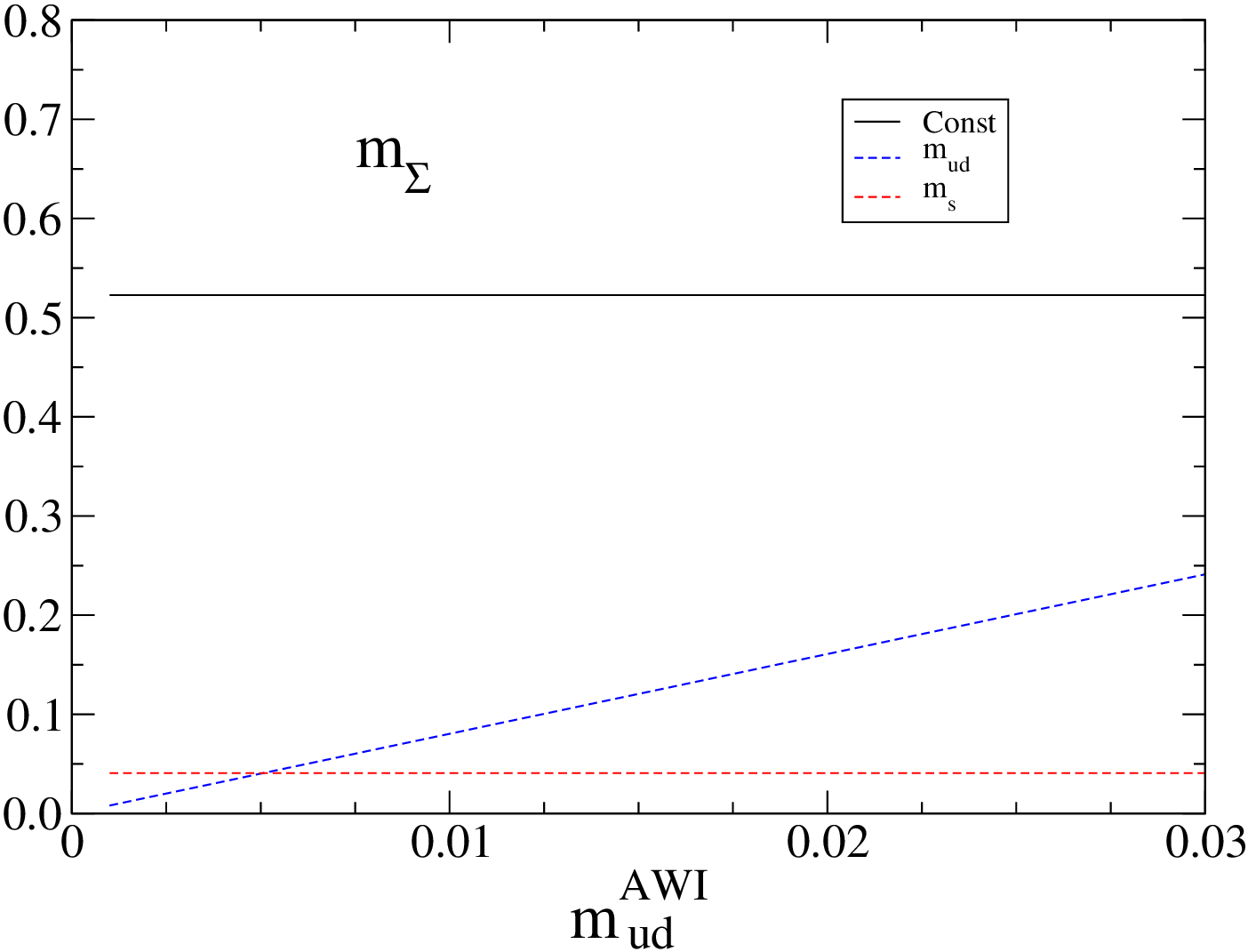}
\hspace*{2mm}&\hspace*{2mm}
\includegraphics[width=85mm,angle=0]{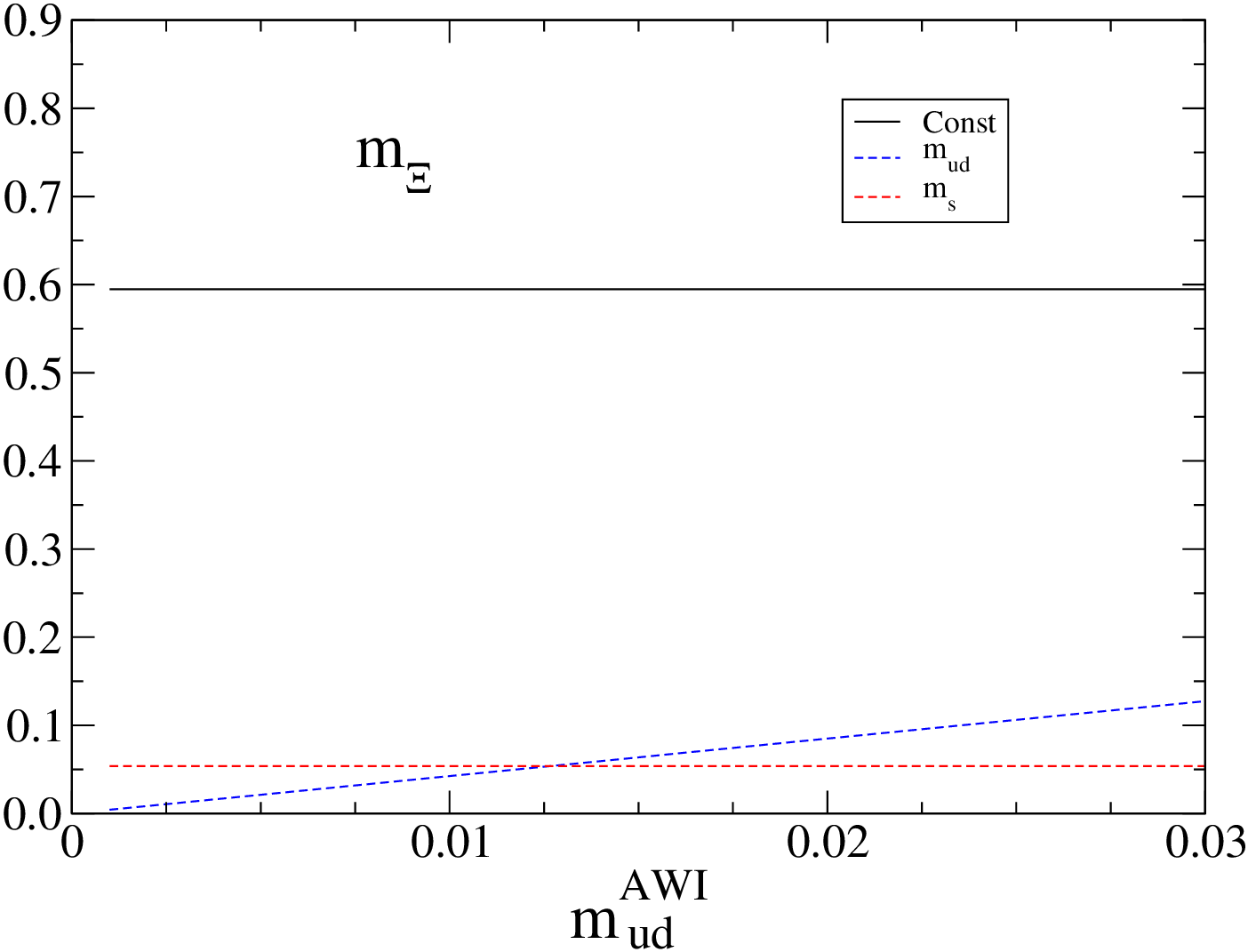}
\end{tabular}
\end{center}
\vspace{-.5cm}
\caption{\label{fig:linear_su2_conv_o}Convergence behavior for the
 octet baryon masses with the linear formula Eq.~(\ref{eq:linear_su2}).
$m_{\rm s}$ is fixed at the measured value 
at $(\kappa_{\rm ud},\kappa_{\rm s})=(0.13754,0.13640)$.}
\end{figure*}

\begin{figure*}[h]
\vspace{13mm}
\begin{center}
\begin{tabular}{cc}
\includegraphics[width=85mm,angle=0]{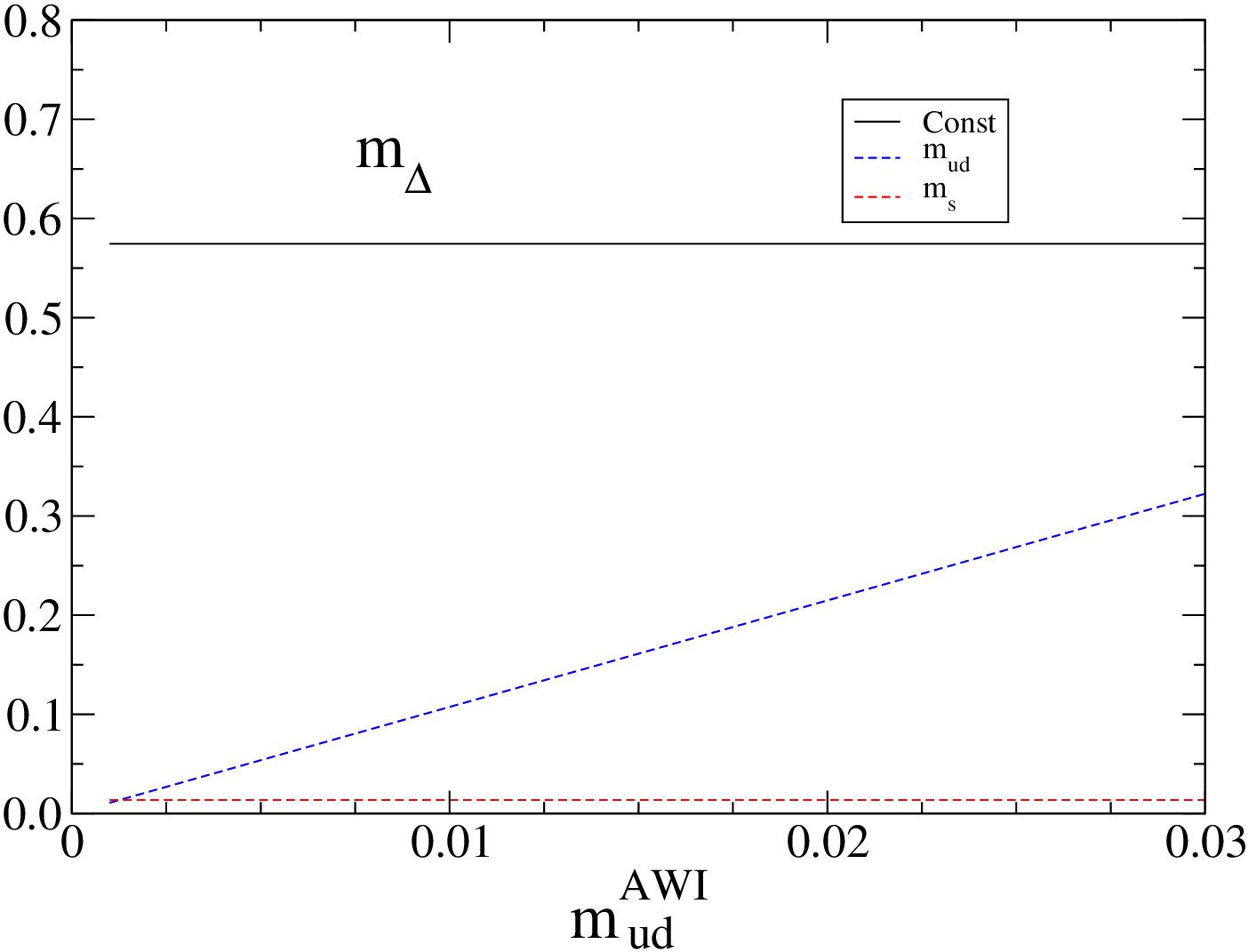}
\hspace*{2mm}&\hspace*{2mm}
\includegraphics[width=85mm,angle=0]{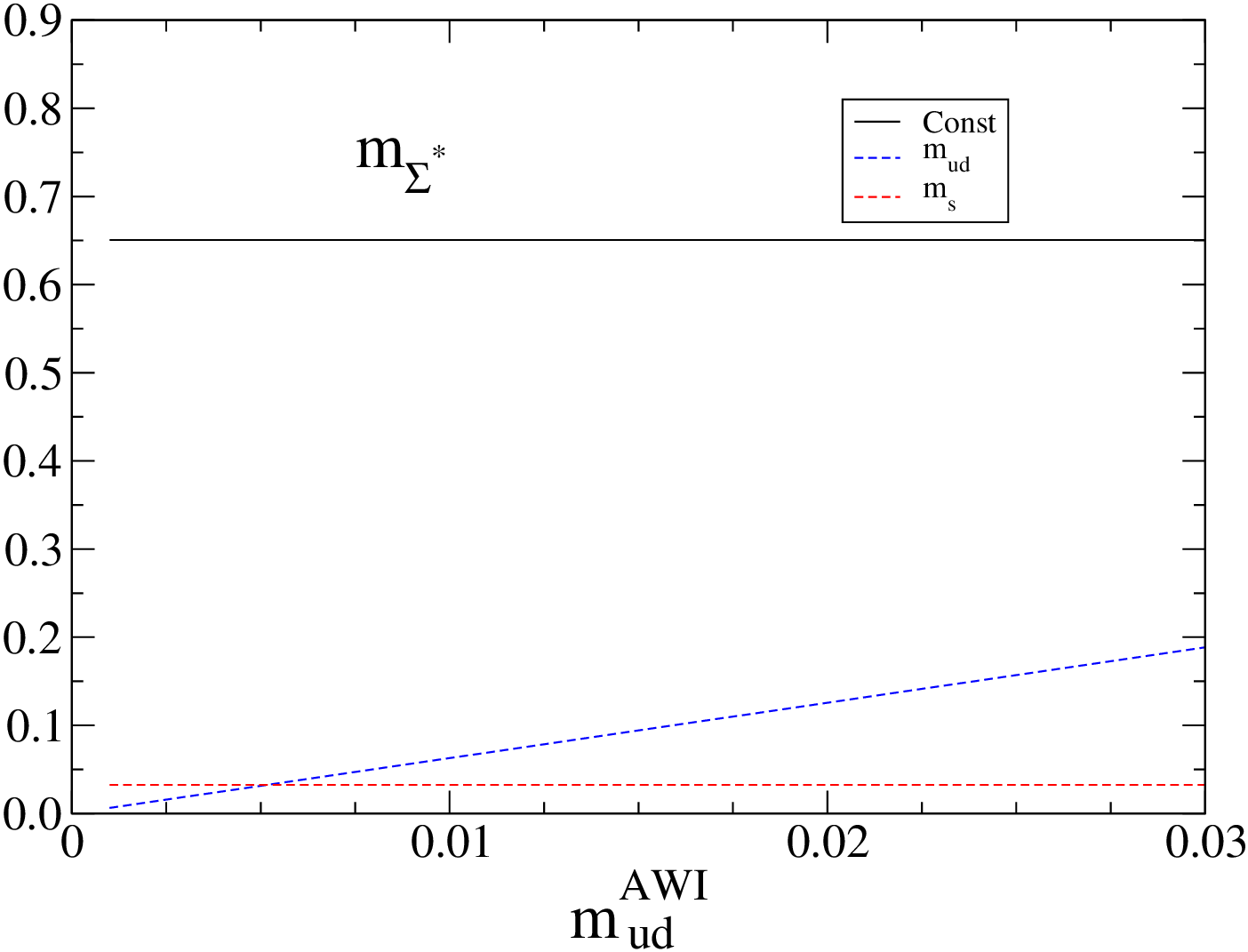}\\
\vspace*{7mm}& \\
\includegraphics[width=85mm,angle=0]{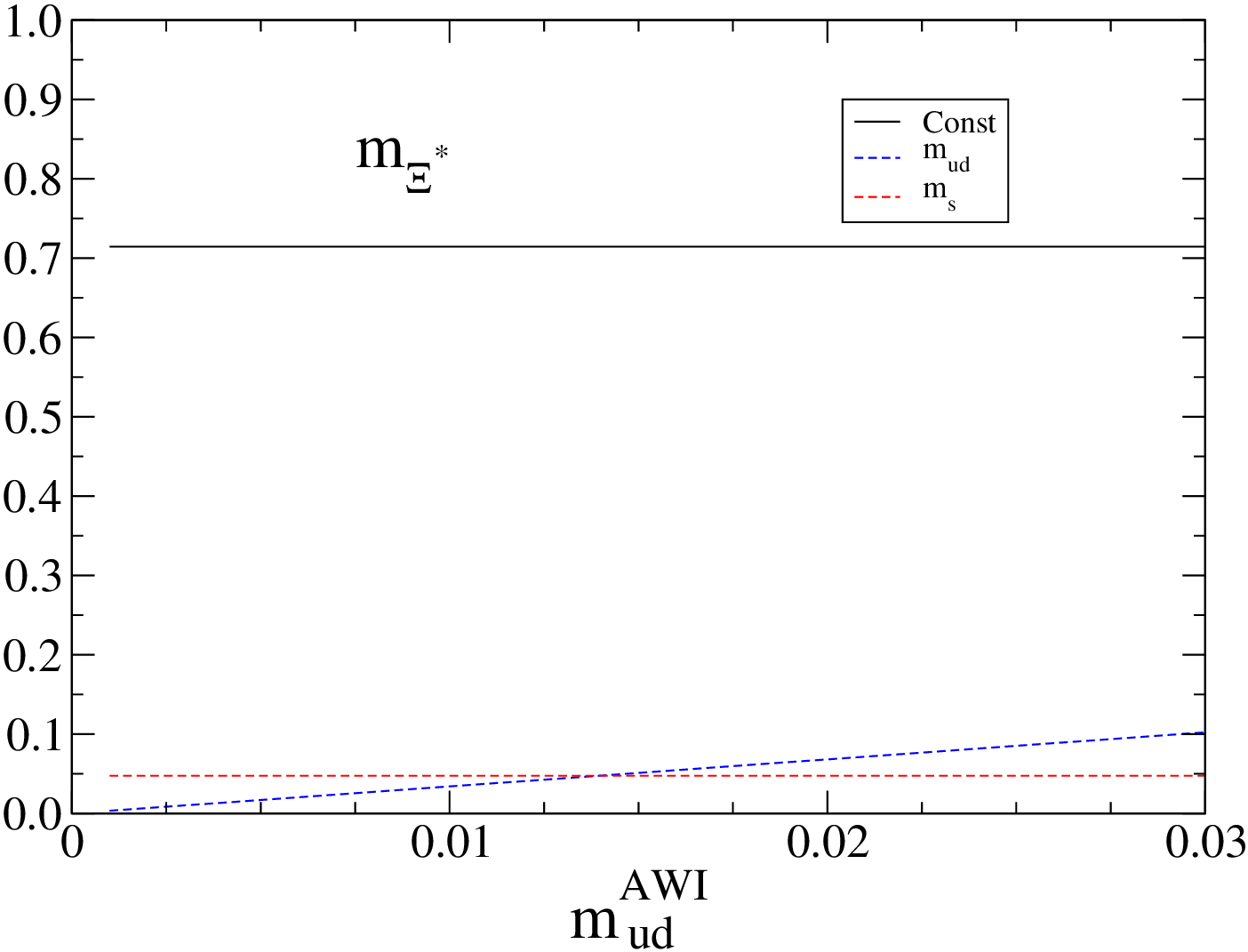}
\hspace*{2mm}&\hspace*{2mm}
\includegraphics[width=85mm,angle=0]{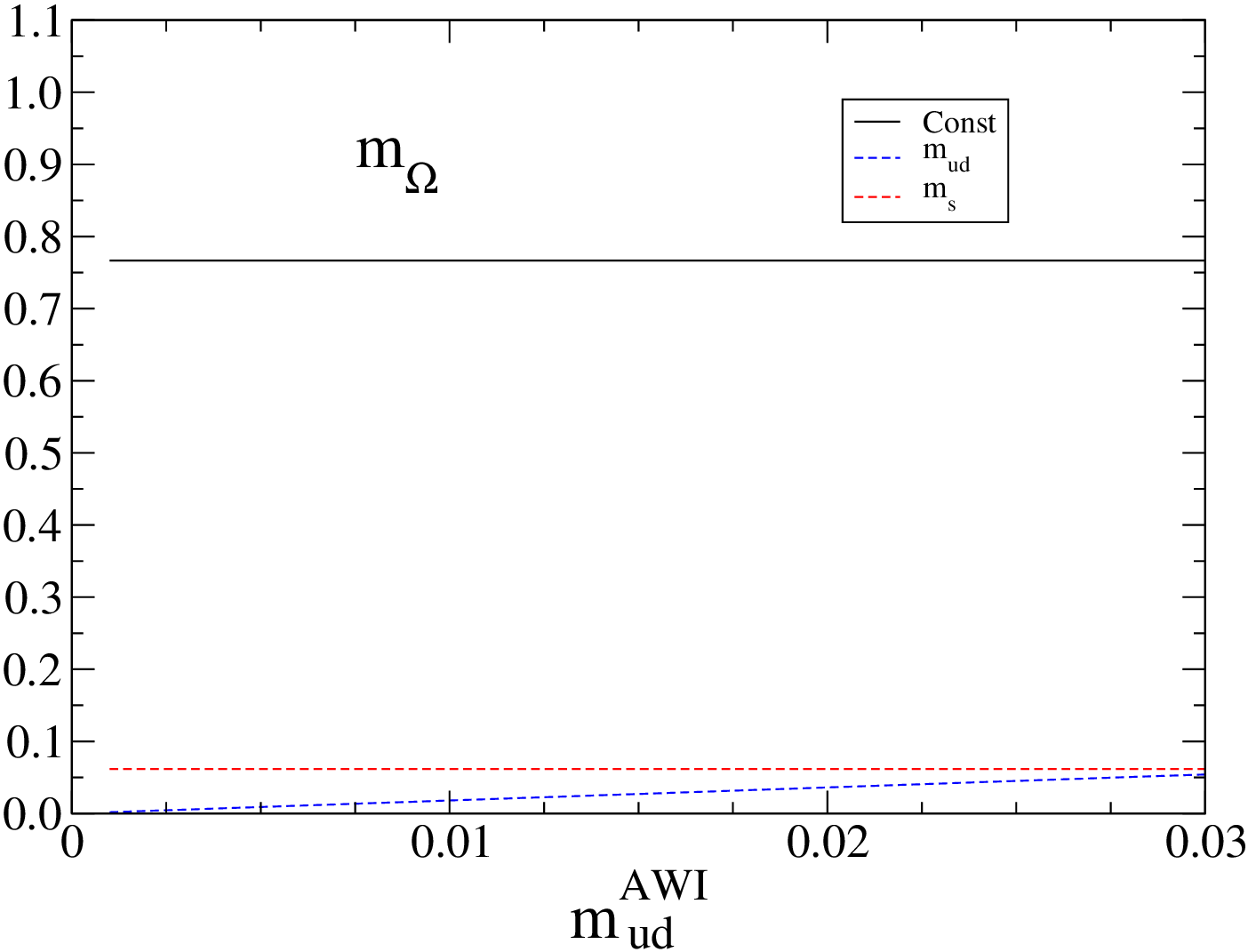}
\end{tabular}
\end{center}
\vspace{-.5cm}
\caption{\label{fig:linear_su2_conv_d}Convergence behavior for the
 decuplet baryon masses with the linear 
formula Eq.~(\ref{eq:linear_su2}). $m_{\rm s}$ is fixed at the measured value 
at $(\kappa_{\rm ud},\kappa_{\rm s})=(0.13754,0.13640)$.}
\end{figure*}

The simulation points and the measured hadron masses are given in 
Table I and Table III in Ref.~\cite{pacscs_nf3}, respectively.
The fit range is the lightest four points at 
$\kappa_{\rm ud}\ge 0.13754$, where 
the pion mass varies from 156 MeV to 410 MeV and 
$m_{\rm ud}^{\rm AWI}$ from 3 MeV to 24 MeV in the $\msbar$ scheme.
In Figs.~\ref{fig:linear_su2_o} and \ref{fig:linear_su2_d} 
we present the fit results 
for the octet and the decuplet baryon masses, respectively.
Star symbols denote the extrapolated values at the physical point
whose numerical values 
are listed in Tables~\ref{tab:m_o_ph} and \ref{tab:m_d_ph}.
The physical point together with the lattice cutoff is 
determined with $m_\pi,m_K,m_\Omega$ inputs by applying
the NLO SU(2) ChPT fit to the PS meson sector\cite{pacscs_nf3}.
The data are reasonably described by the linear function Eq.~(\ref{eq:linear_su2}).
The values for $\alpha,\beta,\gamma$ and $\chi^2/{\rm dof}$
are summarized in Tables \ref{tab:linear_su2_o} and \ref{tab:linear_su2_d}.
In oder to investigate the convergence behavior 
the contribution of each term in Eq.~(\ref{eq:linear_su2_conv}) is drawn in 
Fig.~\ref{fig:linear_su2_conv_o} for the octet baryon
masses, where $m_{\rm s}^{\rm AWI}$ is fixed at the measured value 
for $(\kappa_{\rm ud},\kappa_{\rm s})=(0.13754,0.13640)$. 
We observe that the $O(m_{\rm s}-m_{\rm s, ph})$ terms are small, and that  
$O(m_{\rm ud})$ contributions are 
less than 20\% of that of $\alpha^\prime$ for $m_{\rm ud}^{\rm AWI}\simlt 0.01$.
This suggests that higher order terms in $m_{\rm ud}$ and 
$m_{\rm s}-m_{\rm s, ph}$ would be small. 
Similar trends are found for the decuplet baryon masses 
in Fig.~\ref{fig:linear_su2_conv_d}.

\section{SU(3) HBChPT analyses on baryon masses}
\label{sec:hbchpt}

\subsection{Lagrangian to leading order}
\label{subsec:lagrangian}

We use the continuum HBChPT in this article, leaving 
the development of the Wilson HBChPT, 
which incorporates the chiral
symmetry breaking effects of the Wilson-type quark action,
for future work. 
We follow the notation described in Ref.~\cite{walker-loud}.
The lagrangian of HBChPT is written in terms of
velocity-dependent baryon fields with a perturbative derivative
expansion.
To fix the notation we present the leading order terms:
\begin{widetext}
\ben
{\cal L}&=&\left({\bar B}iv\cdot DB\right)+2\alpha_M\left({\bar B}B{\cal
M}_+\right)+2\beta_M\left({\bar B}{\cal M}_+B\right)
+2\sigma_M\left({\bar B}B\right)\tr\left({\cal M}_+\right)\nn\\
&&-\left({\bar T}^\mu\left\{iv\cdot D-\Delta\right\}T_\mu\right)
+2\gamma_M\left({\bar T}^\mu {\cal M}_+ T_\mu\right)
-2{\bar \sigma}_M\left({\bar T}^\mu T_\mu\right)\tr\left({\cal
M}_+\right)\nn\\
&&+2\alpha \left({\bar B}S^\mu B A_\mu \right)+
2\beta \left({\bar B}S^\mu  A_\mu B\right)
+2{\cal H}\left({\bar T}^\nu S^\mu A_\mu T_\nu\right)\nn\\
&&+\sqrt{\frac{3}{2}}{\cal C}\left[\left({\bar T}^\nu A_\nu B\right)
+\left({\bar B}A_\nu T^\nu\right)\right],
\label{eq:lagrangian}
\een
\end{widetext}
where $B$ and $T$ represent the velocity-dependent octet and decuplet
baryon fields with the four-velocity $v_\mu$, respectively.
This lagrangian contains 9 LECs: $\alpha_M,\beta_M,\sigma_M,\gamma_M,{\bar
\sigma}_M,\alpha,\beta,{\cal H},{\cal C}$. Hereafter we use the axial 
couplings $F, D$ instead of $\alpha, \beta$. They are related with
\ben
\alpha&=&\frac{2}{3}D+2F,\\
\beta&=&-\frac{5}{3}D+F.
\een
The decuplet-octet mass difference denoted by $\Delta=m_T-m_B$ 
has a comparable magnitude to inter-multiplet mass splittings 
of both the octet and the decuplet. We are allowed to treat it as 
a perturbation.
The octet pseudoscalar mesons couple derivatively to the baryon fields
through the vector combination $V_\mu$, the axial vector one $A_\mu$ and
the chiral covariant derivative $D_\mu$. The light quark masses 
$m_{\rm u},m_{\rm d},m_{\rm s}$ are contained in ${\cal M}_+$.
We refer to Ref.~\cite{walker-loud} for the explicit expression 
of $V_\mu,A_\mu,D_\mu,{\cal M}_+$.

In this section the physical $m_{\rm ud}$ and $m_{\rm s}$ are determined 
by the SU(3) ChPT analyses in the pseudoscalar meson sector 
including $m_\pi$ and $m_K$; $m_\Omega$ is an additional input for 
the lattice cutoff\cite{pacscs_nf3}.

\subsection{Octet baryons}
\label{subsec:octet_su3}

In the SU(3) HBChPT the renormalized mass 
of the $i$-th octet baryon is expanded as
\ben
 m_{B_i}=m_{B}-m_{B_i}^{(1)}-m_{B_i}^{(3/2)}+\cdots,
\een
where $m_B$ is the octet baryon mass in the chiral limit 
under the SU(3) flavor symmetry, and $m_{B_i}^{(n)}$ is the $O(m_q^{(n)})$
contribution to the $i$-th octet baryon.
The LO corrections are given by
\ben
 m_{N}^{(1)} &=& (2\alpha_M+2\beta_M+4\sigma_M)m_{\rm ud} +2\sigma_M
 m_{\rm s}, \\
  m_{\Lambda}^{(1)} &=& (\alpha_M+2\beta_M+4\sigma_M)m_{\rm ud}
  +(\alpha_M+2\sigma_M) m_{\rm s}, \\
  m_{\Sigma}^{(1)} &=&
  \left(\frac{5}{3}\alpha_M+\frac{2}{3}\beta_M+4\sigma_M\right)m_{\rm ud} \nonumber \\
             &&
	     +\left(\frac{1}{3}\alpha_M+\frac{4}{3}\beta_M+2\sigma_M\right) m_{\rm s}, \\
    m_{\Xi}^{(1)} &=&
    \left(\frac{1}{3}\alpha_M+\frac{4}{3}\beta_M+4\sigma_M\right)m_{\rm ud} \nonumber \\
    && +\left(\frac{5}{3}\alpha_M+\frac{2}{3}\beta_M+2\sigma_M\right)
    m_{\rm s}
\een
with $m_{\rm ud}$ the averaged up-down quark mass, $m_{\rm s}$
the strange quark mass, and 
$\alpha_M, \beta_M, \sigma_M$ are LECs in the lagrangian (\ref{eq:lagrangian}).

The NLO contributions, which are obtained from one-loop diagrams, 
are written as
\begin{widetext}
\ben
m_{B_i}^{(3/2)}  &=&  \frac{2}{(4\pi f_0)^2}
\sum_{\phi=\pi,K,\eta}\left[A_\phi^{B_i} {\cal F}(m_{\phi},0,\mu)
     + {\cal C}^2 B_\phi^{B_i} {\cal F}(m_{\phi},\Delta,\mu) \right],
\een  
\end{widetext}
where $\Delta$ is the octet-decuplet baryon mass difference in the
SU(3) chiral limit. The pion decay constant $f_0$ is also defined in the
SU(3) chiral limit with the convention of $f_\pi=130.4$ MeV. 
The function ${\cal F}$ is expressed as
\begin{widetext}
\ben
{\cal F}({m_{\phi},\Delta,\mu})&=&(m_{\phi}^2-\Delta^2) 
\left[ \sqrt{\Delta^2-m_{\phi}^2+i\epsilon} 
  \ln\left(\frac{\Delta-\sqrt{\Delta^2-m_{\phi}^2+i\epsilon}}{\Delta+\sqrt{\Delta^2-m_{\phi}^2+i\epsilon}}\right) 
  -\Delta\ln\left(\frac{m_{\phi}^2}{\mu^2}\right) \right] 
  -\frac{1}{2}  \Delta
 m_{\phi}^2\ln\left(\frac{m_{\phi}^2}{\mu^2}\right)
\label{eq:func_f}
\een
\end{widetext}
with ${\cal F}({m_{\phi},0,\mu})=\pi m_\phi^3$, and $\mu$ the
renormalization scale. 
This formula assumes $\Delta\ge m_\phi$. 
For $\Delta< m_\phi$ we apply the analytic continuation: 
\ben
 \sqrt{\Delta^2-m_{\phi}^2+i\epsilon} 
  \ln\left(\frac{\Delta-\sqrt{\Delta^2-m_{\phi}^2+i\epsilon}}{\Delta+\sqrt{\Delta^2-m_{\phi}^2+i\epsilon}}\right) \nonumber \\
\rightarrow \arccos\left(\frac{\Delta}{m_\phi}\right).
\een
We revisit the function ${\cal F}$
later in Sec.~\ref{sec:fse} to discuss finite size effects.
The contributions from the octet-octet- and the
decuplet-octet-axial couplings are factored out by
$A_{\phi}^{B_i}$ and ${\cal C}^2 B_\phi^{B_i}$, respectively.
We summarize their values in Table~\ref{tab:ac_octet}.
The LECs are phenomenologically estimated as\cite{hsueh,pdg,c_ph}
\ben
 D=0.80, \; \; \; \; F=0.47, \; \; \; \; {\cal C}=1.5.
\label{eq:dfc_ph}
\een

\begin{figure*}[h]
\vspace{13mm}
\begin{center}
\begin{tabular}{cc}
\includegraphics[width=85mm,angle=0]{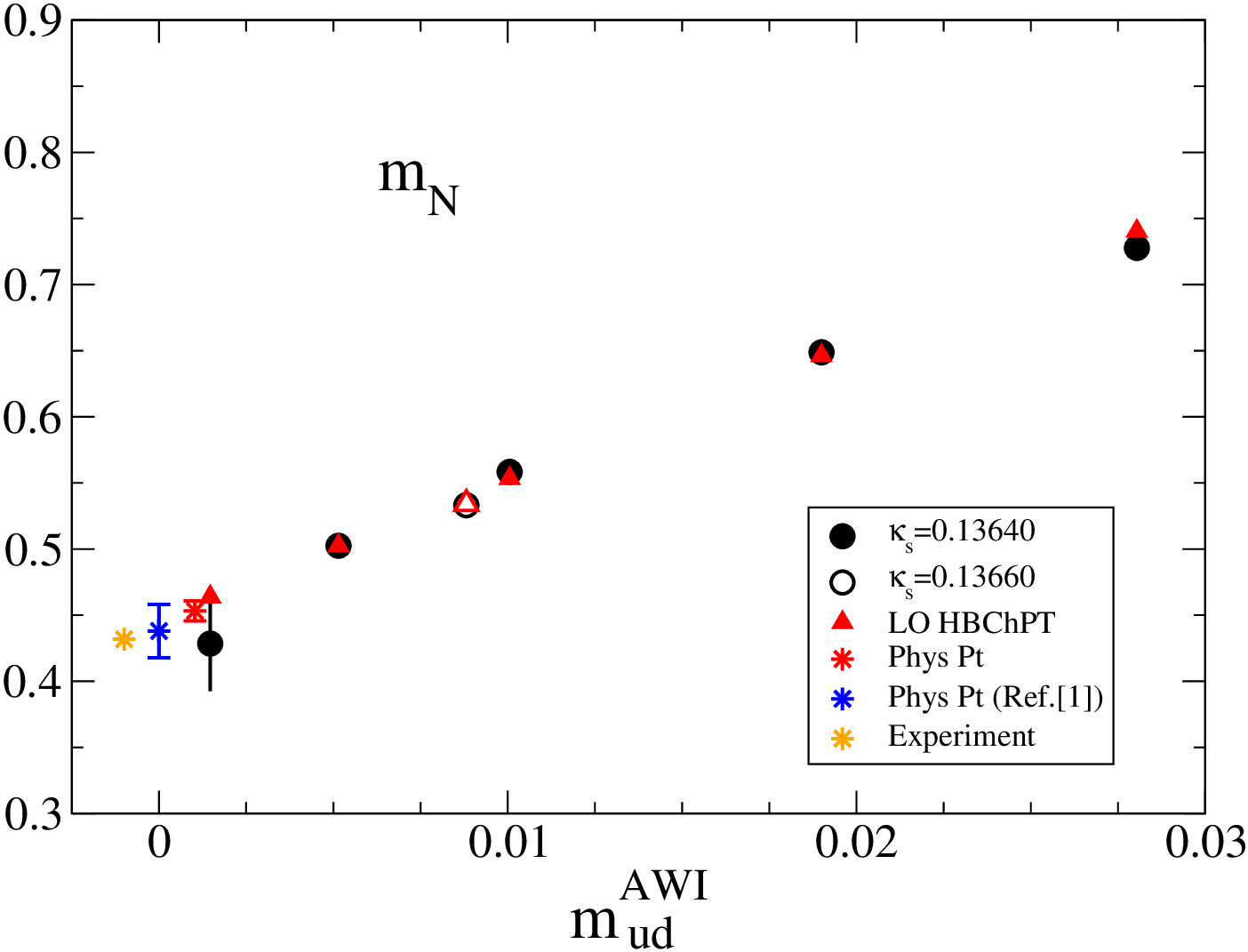}
\hspace*{2mm}&\hspace*{2mm}
\includegraphics[width=85mm,angle=0]{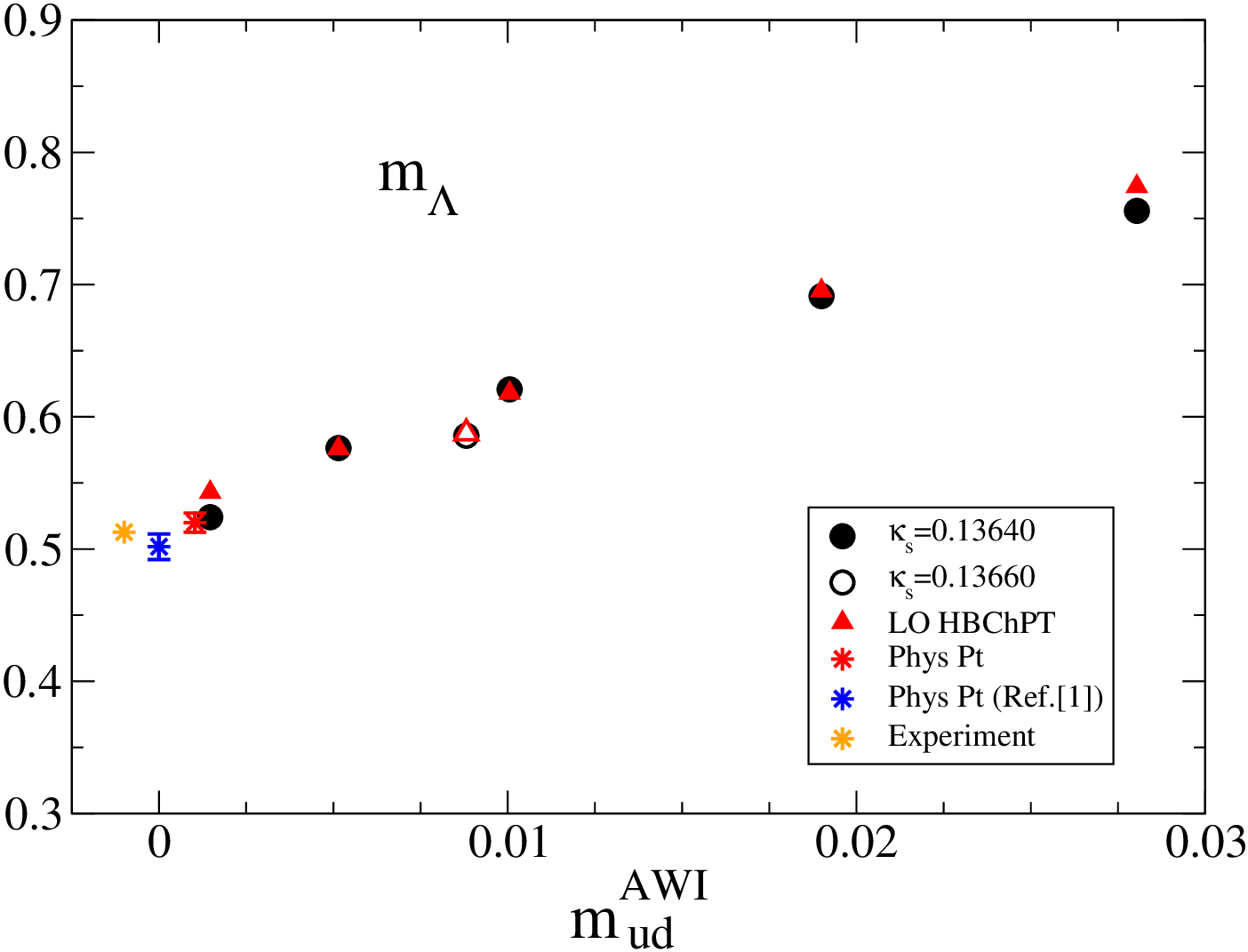}\\
\vspace*{7mm}& \\
\includegraphics[width=85mm,angle=0]{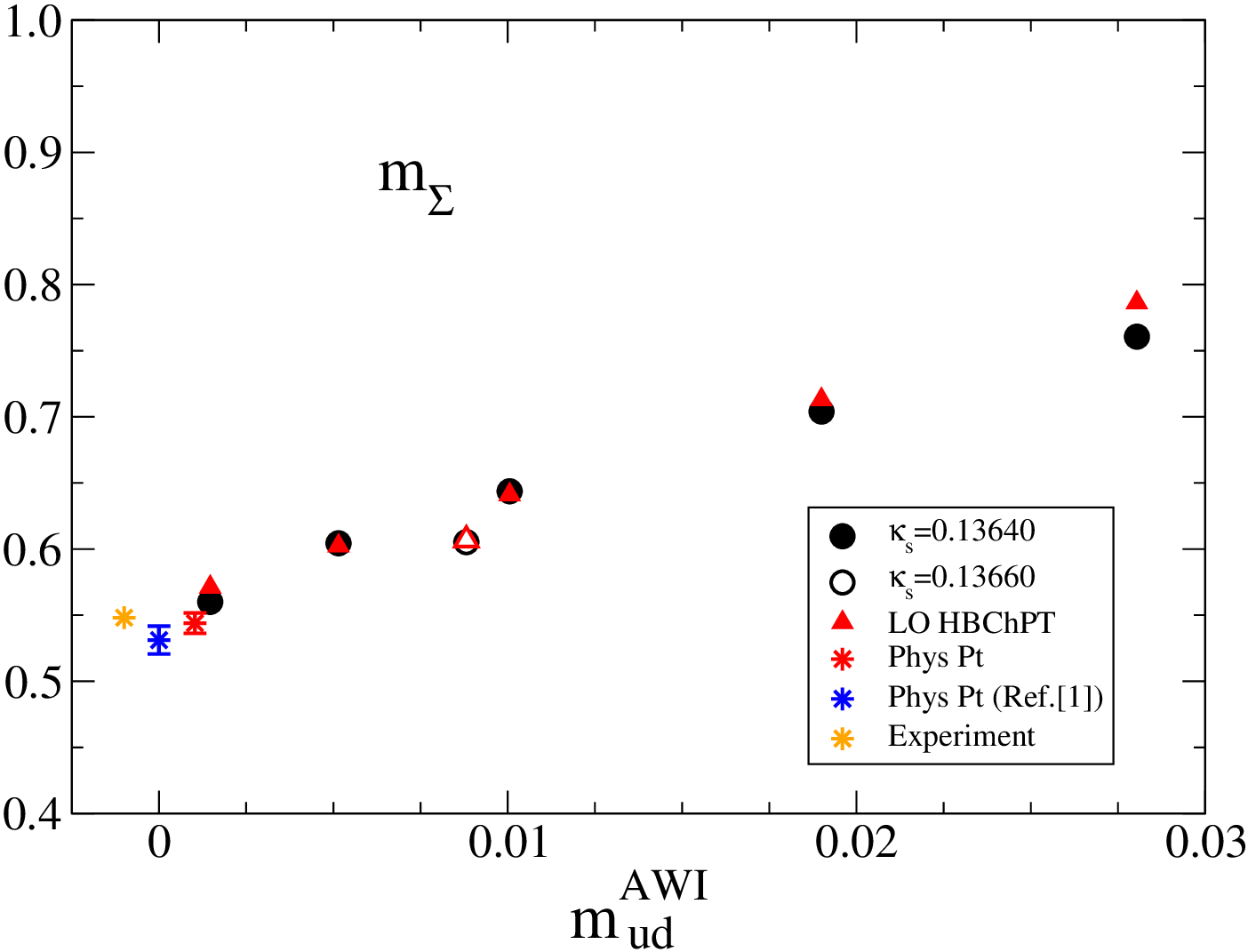}
\hspace*{2mm}&\hspace*{2mm}
\includegraphics[width=85mm,angle=0]{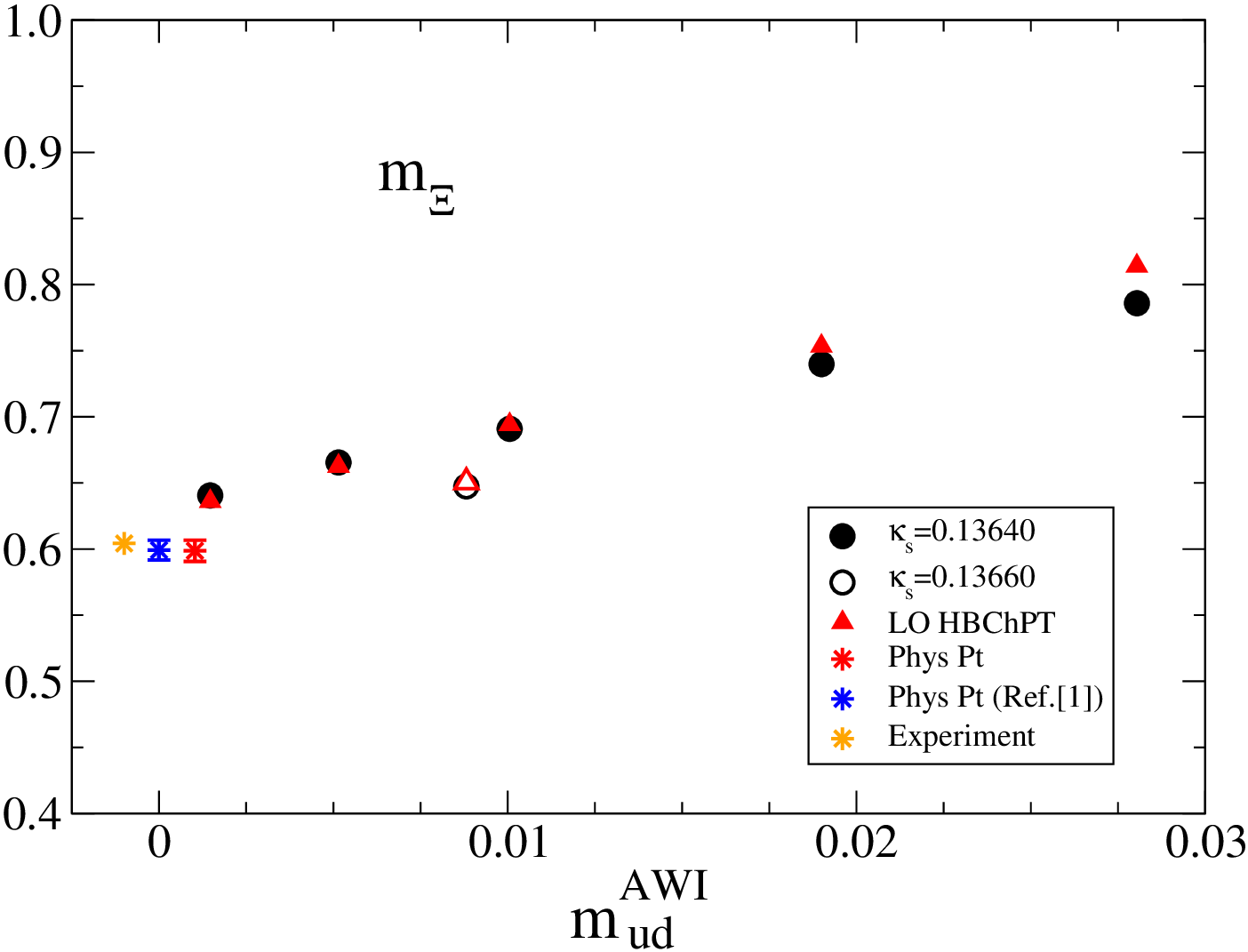}
\end{tabular}
\end{center}
\vspace{-.5cm}
\caption{\label{fig:hbchpt_su3_lo_o}Fit results with the SU(3) HBChPT 
up to LO for the octet baryon masses.
Experimental values are given in lattice units with $a^{-1}=2.176$ GeV
in Ref.~\cite{pacscs_nf3}.} 
\end{figure*}

\begin{figure*}[h]
\vspace{13mm}
\begin{center}
\begin{tabular}{cc}
\includegraphics[width=85mm,angle=0]{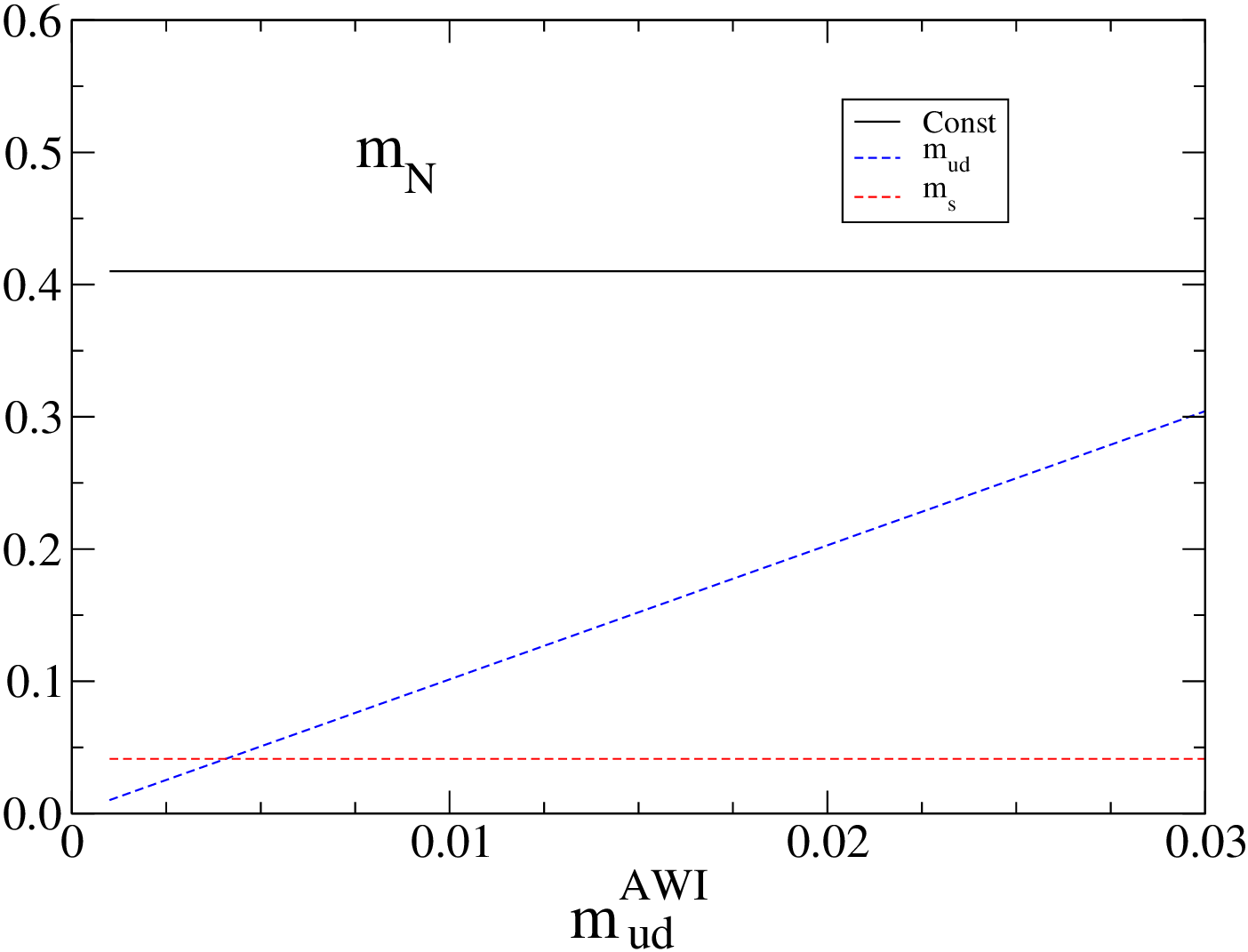}
\hspace*{2mm}&\hspace*{2mm}
\includegraphics[width=85mm,angle=0]{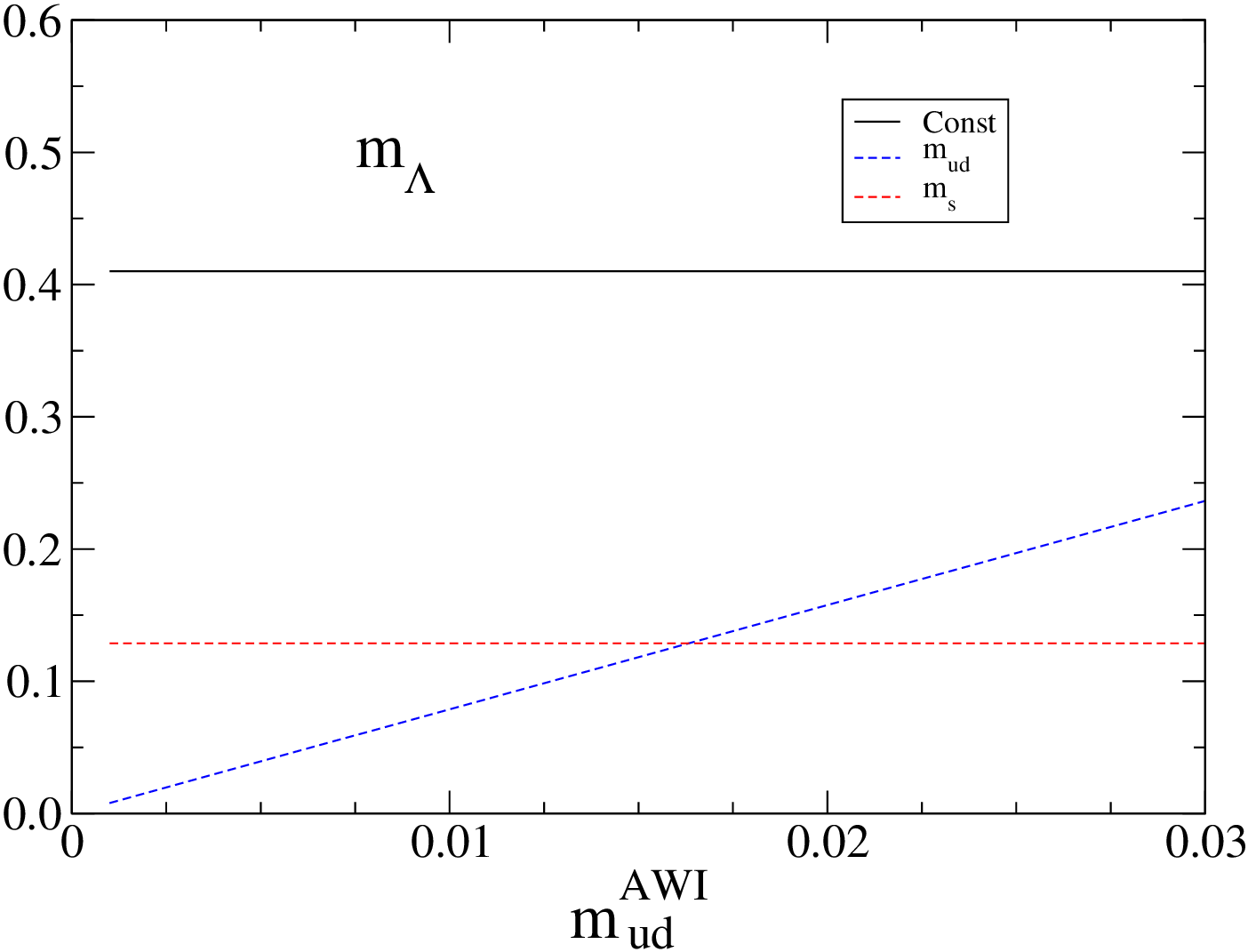}\\
\vspace*{7mm}& \\
\includegraphics[width=85mm,angle=0]{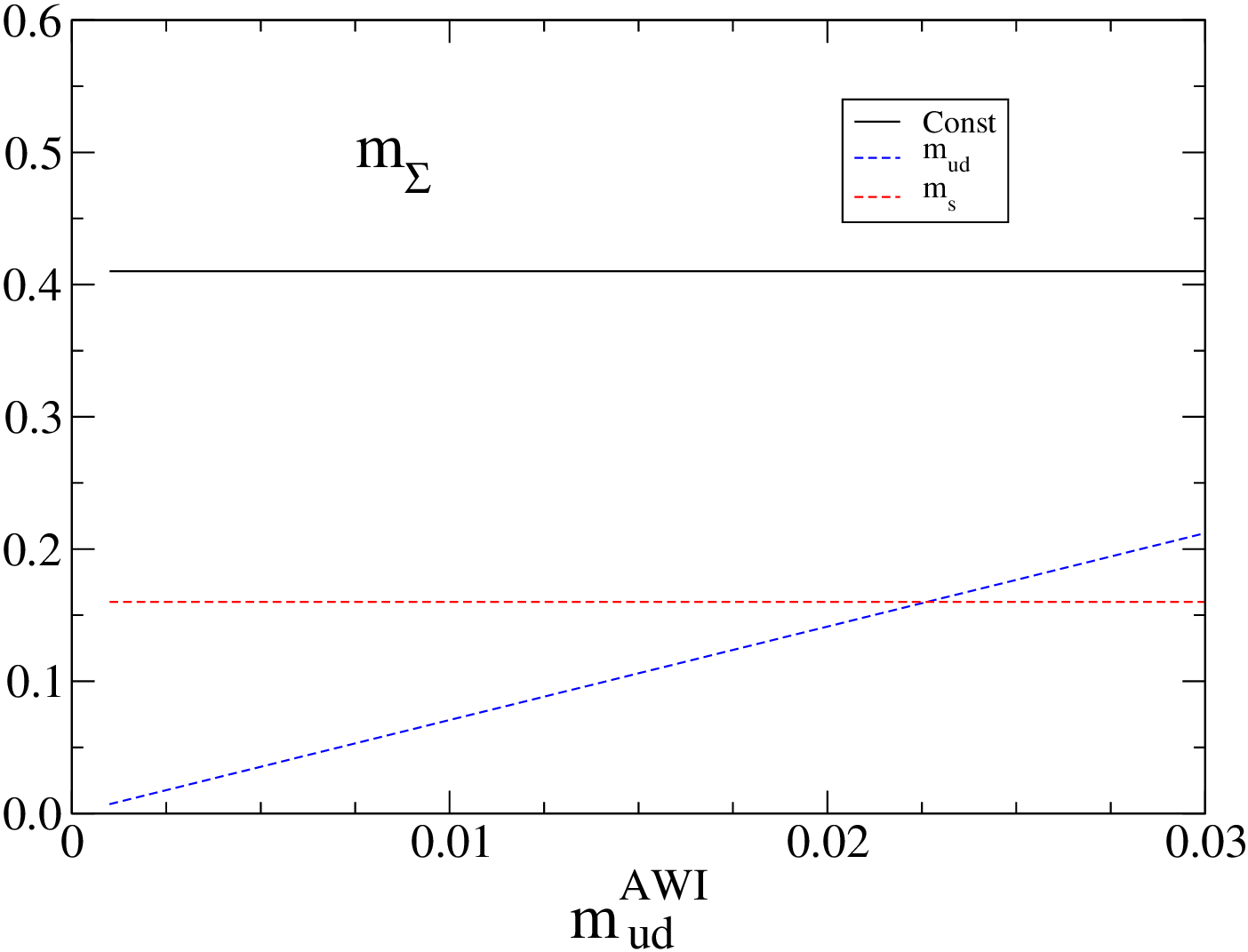}
\hspace*{2mm}&\hspace*{2mm}
\includegraphics[width=85mm,angle=0]{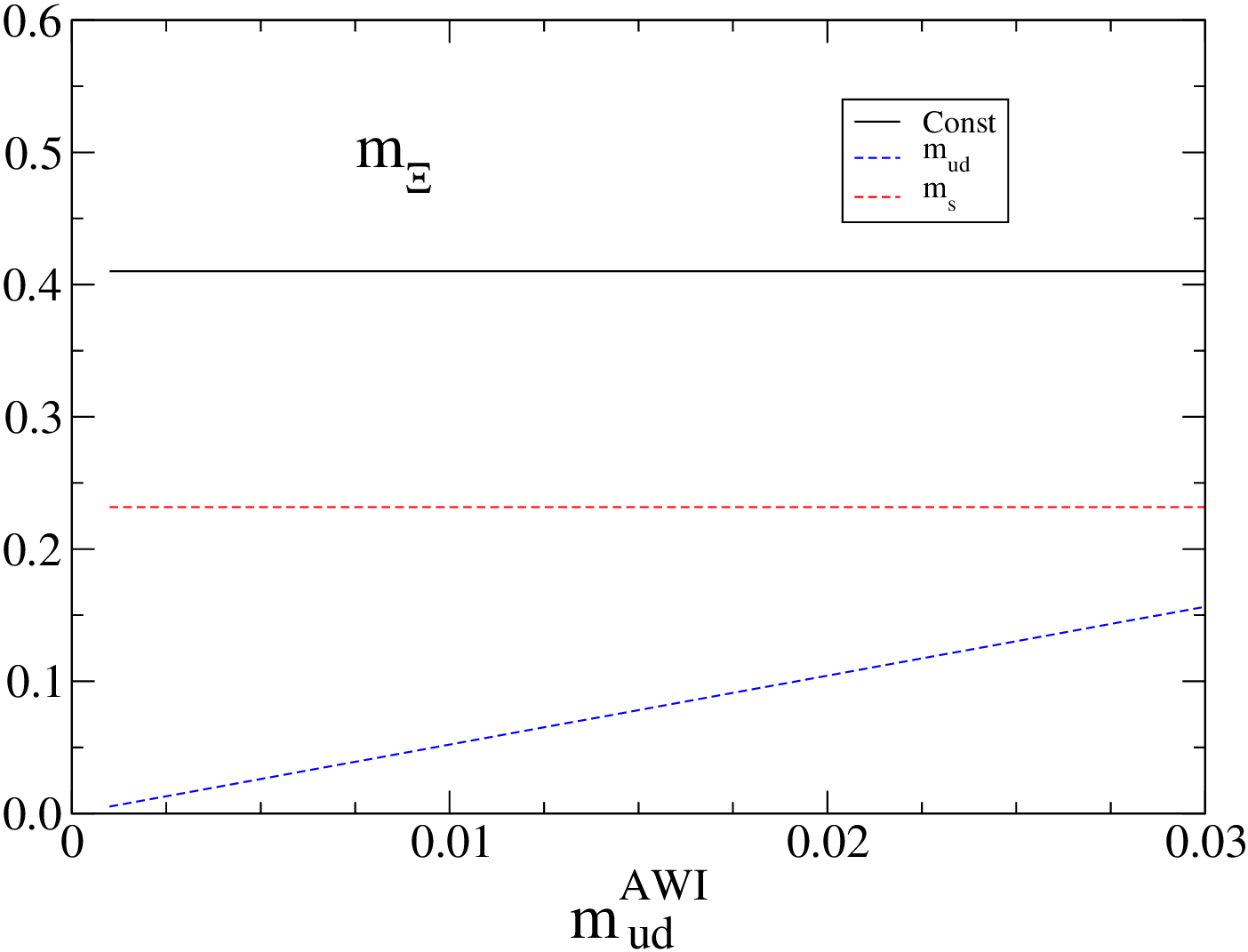}
\end{tabular}
\end{center}
\vspace{-.5cm}
\caption{\label{fig:hbchpt_su3_lo_conv_o}Convergence behavior for the
 octet masses with the SU(3) HBChPT up to LO. 
$m_{\rm s}$ is fixed with the measured value 
at $(\kappa_{\rm ud},\kappa_{\rm s})=(0.13754,0.13640)$.}
\end{figure*}

We first present the fit results up to leading order employing
the formula $m_{B_i}=m_{B}-m_{B_i}^{(1)}$. 
The values for 
$m_B,\alpha_M,\beta_M,\sigma_M$ are given in Table~\ref{tab:hbchpt_su3_o}\ 
together with $\chi^2$/dof. We also present the extrapolated values 
at the physical point in Table~\ref{tab:m_o_ph} in comparison with
the results of the SU(2) linear fit Eq.~(\ref{eq:linear_su2}). 
The two set of results are consistent within 2$\sigma$ error.
Figure~\ref{fig:hbchpt_su3_lo_o} shows that the data are reasonably 
described by the formula. 
The convergence behavior in Fig.~\ref{fig:hbchpt_su3_lo_conv_o}, however, is 
disappointing because of the sizable contribution of the $O(m_{\rm s})$ term.
This point is in a sharp contrast to the SU(2) linear expansion 
of Sec.~\ref{sec:linear}, where
the contributions of the $O(m_{\rm ud},m_{\rm s}-m_{\rm s,ph})$ 
terms are well controlled in the range of $m_{\rm ud}\simlt 0.01$.

\begin{figure*}[h]
\vspace{13mm}
\begin{center}
\begin{tabular}{cc}
\includegraphics[width=85mm,angle=0]{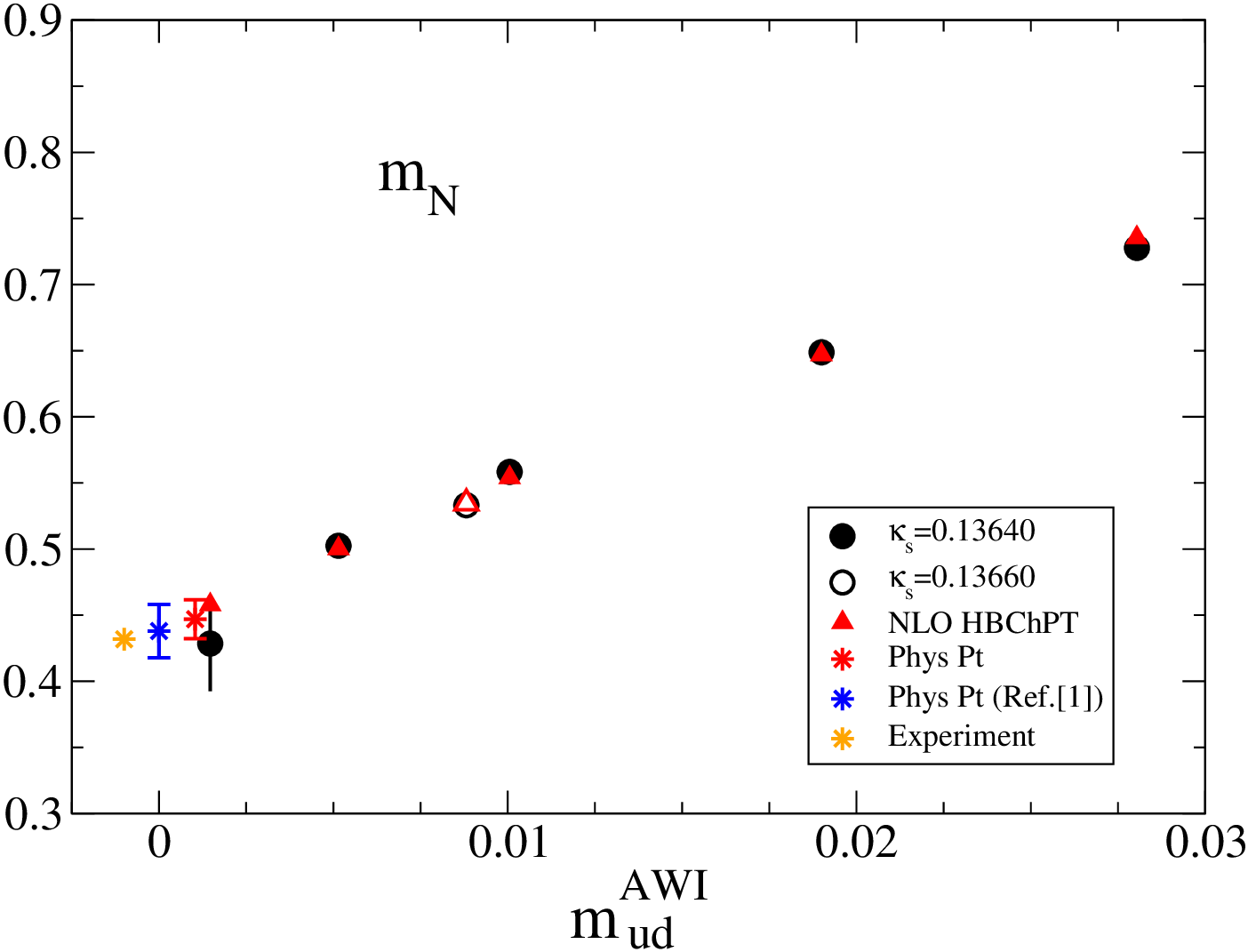}
\hspace*{2mm}&\hspace*{2mm}
\includegraphics[width=85mm,angle=0]{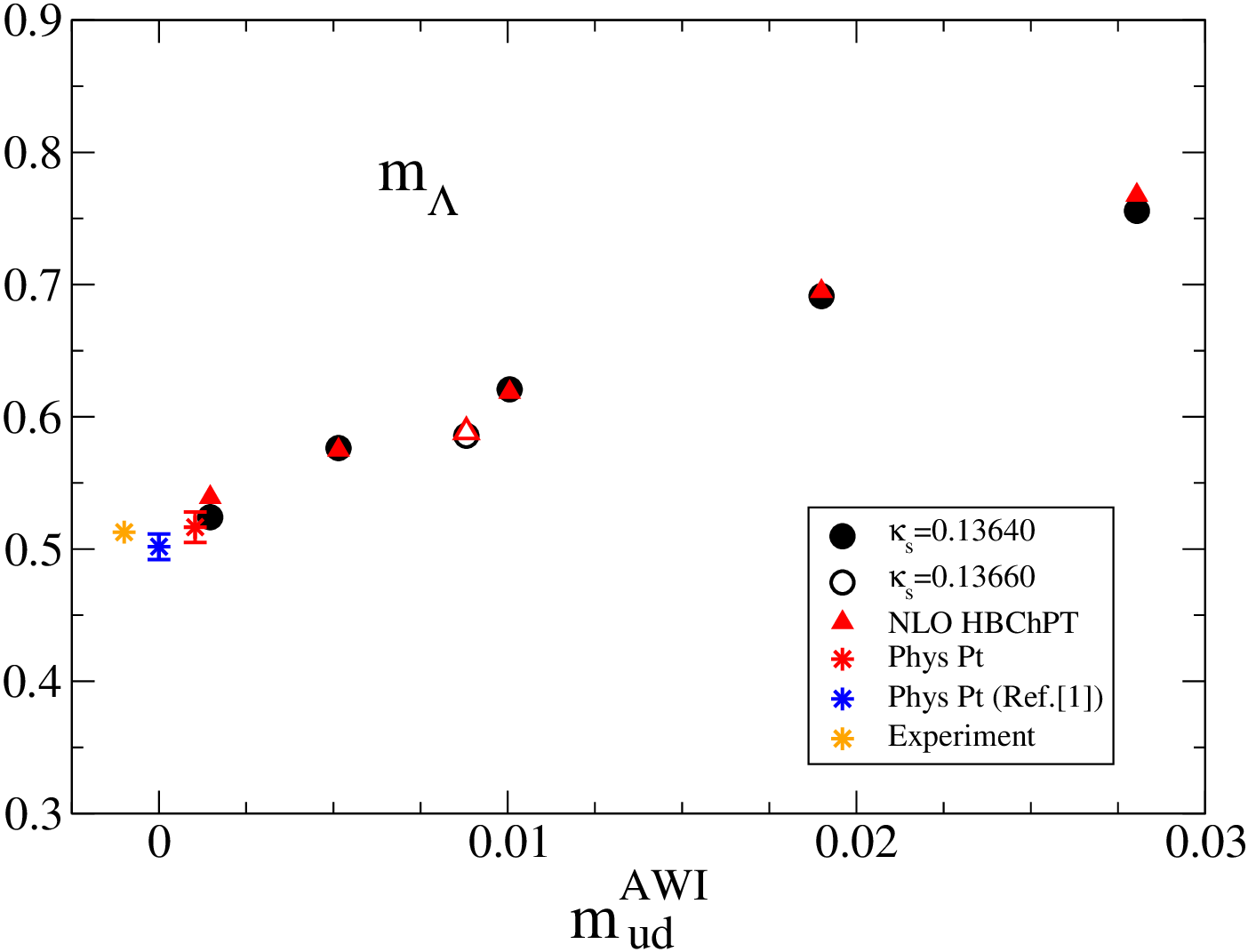}\\
\vspace*{7mm}& \\
\includegraphics[width=85mm,angle=0]{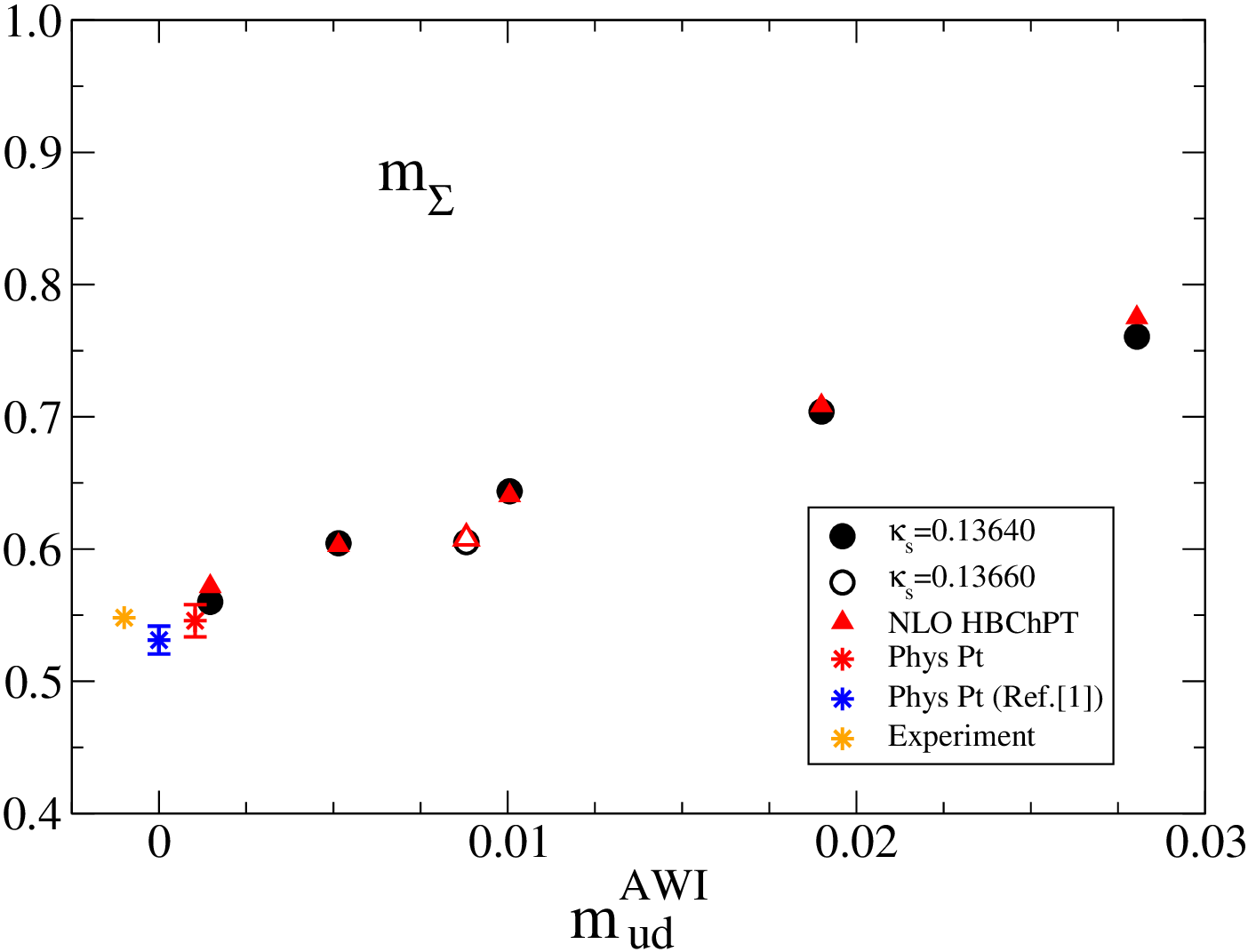}
\hspace*{2mm}&\hspace*{2mm}
\includegraphics[width=85mm,angle=0]{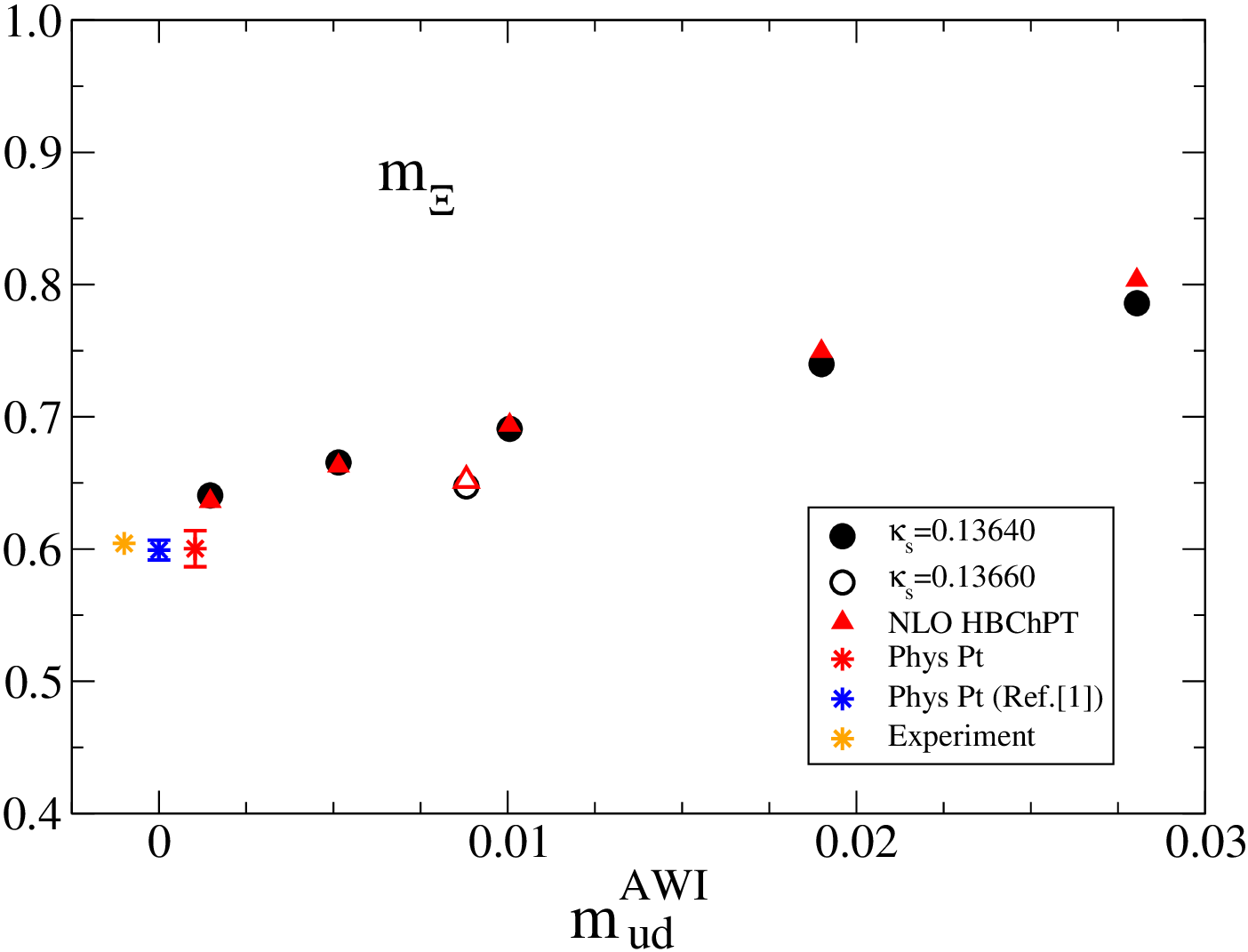}
\end{tabular}
\end{center}
\vspace{-.5cm}
\caption{\label{fig:hbchpt_su3_nlo_o}Fit results with the SU(3) HBChPT 
up to NLO for the octet baryon masses.
Experimental values are given in lattice units with $a^{-1}=2.176$ GeV
in Ref.~\cite{pacscs_nf3}.} 
\end{figure*}

\begin{figure*}[h]
\vspace{13mm}
\begin{center}
\begin{tabular}{cc}
\includegraphics[width=85mm,angle=0]{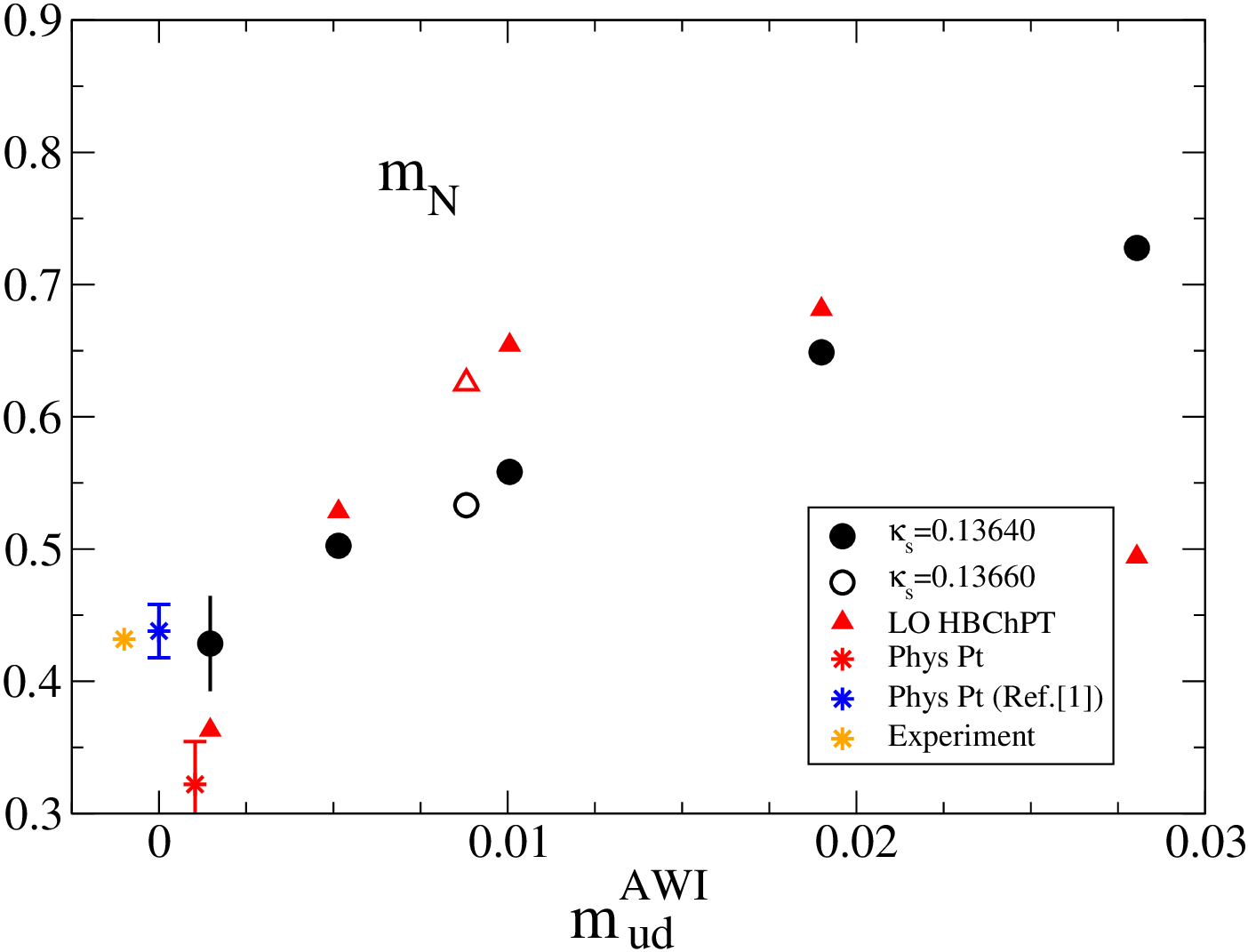}
\hspace*{2mm}&\hspace*{2mm}
\includegraphics[width=85mm,angle=0]{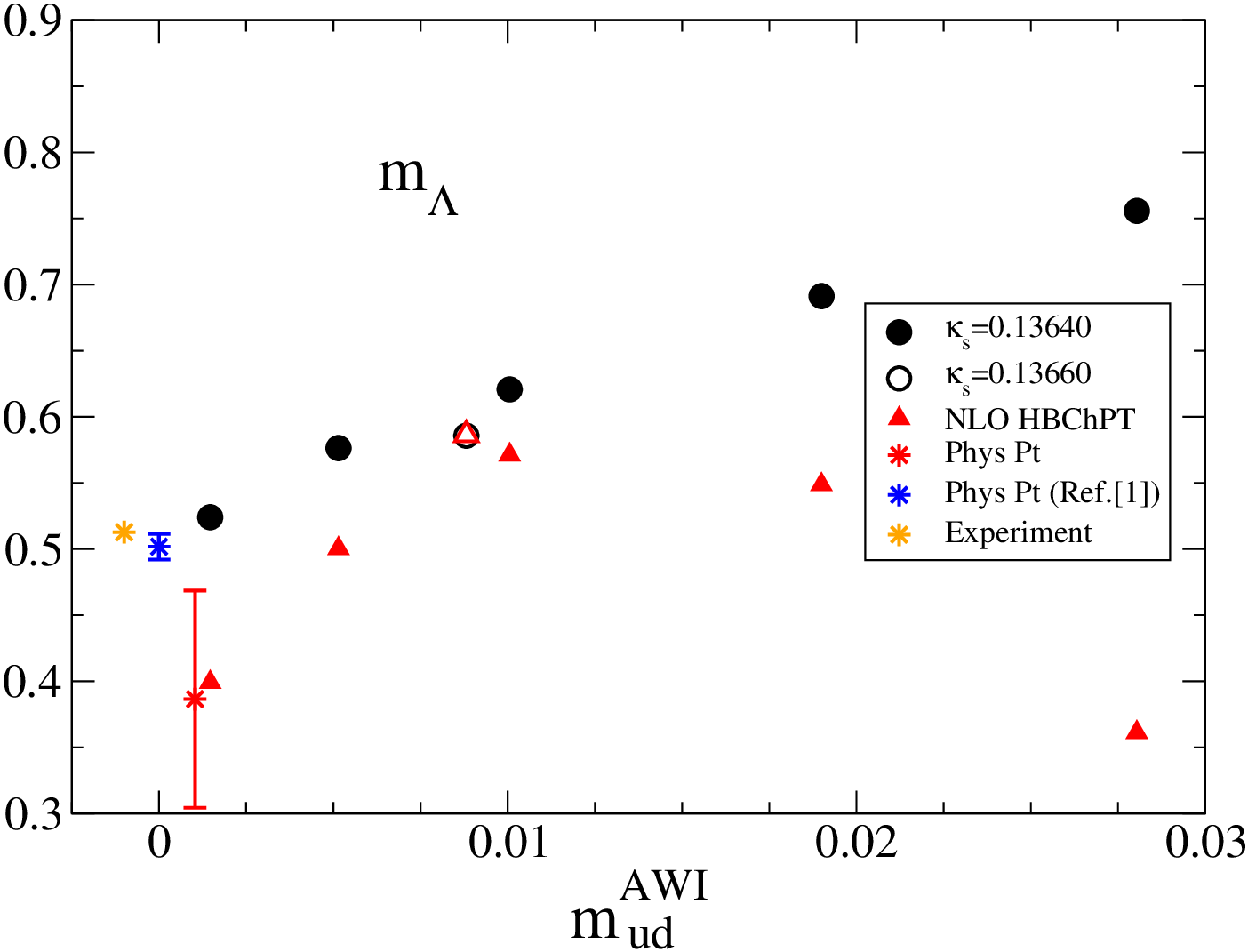}\\
\vspace*{7mm}& \\
\includegraphics[width=85mm,angle=0]{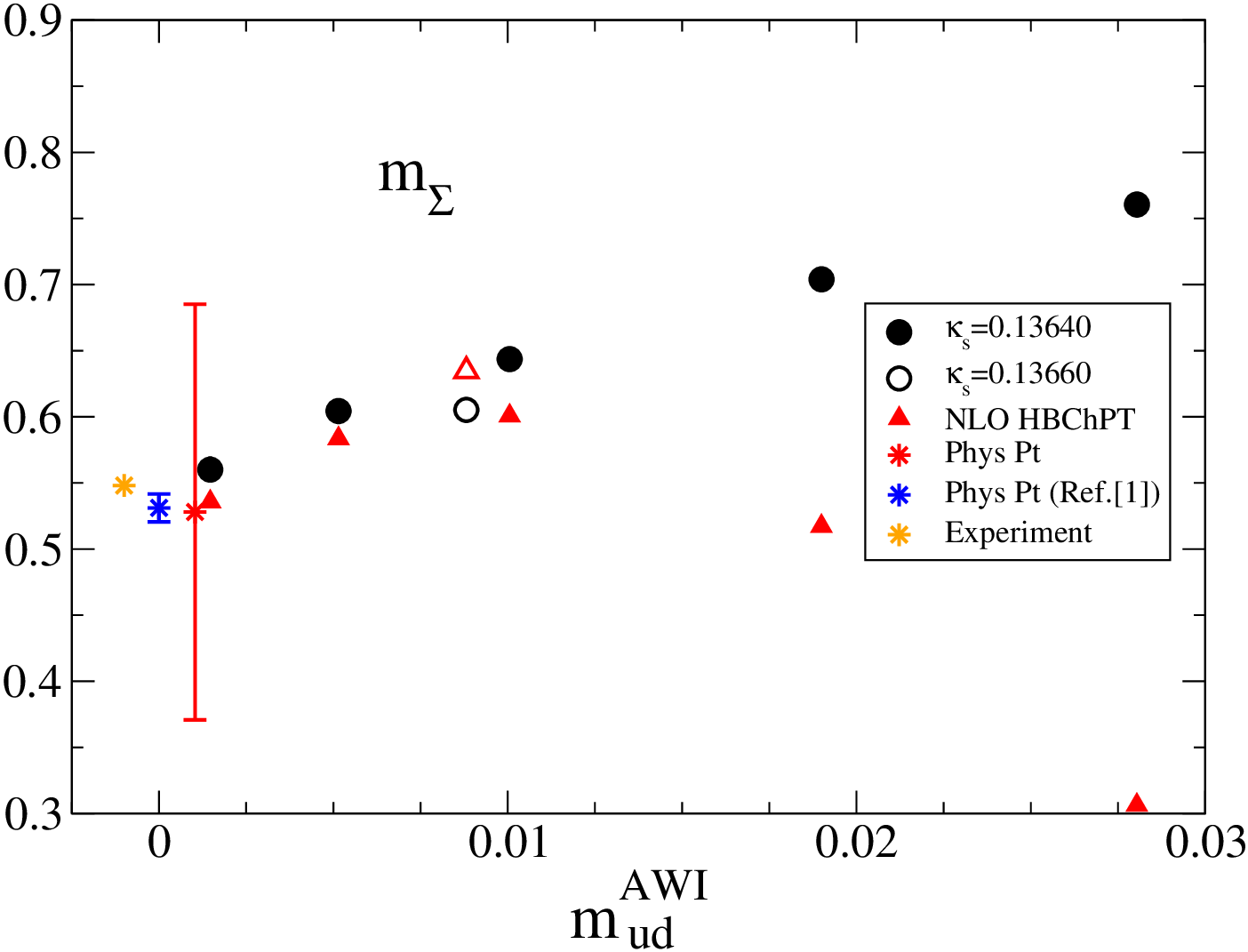}
\hspace*{2mm}&\hspace*{2mm}
\includegraphics[width=85mm,angle=0]{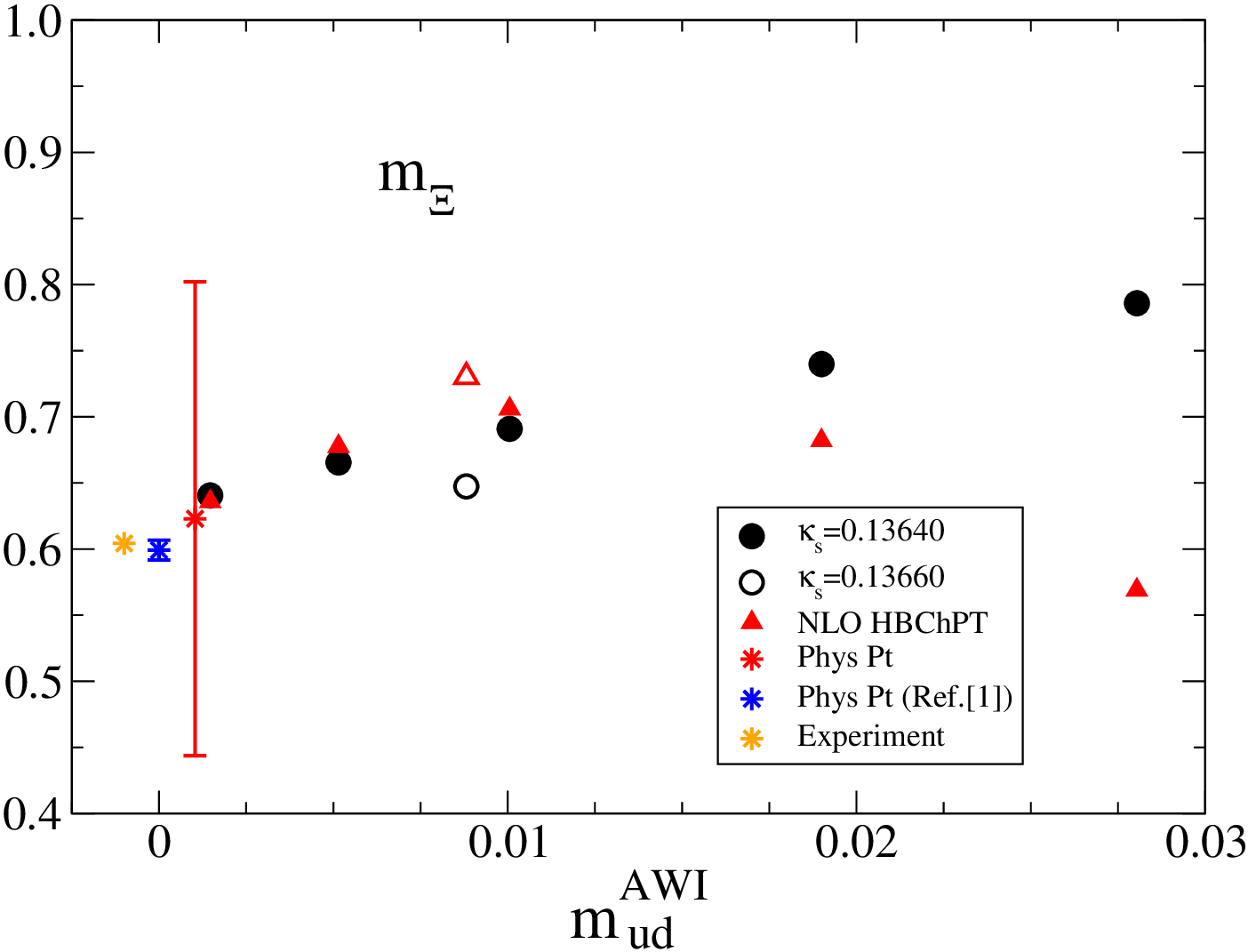}
\end{tabular}
\end{center}
\vspace{-.5cm}
\caption{\label{fig:hbchpt_su3_nlo_fix_o}Fit results with the SU(3) HBChPT 
up to NLO for the octet baryon masses. $D,F,{\cal C}$ are fixed at the
 phenomenological estimates.
Experimental values are given in lattice units with $a^{-1}=2.176$ GeV
in Ref.~\cite{pacscs_nf3}.} 
\end{figure*}

\begin{figure*}[h]
\vspace{13mm}
\begin{center}
\begin{tabular}{cc}
\includegraphics[width=85mm,angle=0]{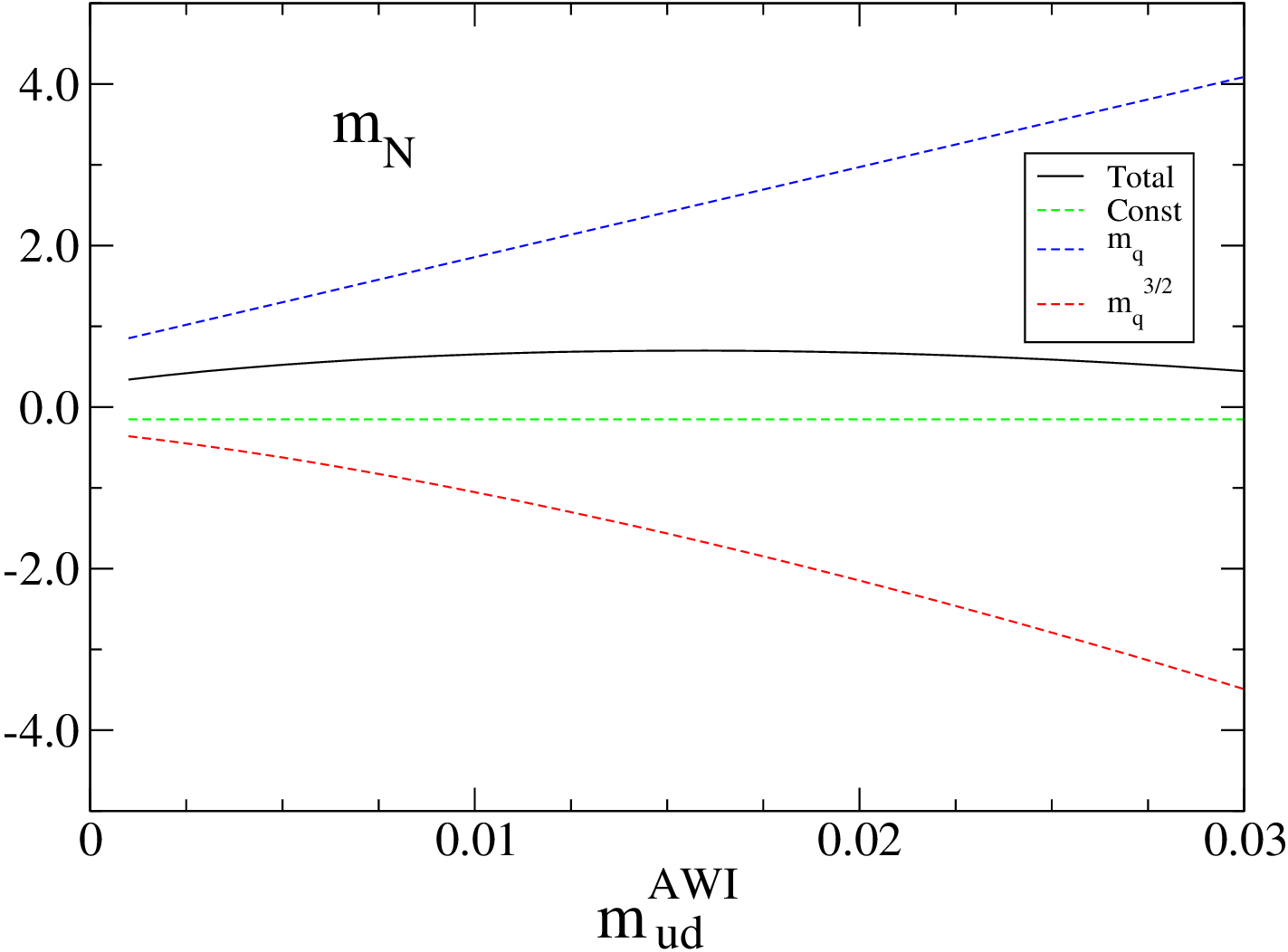}
\hspace*{2mm}&\hspace*{2mm}
\includegraphics[width=85mm,angle=0]{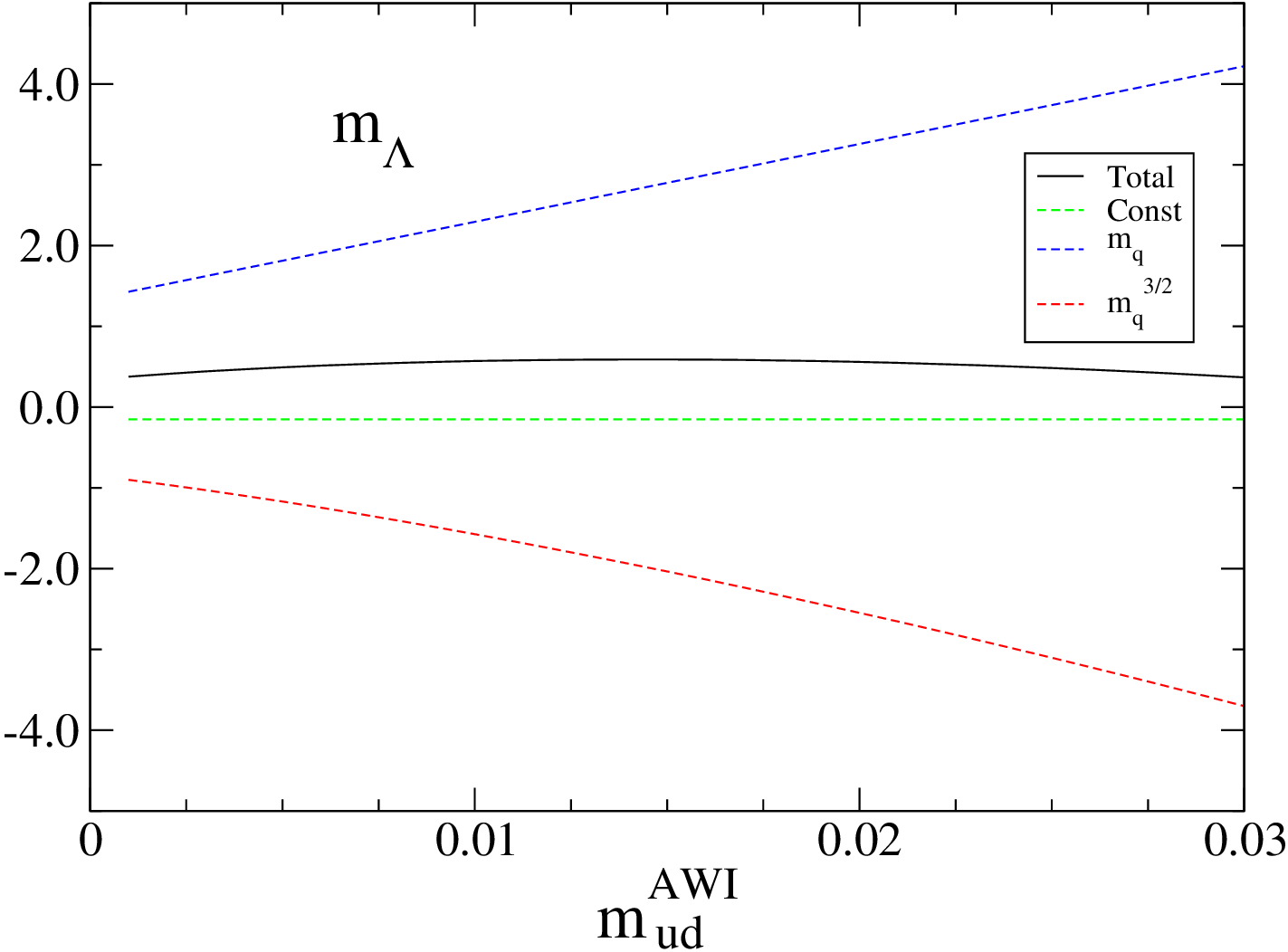}\\
\vspace*{7mm}& \\
\includegraphics[width=85mm,angle=0]{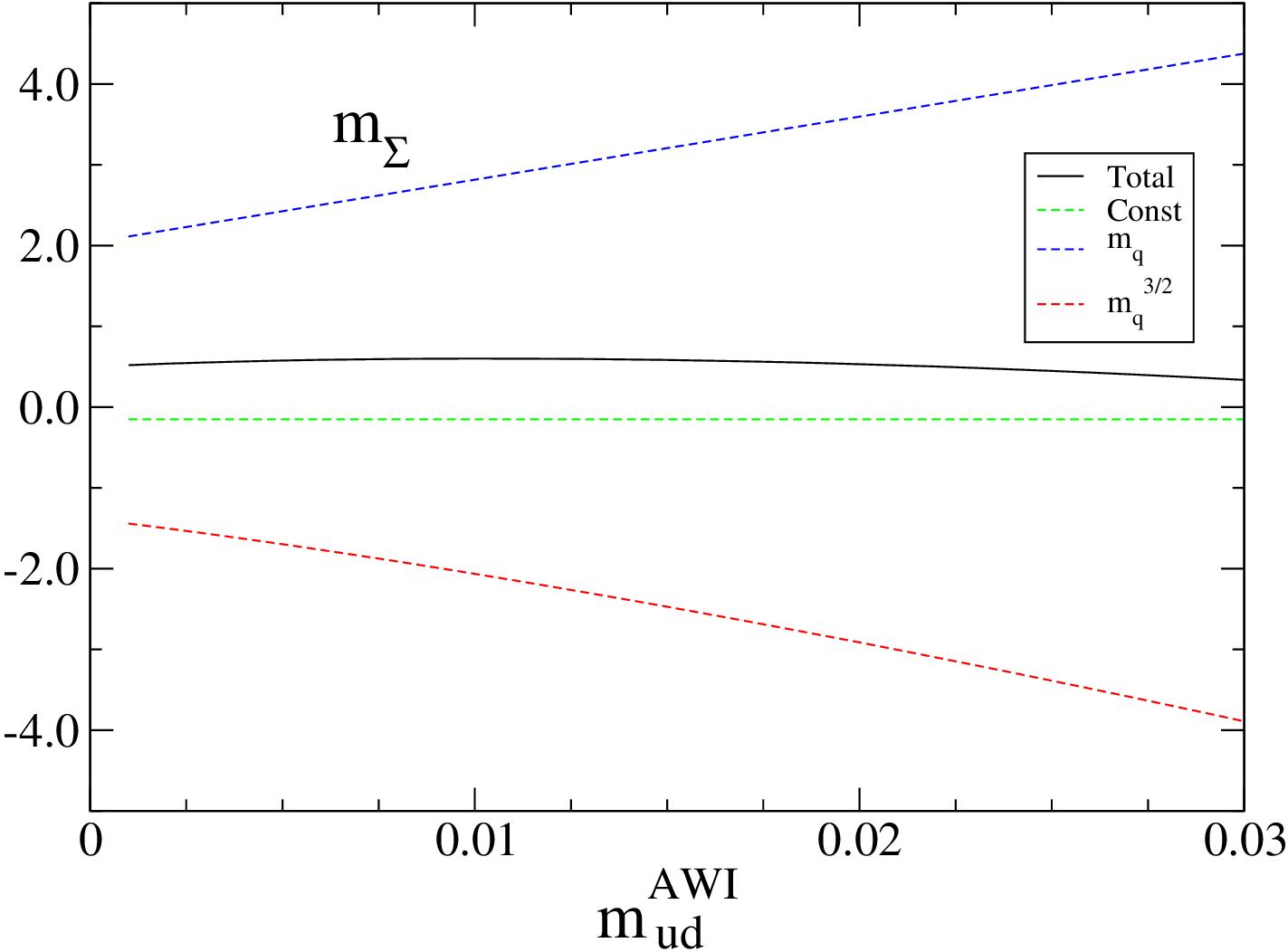}
\hspace*{2mm}&\hspace*{2mm}
\includegraphics[width=85mm,angle=0]{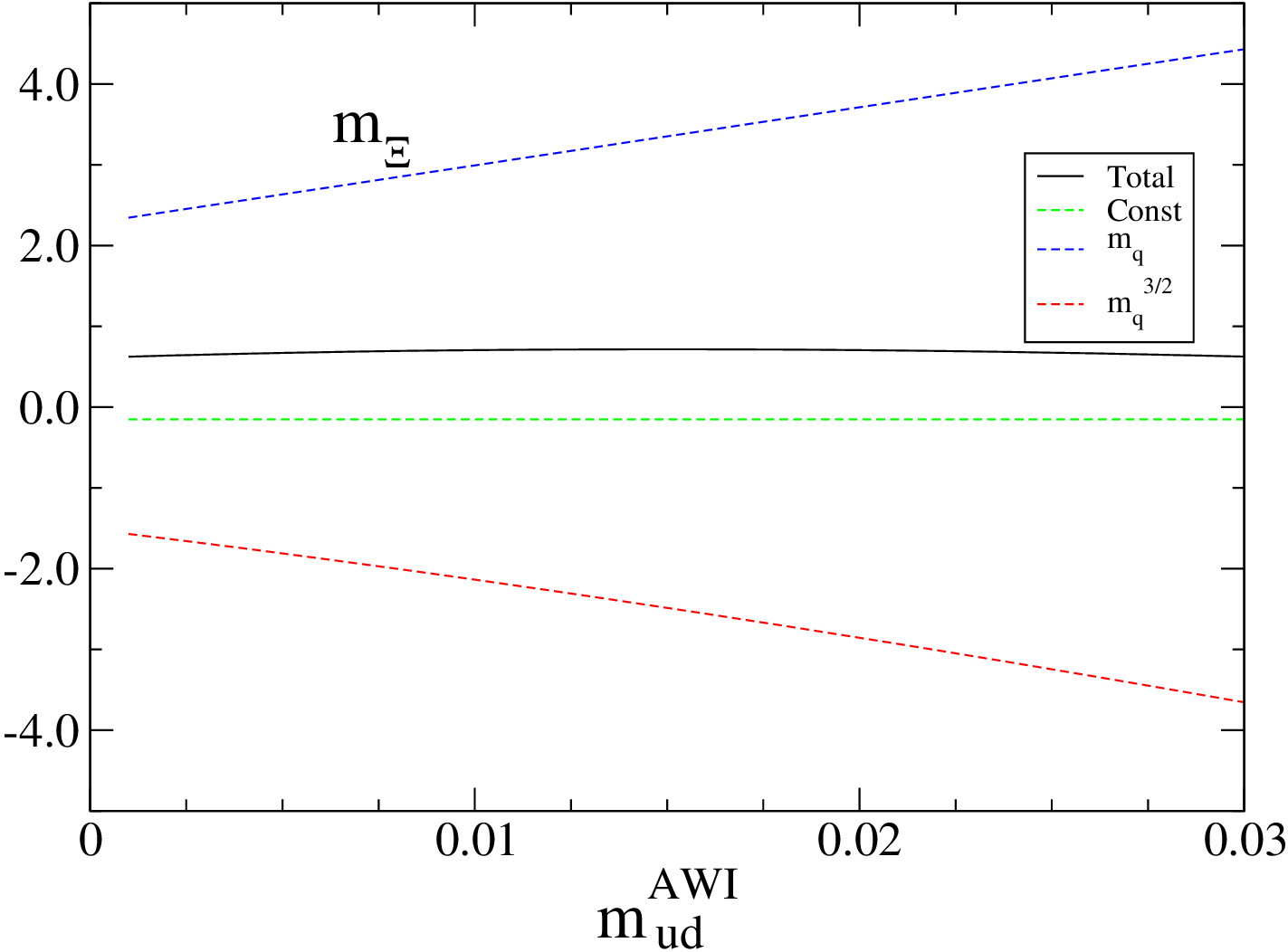}
\end{tabular}
\end{center}
\vspace{-.5cm}
\caption{\label{fig:hbchpt_su3_nlo_fix_conv_o}Convergence behavior for the
 octet masses with the SU(3) HBChPT up to NLO. $D,F,{\cal C}$ are fixed 
at the phenomenological estimates. 
$m_{\rm s}$ is fixed with the measured value 
at $(\kappa_{\rm ud},\kappa_{\rm s})=(0.13754,0.13640)$.}
\end{figure*}

In the NLO fit the number of LECs increases up to seven.
We give the fit results in Fig.~\ref{fig:hbchpt_su3_nlo_o}
and Tables~\ref{tab:hbchpt_su3_o} and \ref{tab:m_o_ph}. 
Although the fit works in a reasonable manner, a critical observation is that
the results for $D,F,{\cal C}$ are essentially consistent with zero
showing a significant deviation from the phenomenological estimates
in Eq.~(\ref{eq:dfc_ph}).
To examine the contributions of the NLO terms, 
we make a fit with $D,F,{\cal C}$ fixed at the phenomenological
estimates. Figure~\ref{fig:hbchpt_su3_nlo_fix_o} 
and Table~\ref{tab:hbchpt_su3_o} show that this 
fit assumption is strongly disfavored because of a prohibitively large value of
$\chi^2$/dof.  The reason is found in
Fig.~\ref{fig:hbchpt_su3_nlo_fix_conv_o}: If 
$D,F,{\cal C}$ are fixed at the phenomenological estimates, the magnitude of 
the NLO contribution is 2$-$5 times larger than that of the data  
even at $m_{\rm ud}=0$. We also remark that the same situation holds if 
either of $D,F,{\cal C}$ is fixed at the phenomenological estimate.

\subsection{Decuplet baryons}
\label{subsec:decuplet_su3}

Let us turn to the decuplet baryons. 
The chiral expansion of the $i$-th decuplet baryon mass is 
written as
\ben
 m_{T_i}=m_{T}+m_{T_i}^{(1)}+m_{T_i}^{(3/2)}+\cdots
\een
with $m_T$ the decuplet baryon mass in the SU(3) chiral limit. 
The LO and NLO corrections are given by
\ben
 m_{\Delta}^{(1)} &=& \frac{2}{3}\gamma_M (3m_{\rm ud}) -2\bar\sigma_M (2m_{\rm ud}+m_{\rm s}),  \\
 m_{\Sigma^*}^{(1)} &=& \frac{2}{3}\gamma_M (2m_{\rm ud}+m_{\rm s}) -2\bar\sigma_M (2m_{\rm ud}+m_{\rm s}),  \\
 m_{\Xi^*}^{(1)} &=& \frac{2}{3}\gamma_M (m_{\rm ud}+2 m_{\rm s}) -2\bar\sigma_M (2m_{\rm ud}+m_{\rm s}), \\
 m_{\Omega}^{(1)} &=& \frac{2}{3}\gamma_M (3m_{\rm s}) 
-2\bar\sigma_M (2m_{\rm ud}+m_{\rm s})
\een
and
\begin{widetext}
\ben
 m_{T_i}^{(3/2)} &=& -\frac{1}{(4\pi f_0)^2}  
  \sum_{\phi=\pi,K,\eta}\left[\frac{10}{9} {\cal H}^2A_{\phi}^{T_i}{\cal F}(m_{\phi},0,\mu)
     + {\cal C}^2B_{\phi}^{T_i} {\cal F}(m_{\phi},-\Delta,\mu) \right]
\een
\end{widetext}
where $\gamma_M, \bar\sigma_M, {\cal H}$ are additional LECs, and 
${\cal H}^2 A_\phi^{T_i}$ and ${\cal C}^2 B_\phi^{T_i}$ denote the
contributions coming from the decuplet-decuplet- and the
decuplet-octet-axial couplings, respectively, whose
complete list is given in Table~\ref{tab:ac_decuplet}.
Note that the function ${\cal F}(m_{\phi},-\Delta,\mu)$ 
yields an imaginary part if $\Delta> m_\phi$\cite{walker-loud}.
Our analyses are made by taking account of the real part only.
In the region of $\Delta> m_\phi$ it might be problematic to apply 
the continuum chiral expansion for infinite spatial volume 
to lattice data. We leave the proper treatment to future work.

\begin{figure*}[h]
\vspace{13mm}
\begin{center}
\begin{tabular}{cc}
\includegraphics[width=85mm,angle=0]{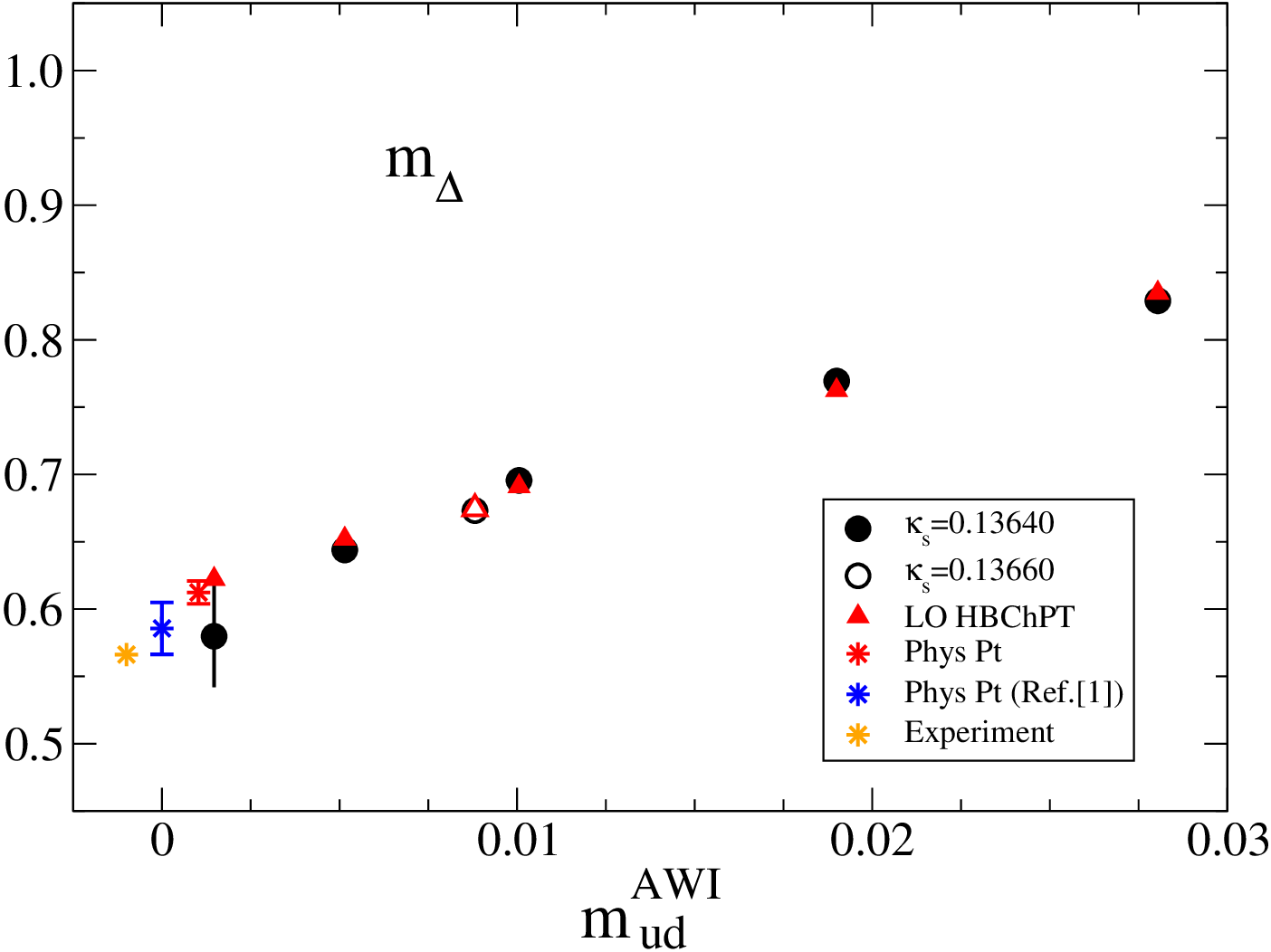}
\hspace*{2mm}&\hspace*{2mm}
\includegraphics[width=85mm,angle=0]{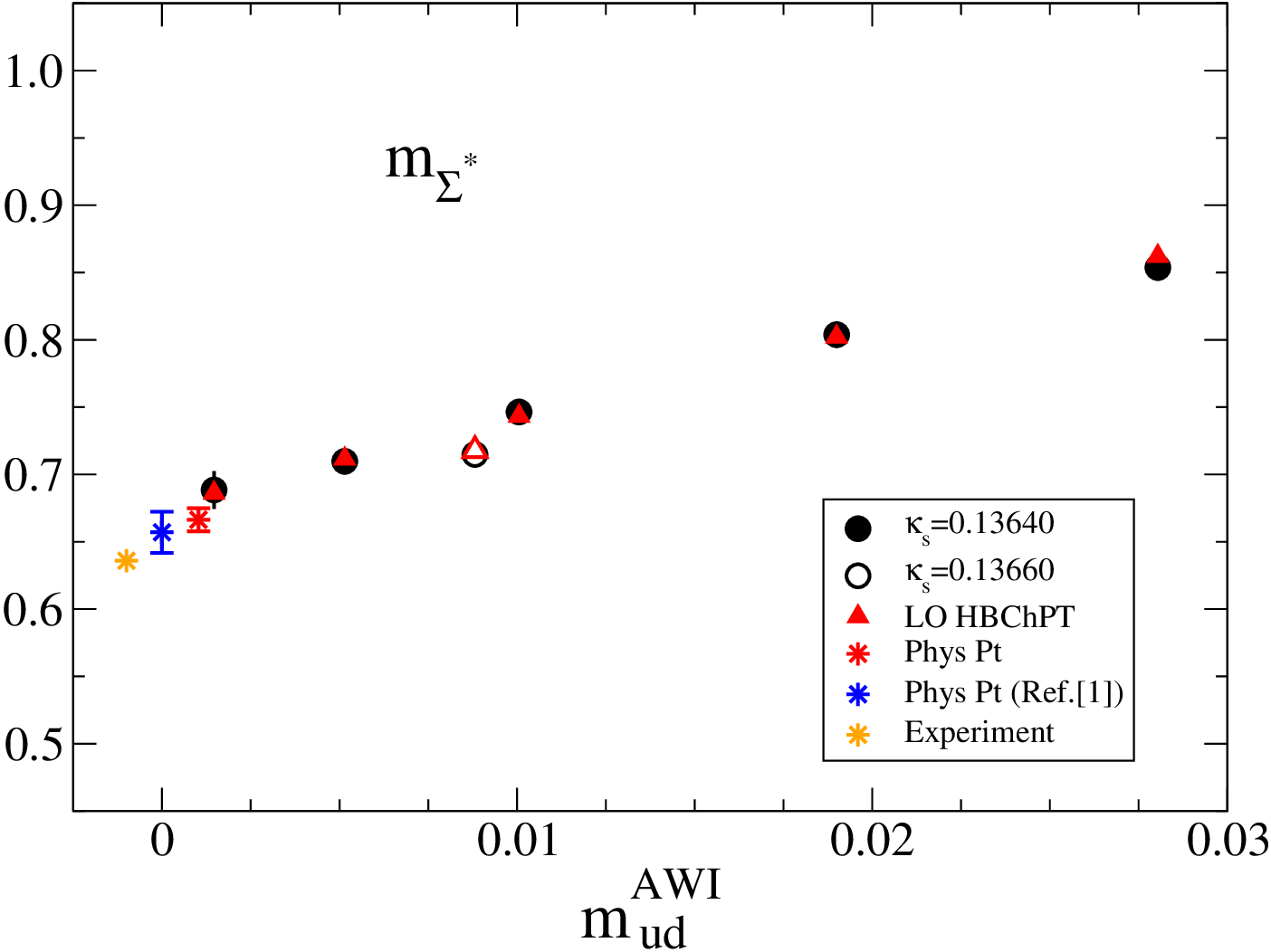}\\
\vspace*{7mm}& \\
\includegraphics[width=85mm,angle=0]{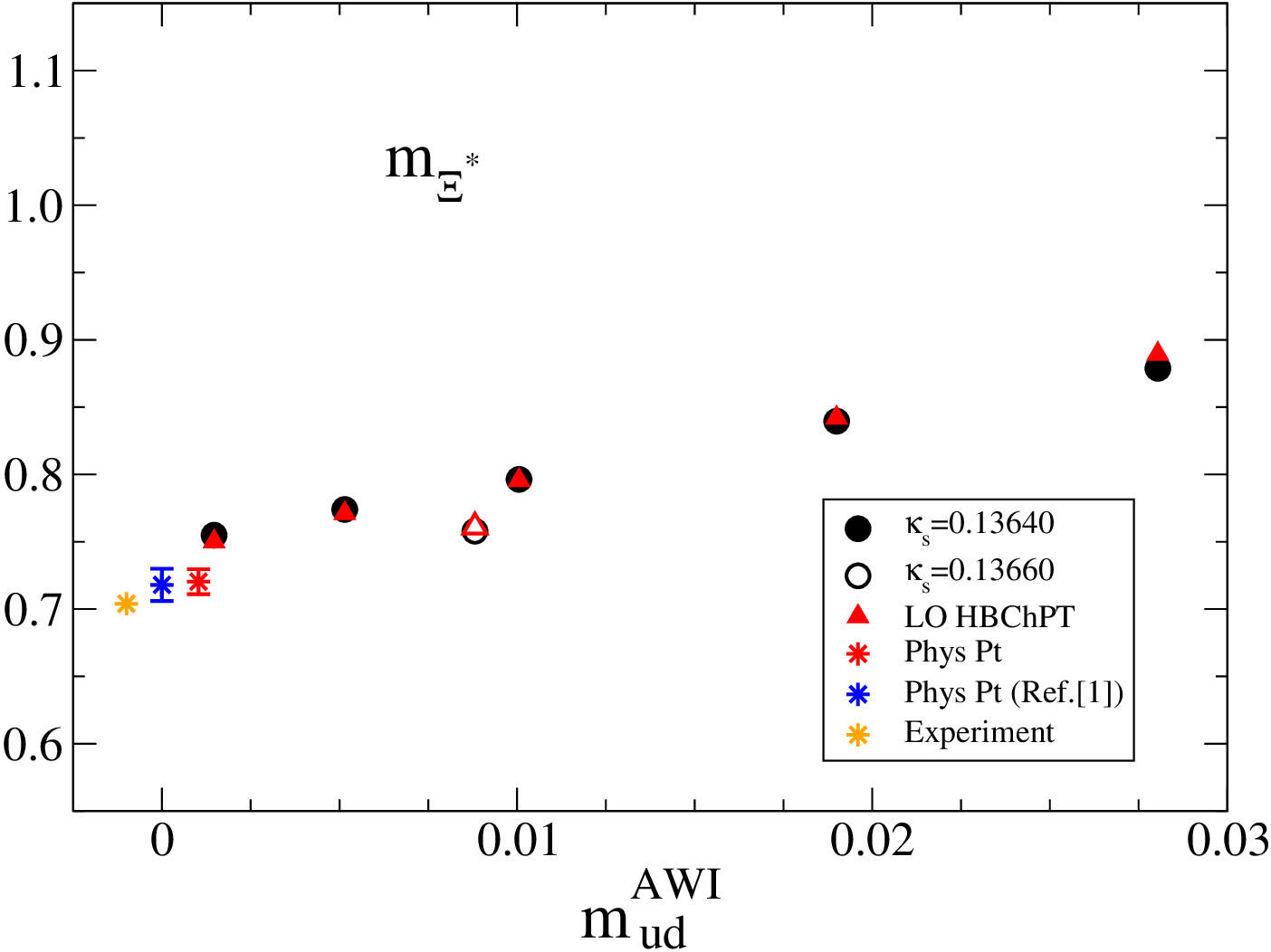}
\hspace*{2mm}&\hspace*{2mm}
\includegraphics[width=85mm,angle=0]{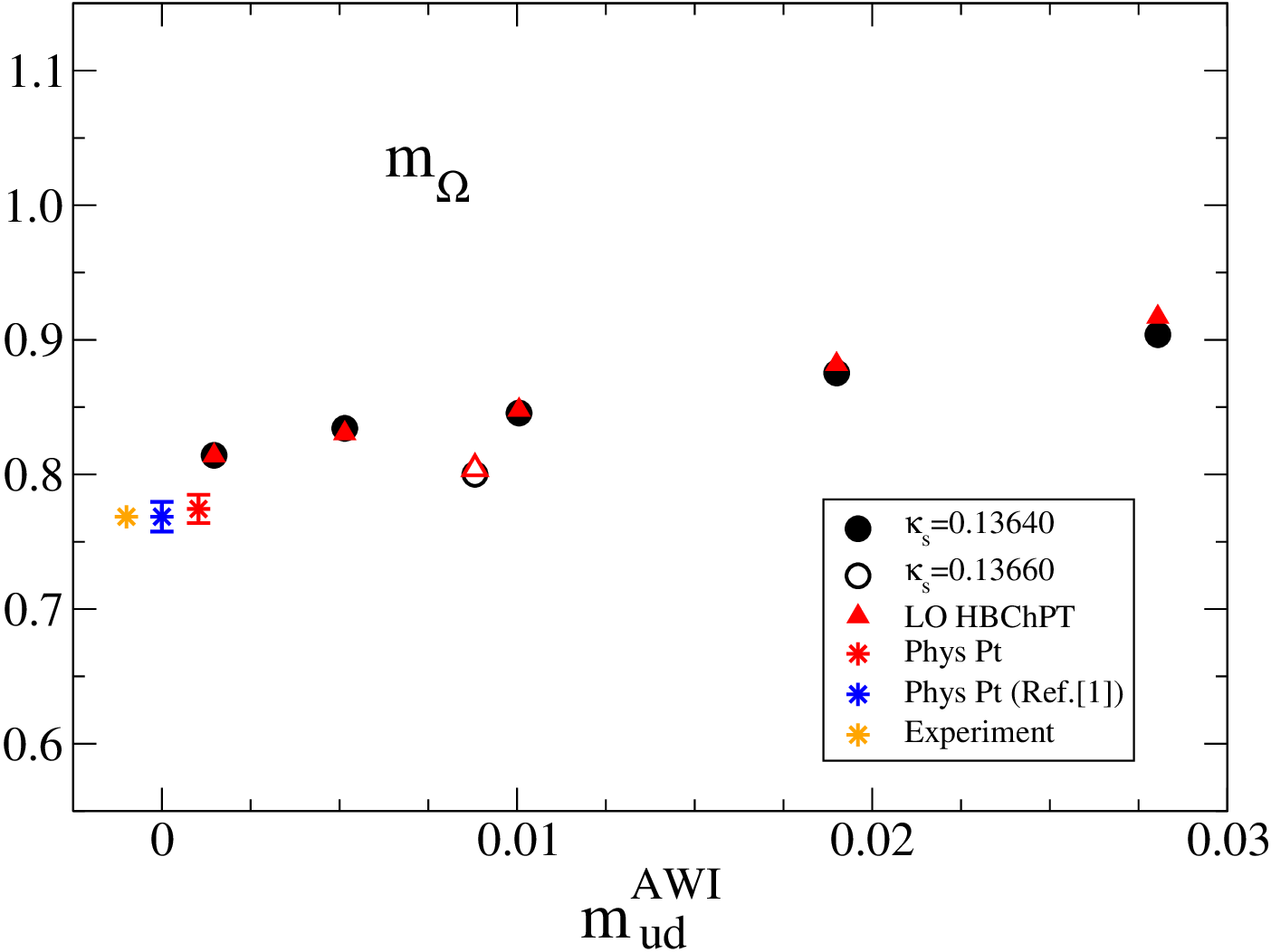}
\end{tabular}
\end{center}
\vspace{-.5cm}
\caption{\label{fig:hbchpt_su3_lo_d}Fit results with the SU(3) HBChPT 
up to LO for the decuplet baryon masses.
Experimental values are given in lattice units with $a^{-1}=2.176$ GeV
in Ref.~\cite{pacscs_nf3}.} 
\end{figure*}

\begin{figure*}[h]
\vspace{13mm}
\begin{center}
\begin{tabular}{cc}
\includegraphics[width=85mm,angle=0]{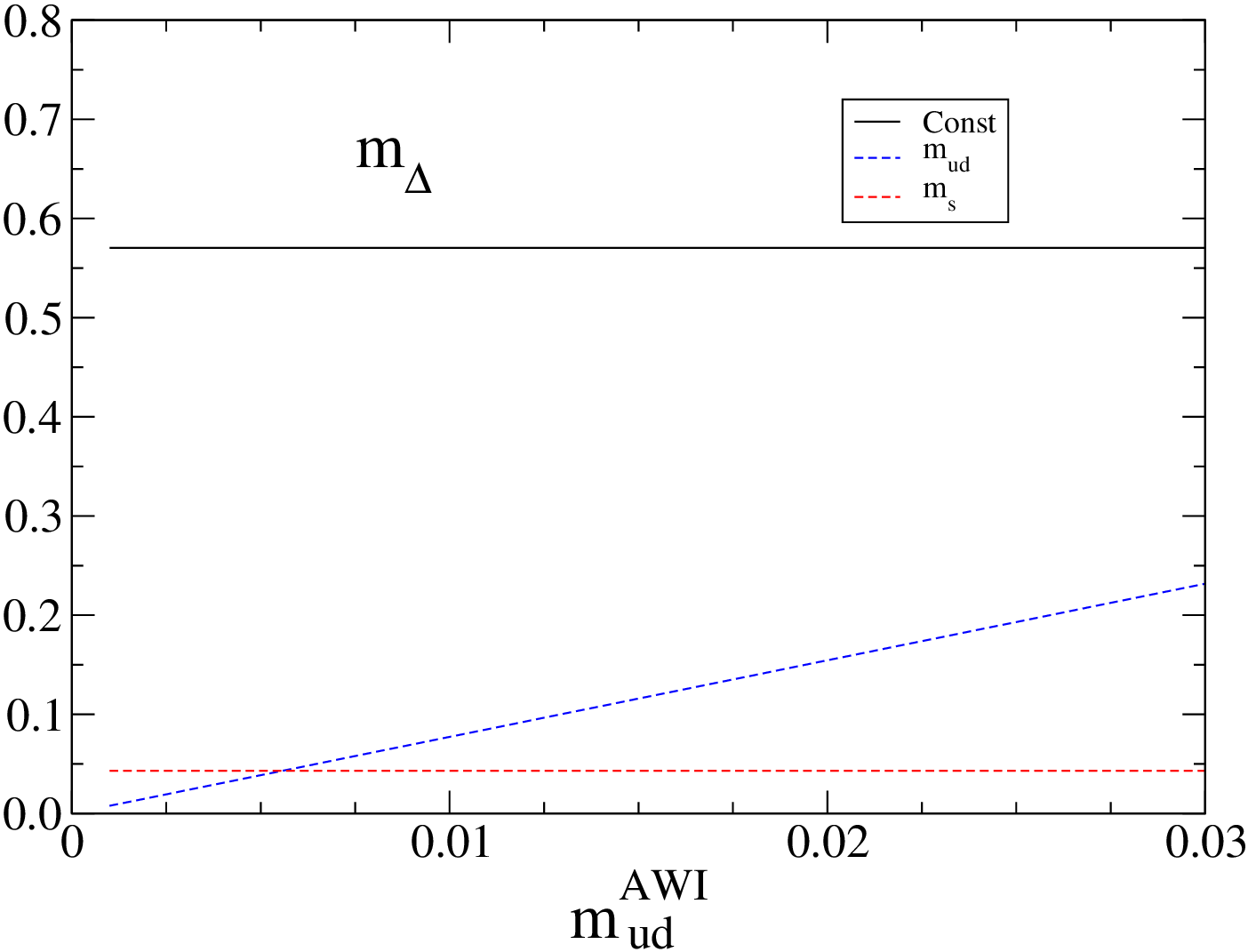}
\hspace*{2mm}&\hspace*{2mm}
\includegraphics[width=85mm,angle=0]{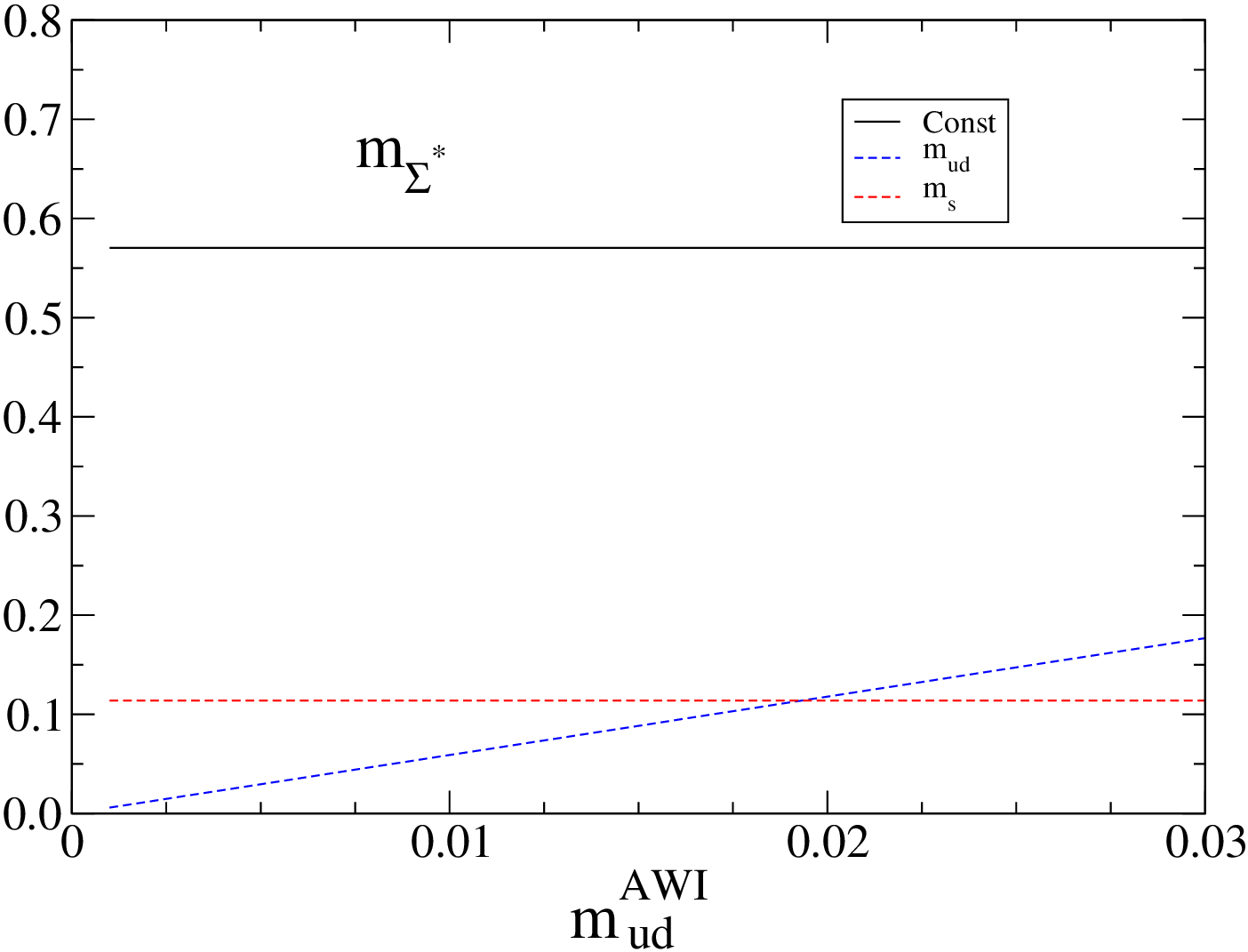}\\
\vspace*{7mm}& \\
\includegraphics[width=85mm,angle=0]{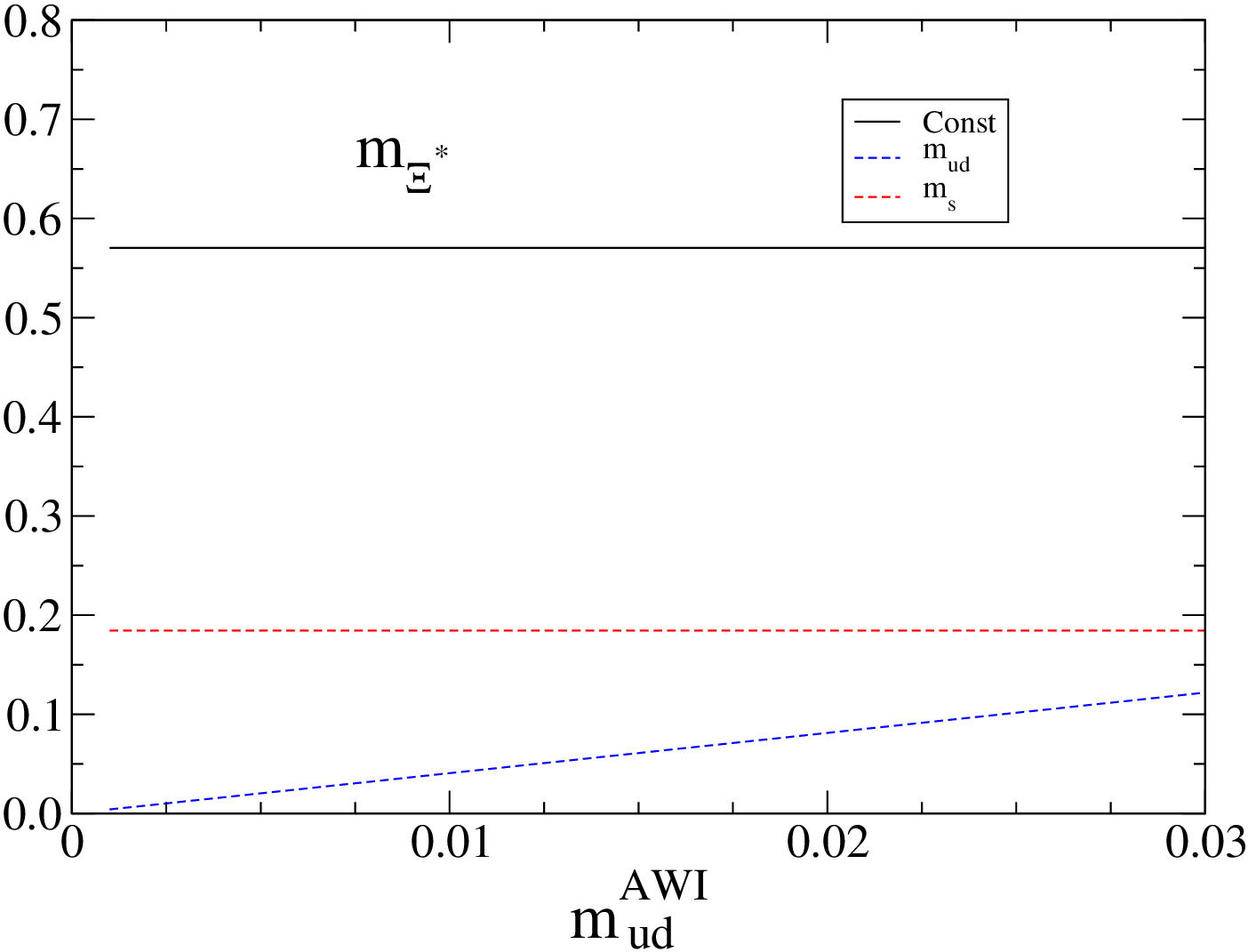}
\hspace*{2mm}&\hspace*{2mm}
\includegraphics[width=85mm,angle=0]{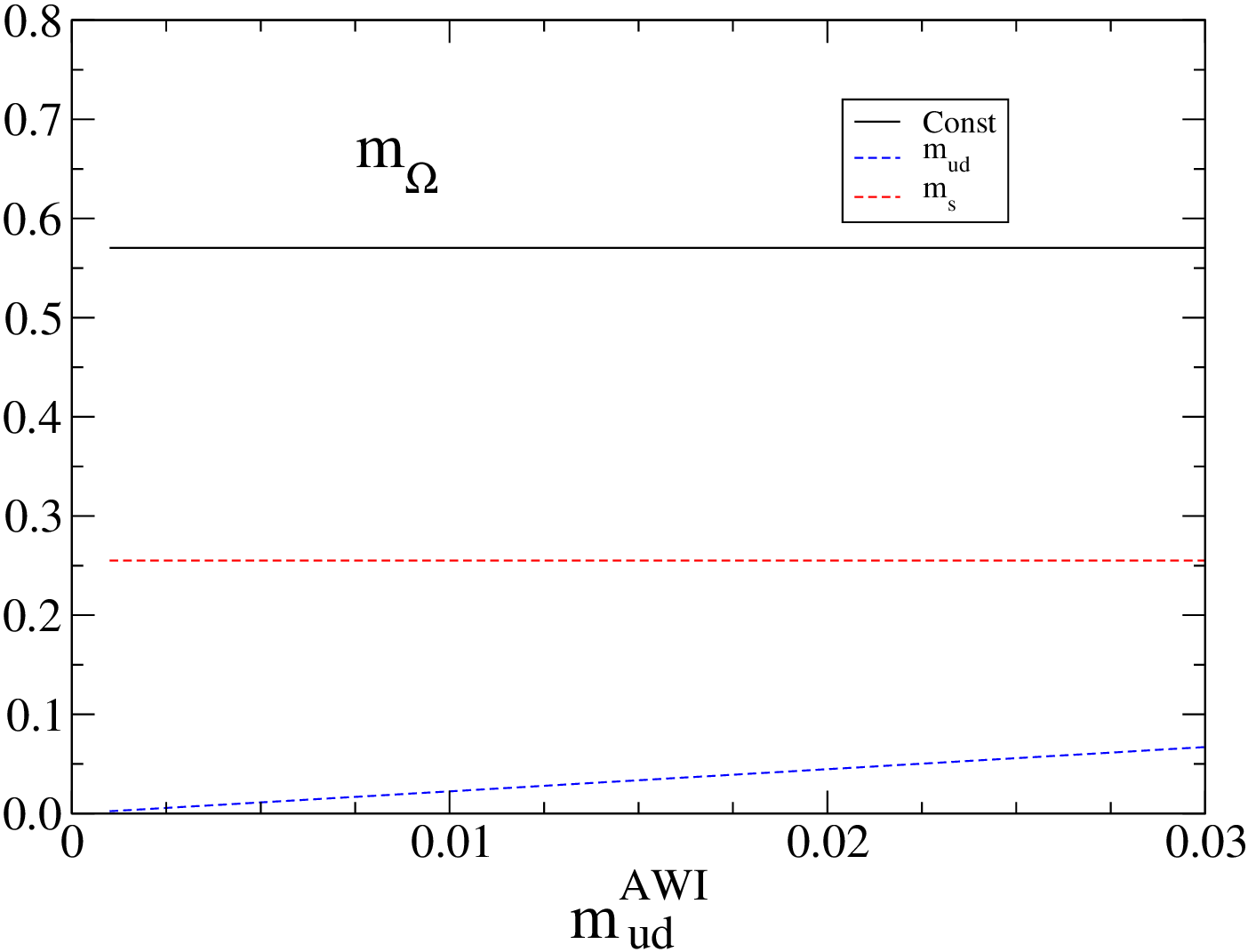}
\end{tabular}
\end{center}
\vspace{-.5cm}
\caption{\label{fig:hbchpt_su3_lo_conv_d}Convergence behavior for the
decuplet masses with the SU(3) HBChPT up to LO.
$m_{\rm s}$ is fixed with the measured value 
at $(\kappa_{\rm ud},\kappa_{\rm s})=(0.13754,0.13640)$.}
\end{figure*}

The LO fit results with the formula $m_{T_i}=m_{T}+m_{T_i}^{(1)}$ 
are presented in Figs.~\ref{fig:hbchpt_su3_lo_d} and  
\ref{fig:hbchpt_su3_lo_conv_d} and Tables~\ref{tab:hbchpt_su3_d} 
and \ref{tab:m_d_ph}.
The situation is similar to the case of the octet baryon masses.
The LO formula successfully describes the quark mass dependence
of the octet baryon masses. The extrapolated values 
at the physical point do not show any sizable deviations beyond error bars 
from those of the SU(2) linear fit Eq.~(\ref{eq:linear_su2}).
The contributions of the $O(m_{\rm s})$ term, however, 
are rather large compared to the magnitude of $m_T$.
The SU(2) linear formula shows a better convergence 
behavior than the SU(3)-symmetric linear formula.

\begin{figure*}[h]
\vspace{13mm}
\begin{center}
\begin{tabular}{cc}
\includegraphics[width=85mm,angle=0]{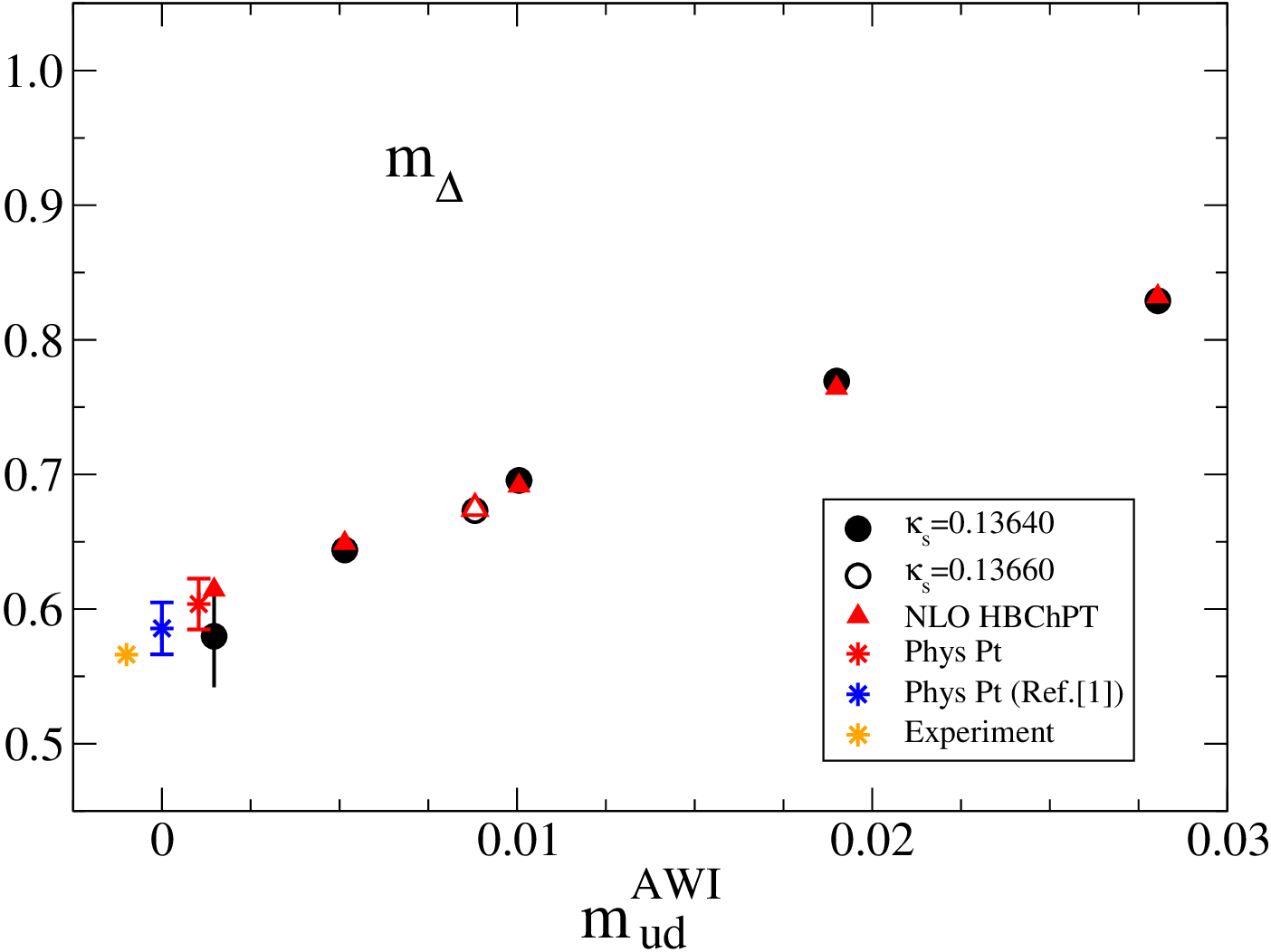}
\hspace*{2mm}&\hspace*{2mm}
\includegraphics[width=85mm,angle=0]{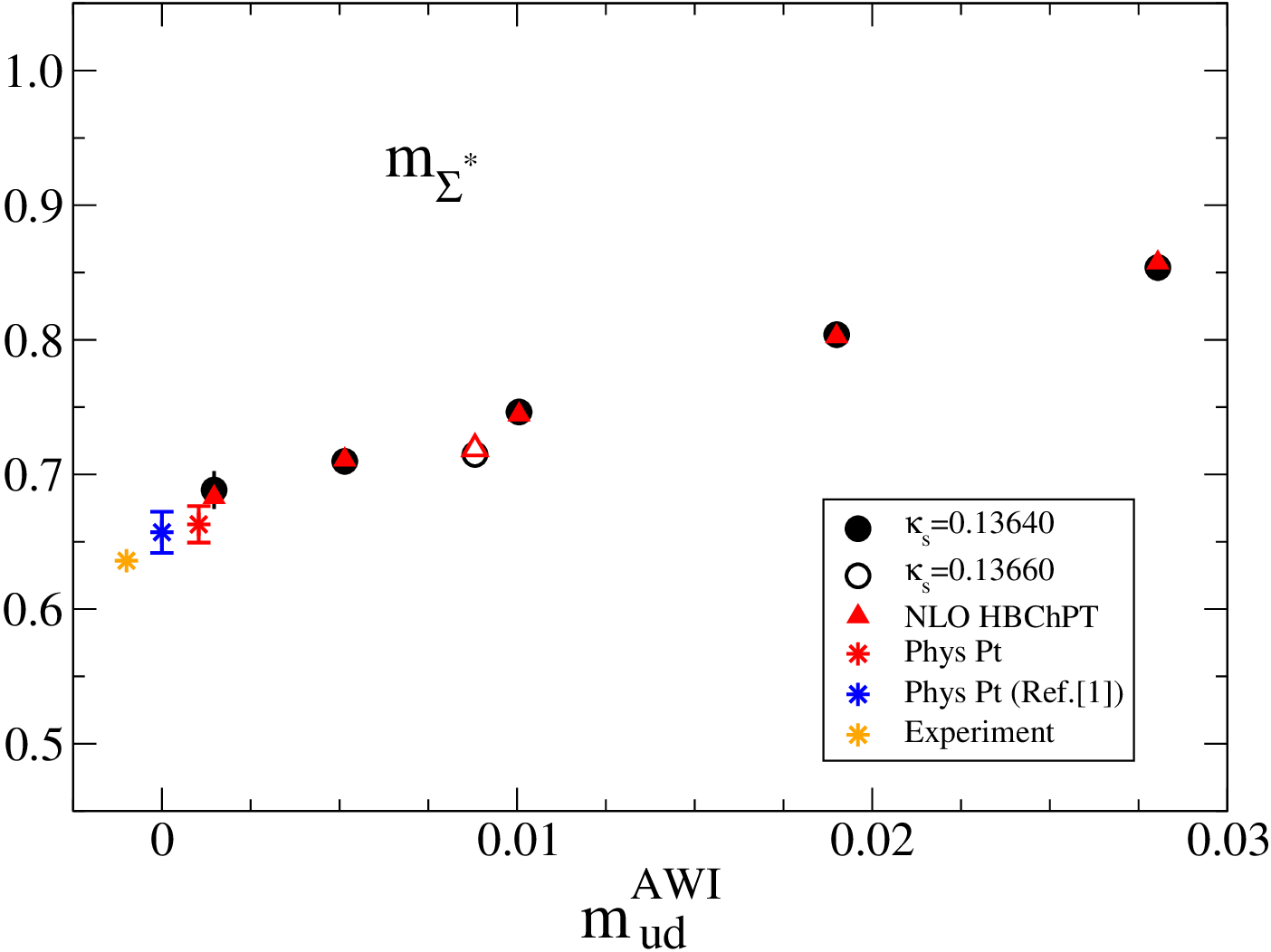}\\
\vspace*{7mm}& \\
\includegraphics[width=85mm,angle=0]{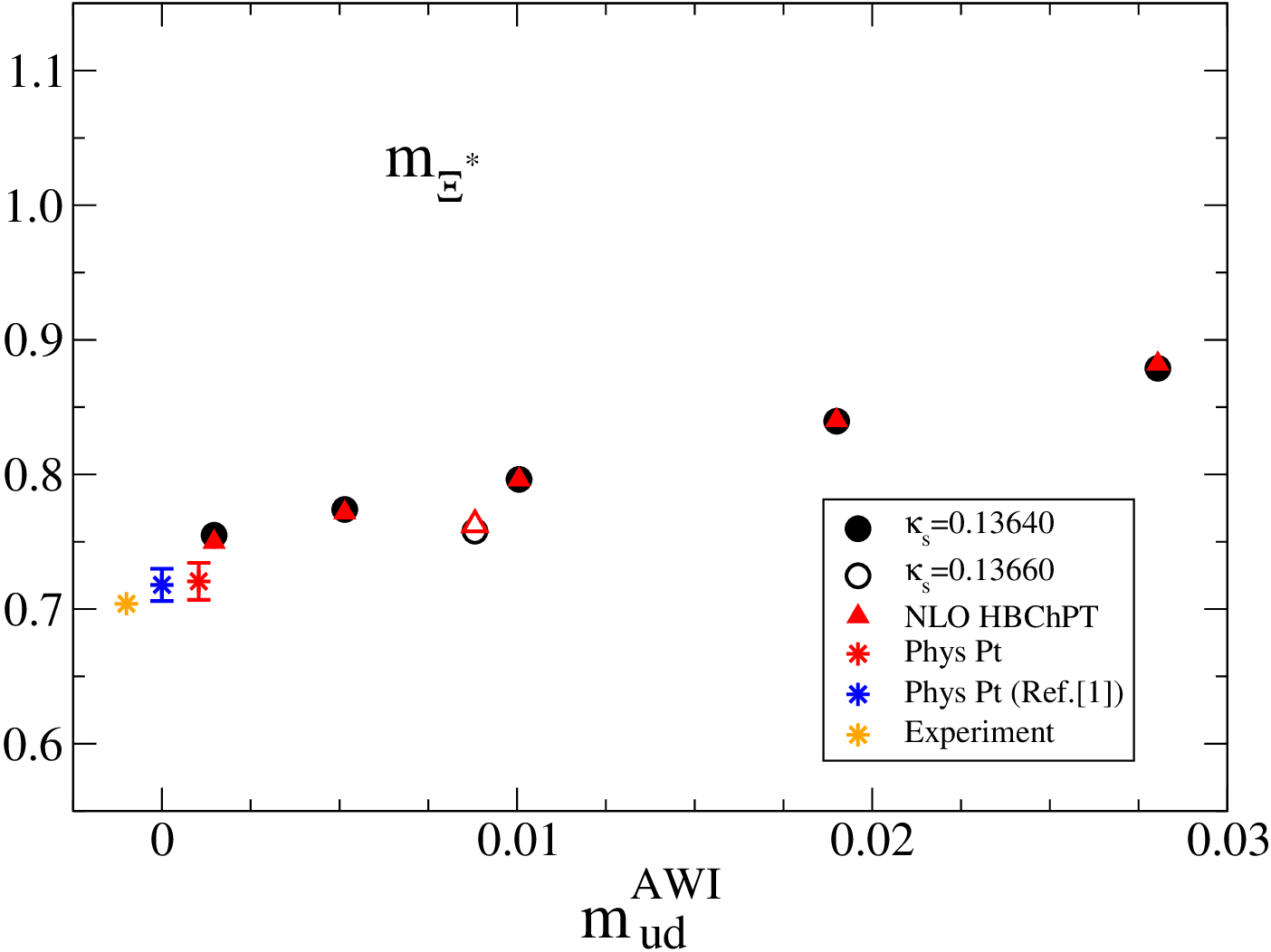}
\hspace*{2mm}&\hspace*{2mm}
\includegraphics[width=85mm,angle=0]{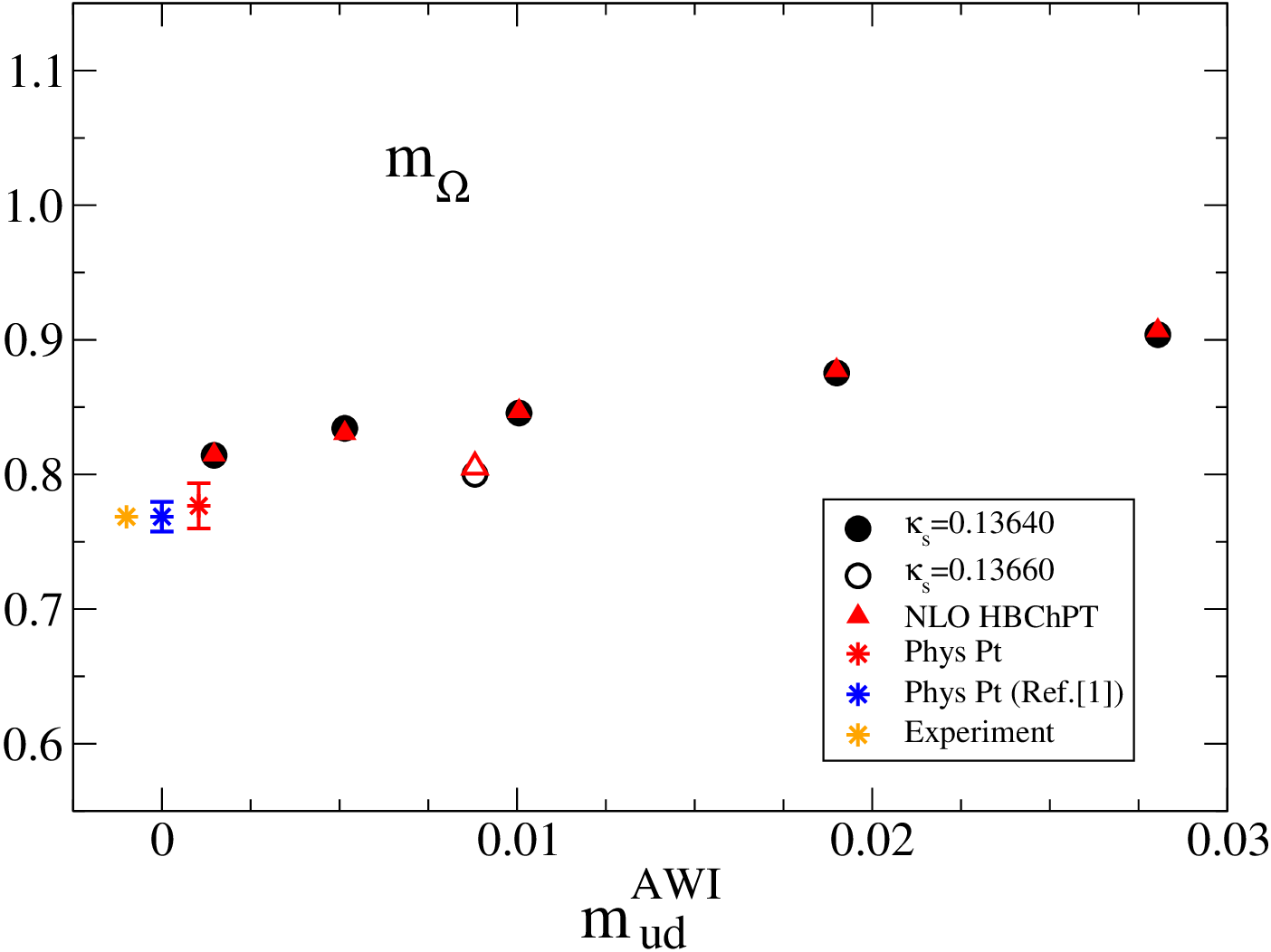}
\end{tabular}
\end{center}
\vspace{-.5cm}
\caption{\label{fig:hbchpt_su3_nlo_d}Fit results with the SU(3) HBChPT 
up to NLO for the decuplet baryon masses.
Experimental values are given in lattice units with $a^{-1}=2.176$ GeV
in Ref.~\cite{pacscs_nf3}.} 
\end{figure*}

\begin{figure*}[h]
\vspace{13mm}
\begin{center}
\begin{tabular}{cc}
\includegraphics[width=85mm,angle=0]{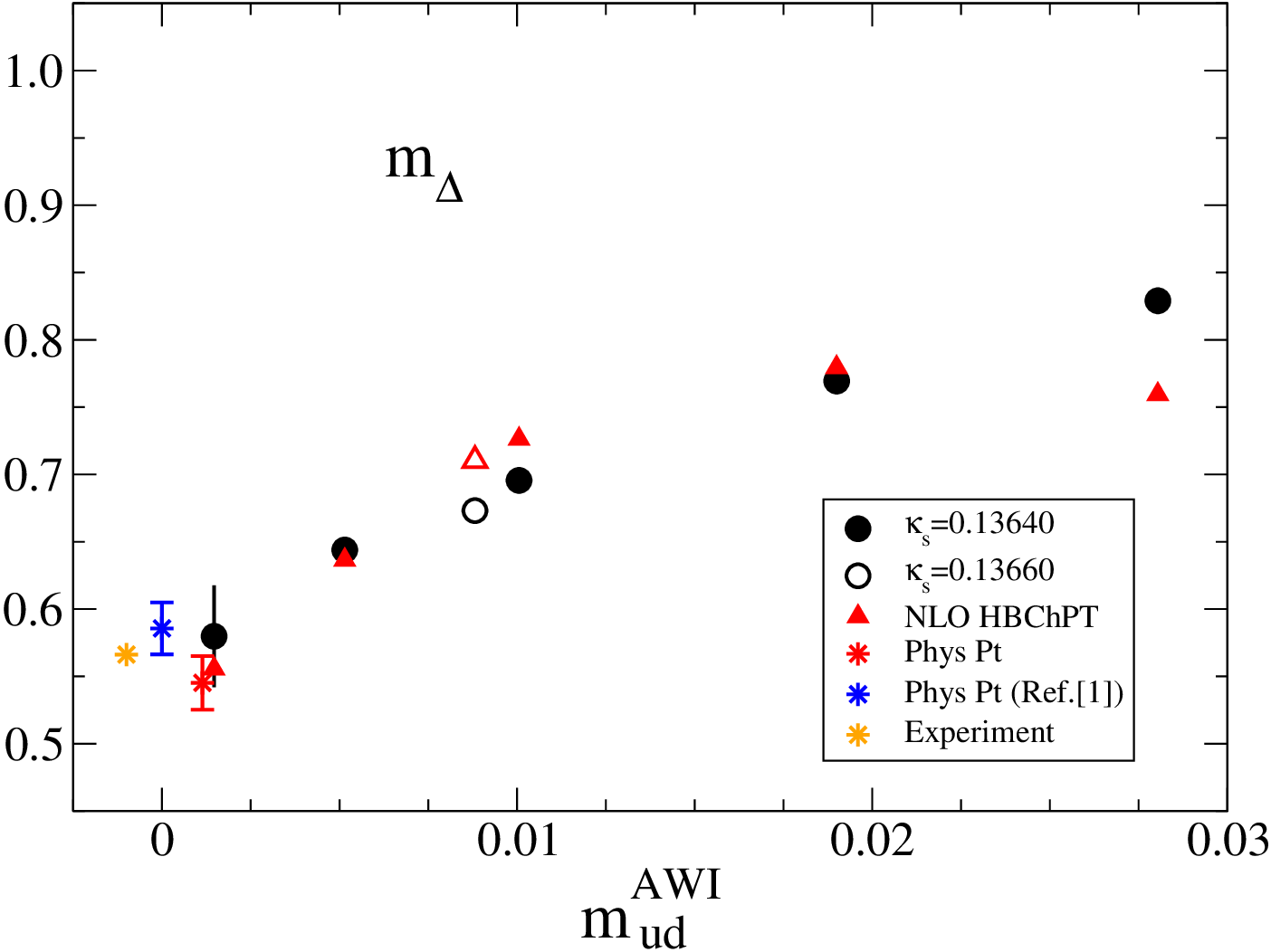}
\hspace*{2mm}&\hspace*{2mm}
\includegraphics[width=85mm,angle=0]{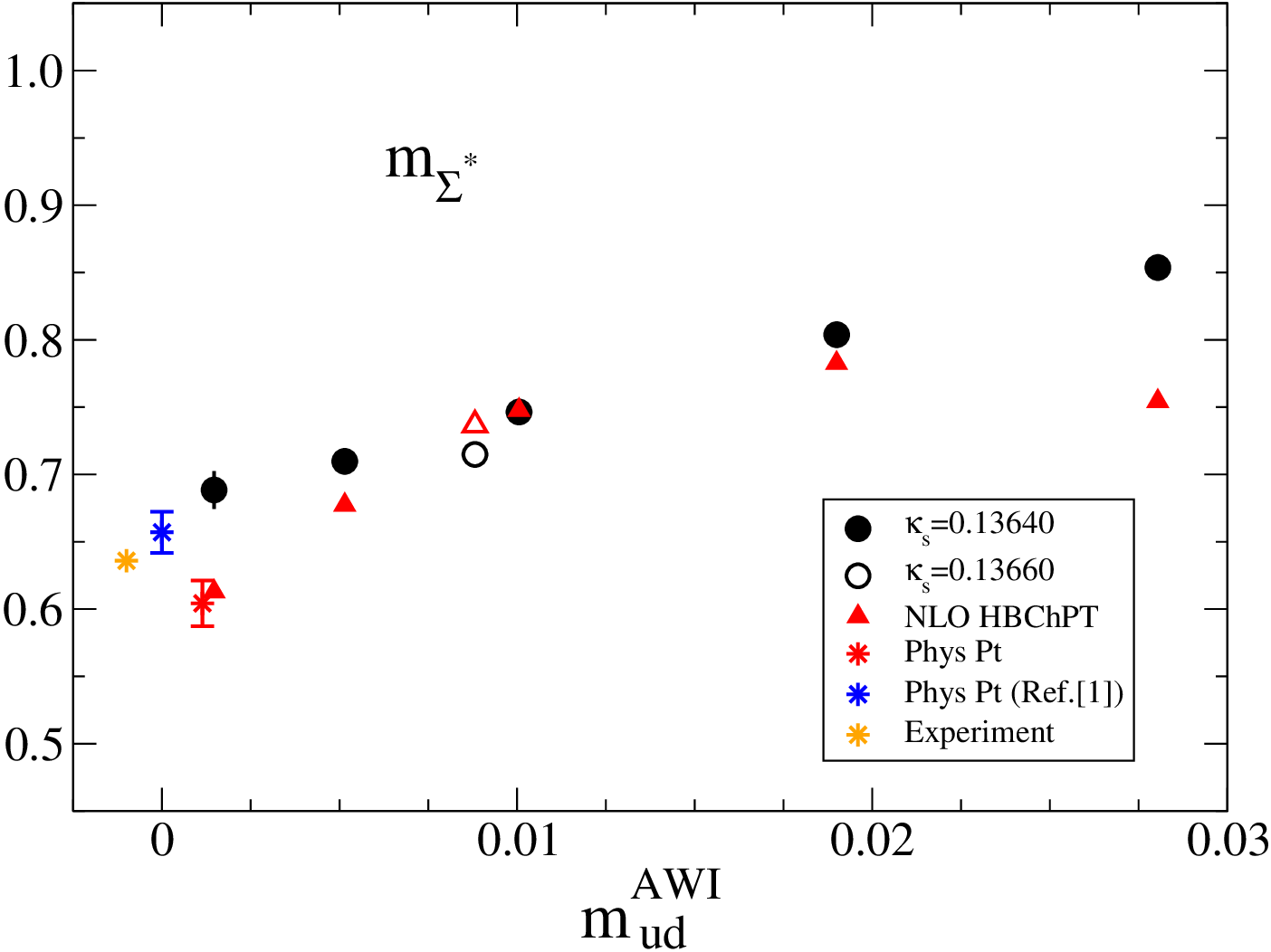}\\
\vspace*{7mm}& \\
\includegraphics[width=85mm,angle=0]{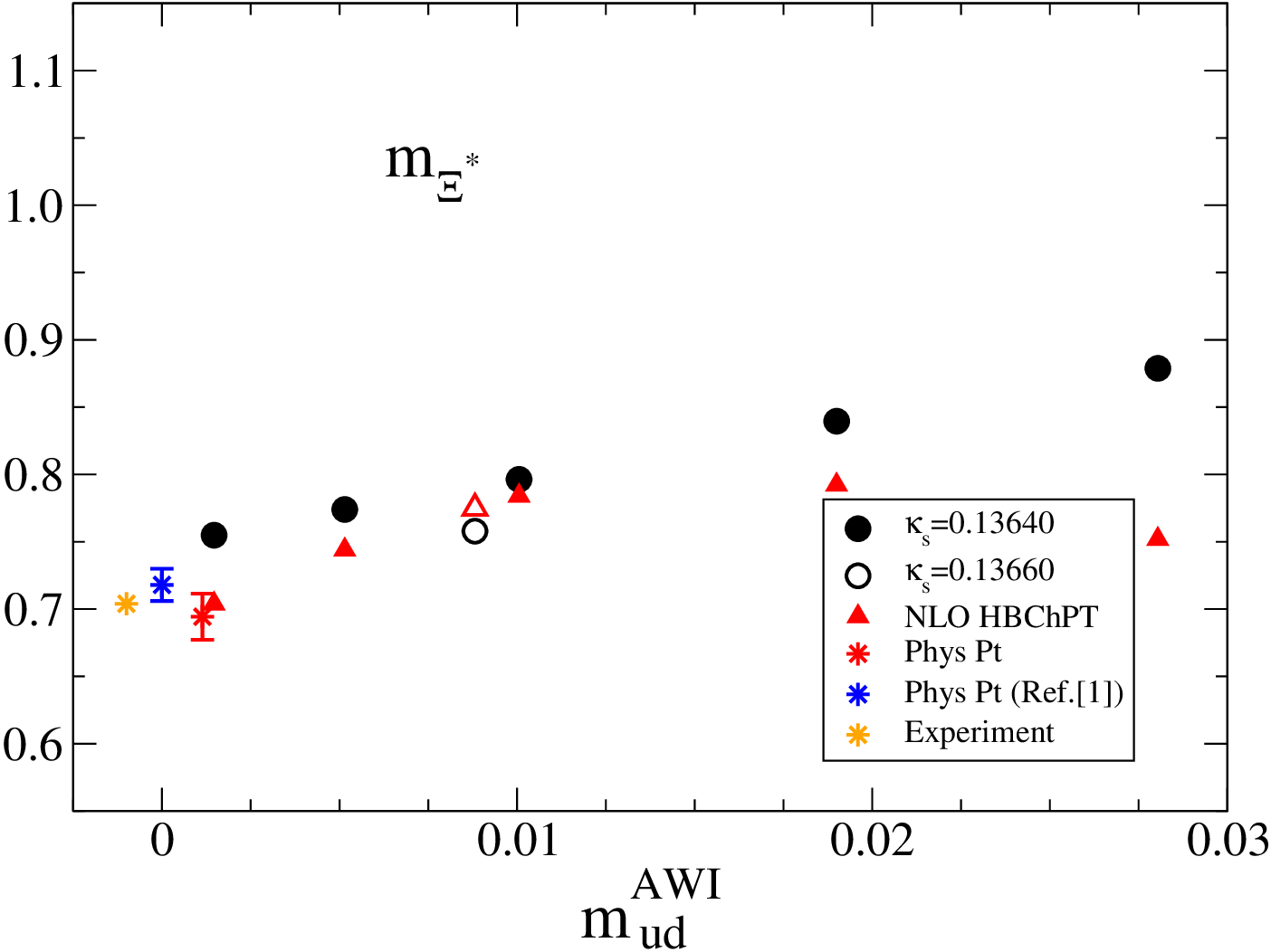}
\hspace*{2mm}&\hspace*{2mm}
\includegraphics[width=85mm,angle=0]{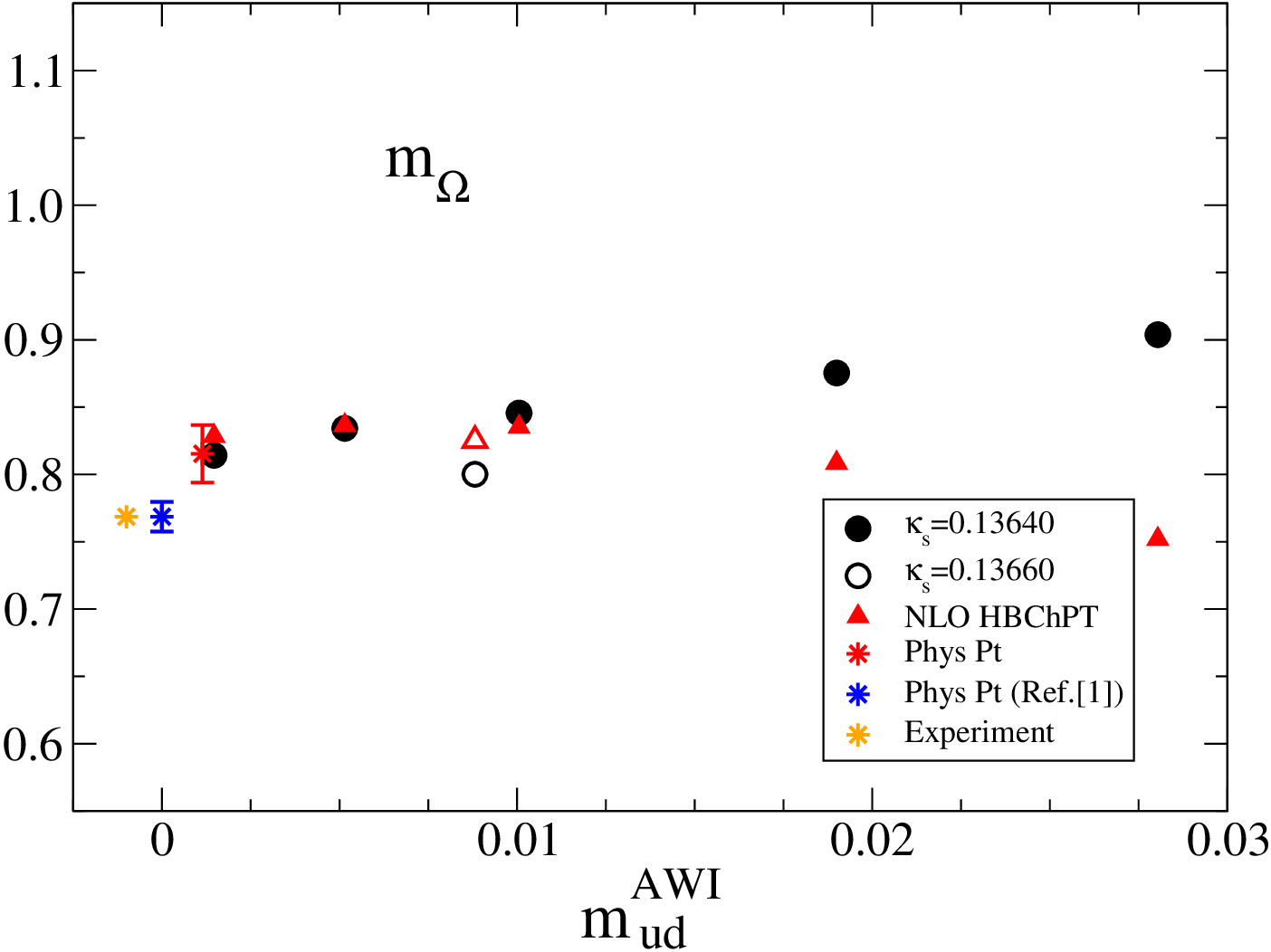}
\end{tabular}
\end{center}
\vspace{-.5cm}
\caption{\label{fig:hbchpt_su3_nlo_fix_d}Fit results with the SU(3) HBChPT 
up to NLO for the decuplet baryon masses. ${\cal C}$ is fixed at the
 phenomenological estimates.
Experimental values are given in lattice units with $a^{-1}=2.176$ GeV
in Ref.~\cite{pacscs_nf3}.} 
\end{figure*}

\begin{figure*}[h]
\vspace{13mm}
\begin{center}
\begin{tabular}{cc}
\includegraphics[width=85mm,angle=0]{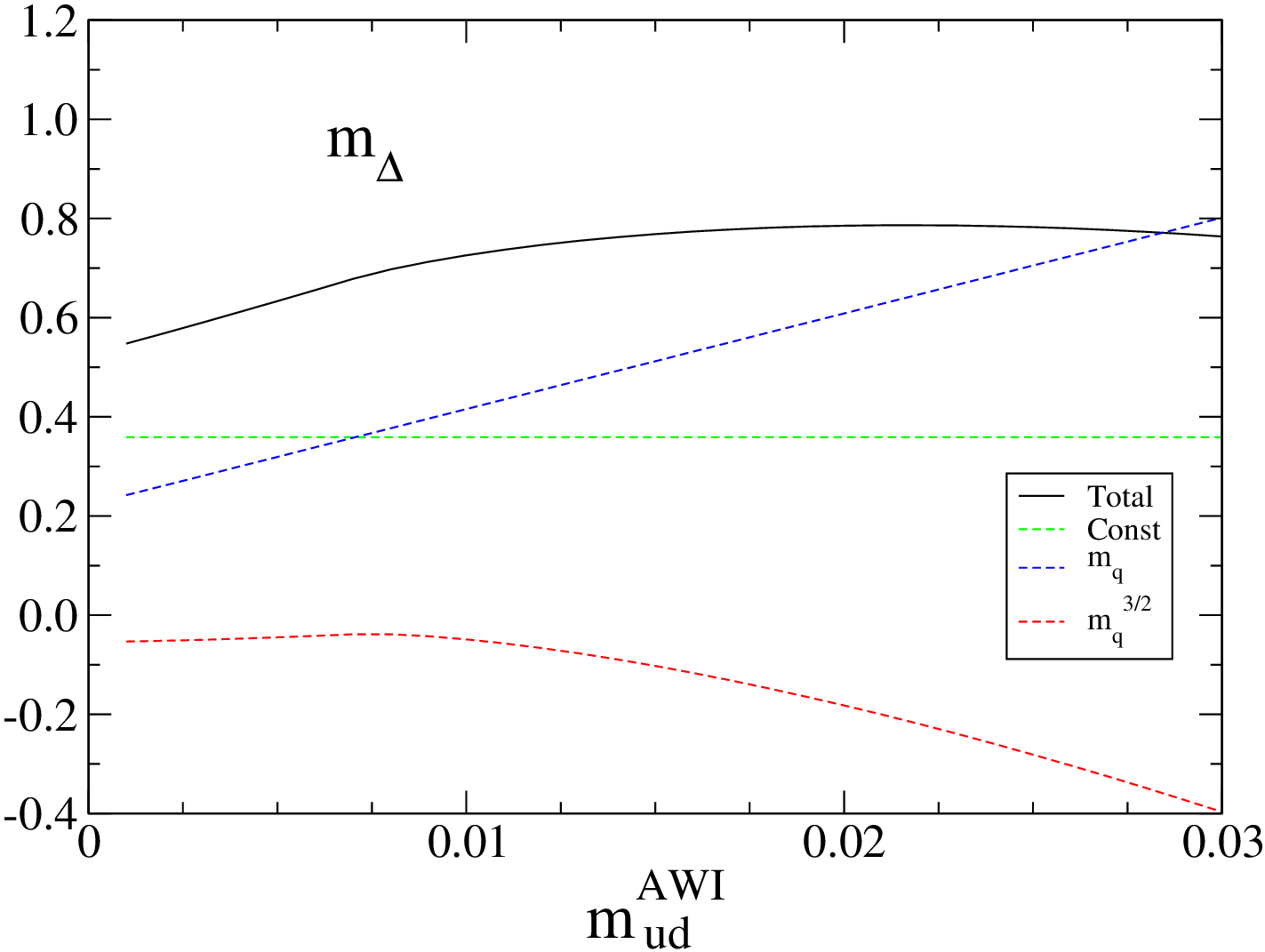}
\hspace*{2mm}&\hspace*{2mm}
\includegraphics[width=85mm,angle=0]{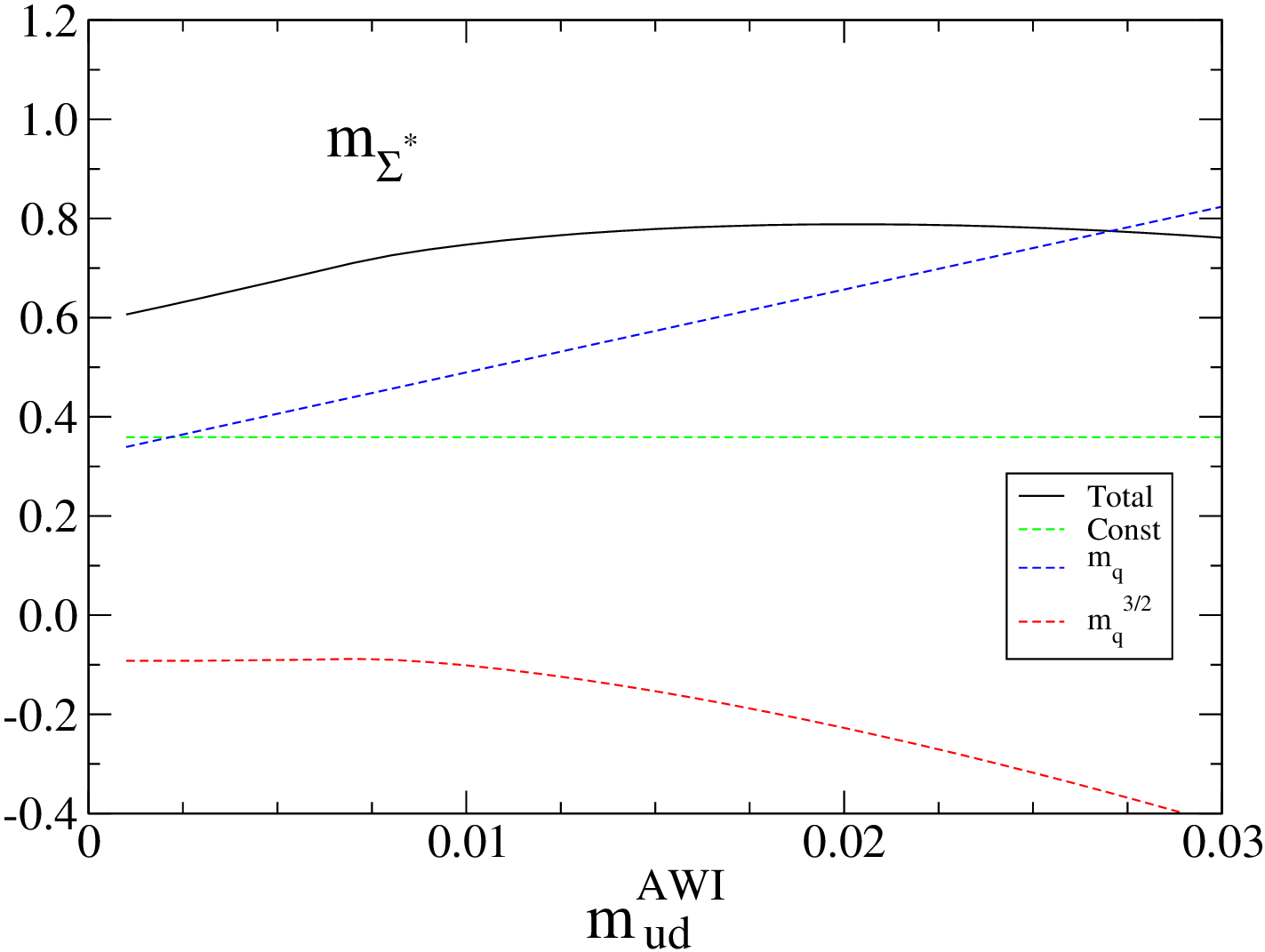}\\
\vspace*{7mm}& \\
\includegraphics[width=85mm,angle=0]{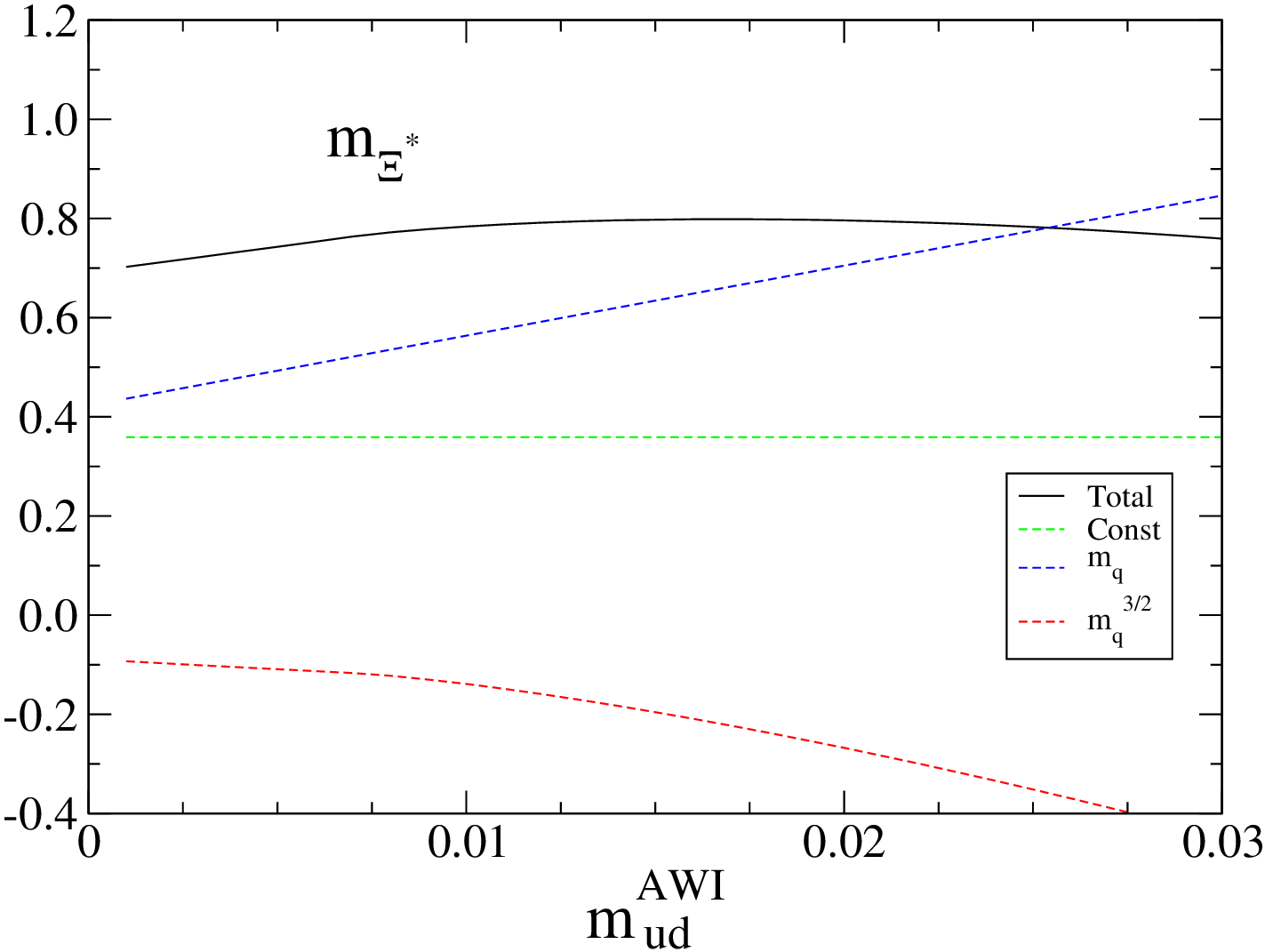}
\hspace*{2mm}&\hspace*{2mm}
\includegraphics[width=85mm,angle=0]{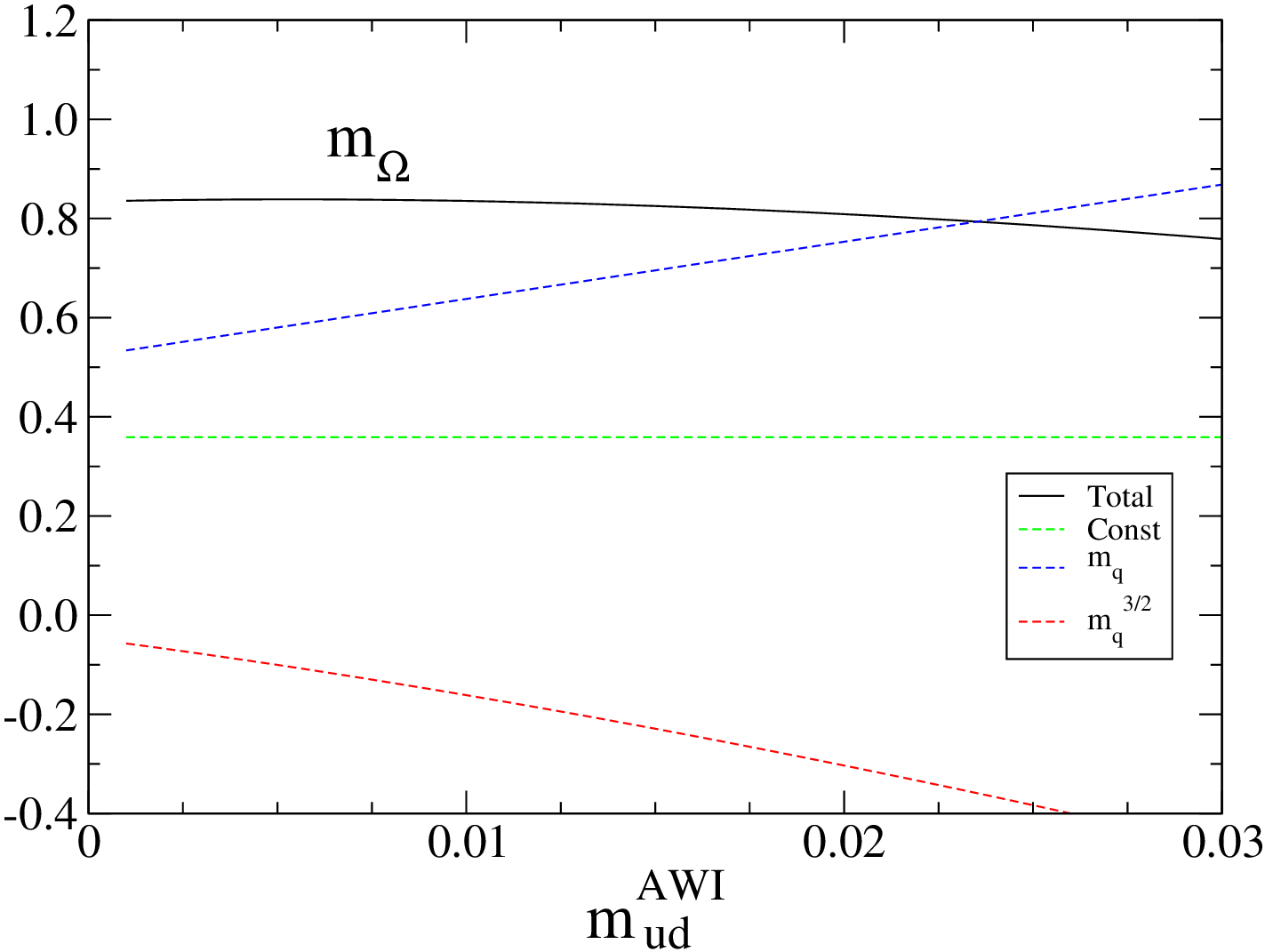}
\end{tabular}
\end{center}
\vspace{-.5cm}
\caption{\label{fig:hbchpt_su3_nlo_fix_conv_d}Convergence behavior for the
decuplet masses with the SU(3) HBChPT up to NLO. ${\cal C}$ is fixed 
at the phenomenological estimates. 
$m_{\rm s}$ is fixed with the measured value 
at $(\kappa_{\rm ud},\kappa_{\rm s})=(0.13754,0.13640)$.}
\end{figure*}

The NLO fit is carried out with and without fixing ${\cal C}$ at the
phenomenological estimate. 
The results for the latter case are
given in Fig.~\ref{fig:hbchpt_su3_nlo_d} and 
Tables~\ref{tab:hbchpt_su3_d} and \ref{tab:m_d_ph}.
We find a reasonable value for $\chi^2$/dof, though  
the vanishing result for ${\cal C}$ shows inconsistency 
with the phenomenological estimate. 
Figure~\ref{fig:hbchpt_su3_nlo_fix_d} illustrates the situation
when ${\cal C}$ is fixed to the phenomenological value: The NLO fit hardly describes 
the quark mass dependence, and suffers from an unacceptable value
of $\chi^2$/dof. 
In Fig.~\ref{fig:hbchpt_su3_nlo_fix_conv_d} 
we find that the convergence is worsened by the NLO contributions.

\section{SU(2) HBChPT analyses on nucleon mass}
\label{sec:nucleon_su2}

In the framework of SU(2) HBChPT the nucleon mass up to $O(m_{\rm ud}^{(3/2)})$
is given by
\ben
 m_N &=& m_0 -4c_1 m_{\pi}^2 -\frac{3g_A^2}{16\pi f^2} m_{\pi}^3,
\label{eq:hbchpt_su2_n}
\een
where $m_0$ and $f$ are defined in the SU(2) chiral limit, 
$c_1$ is a low energy constant and $g_A$ denotes the 
axial vector coupling of the nucleon. 
We further expand $m_0$ around the physical strange quark mass:
\ben
m_0&=&{\bar m}_0+ m_0^\prime (m_{\rm s}-m_{\rm s, ph}).
\een
Since $\vert m_{\rm s}-m_{\rm s, ph}\vert$ is comparable 
to $m_{\rm ud}$ in our simulations, 
the analytic expansion of $c_1, g_A, f$ in terms of  
$m_{\rm s}-m_{\rm s, ph}$ yields higher order
corrections beyond $O(m_{\rm ud}^{(3/2)})$.

\begin{figure*}[h]
\vspace{13mm}
\begin{center}
\begin{tabular}{cc}
\includegraphics[width=85mm,angle=0]{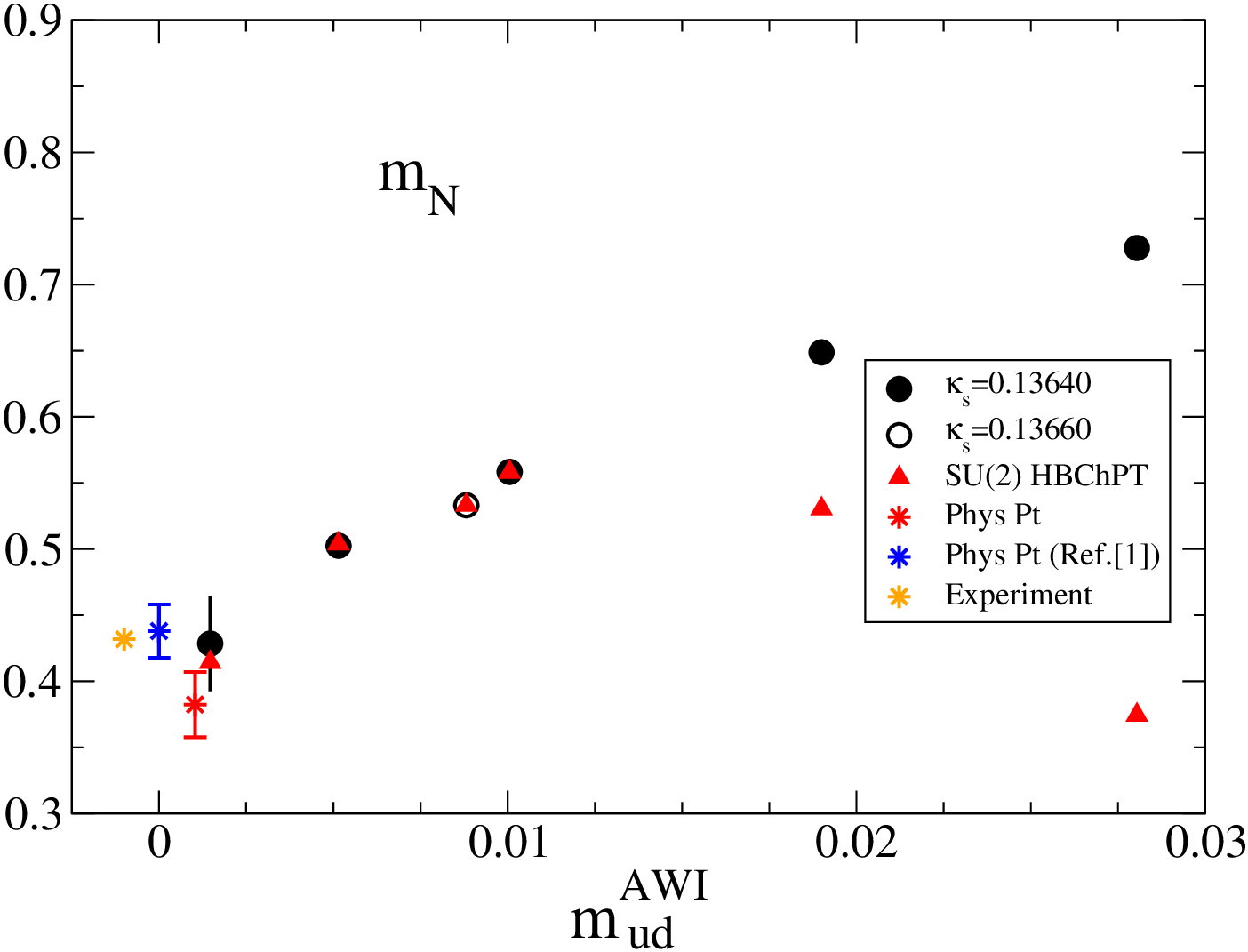}
\hspace*{2mm}&\hspace*{2mm}
\includegraphics[width=85mm,angle=0]{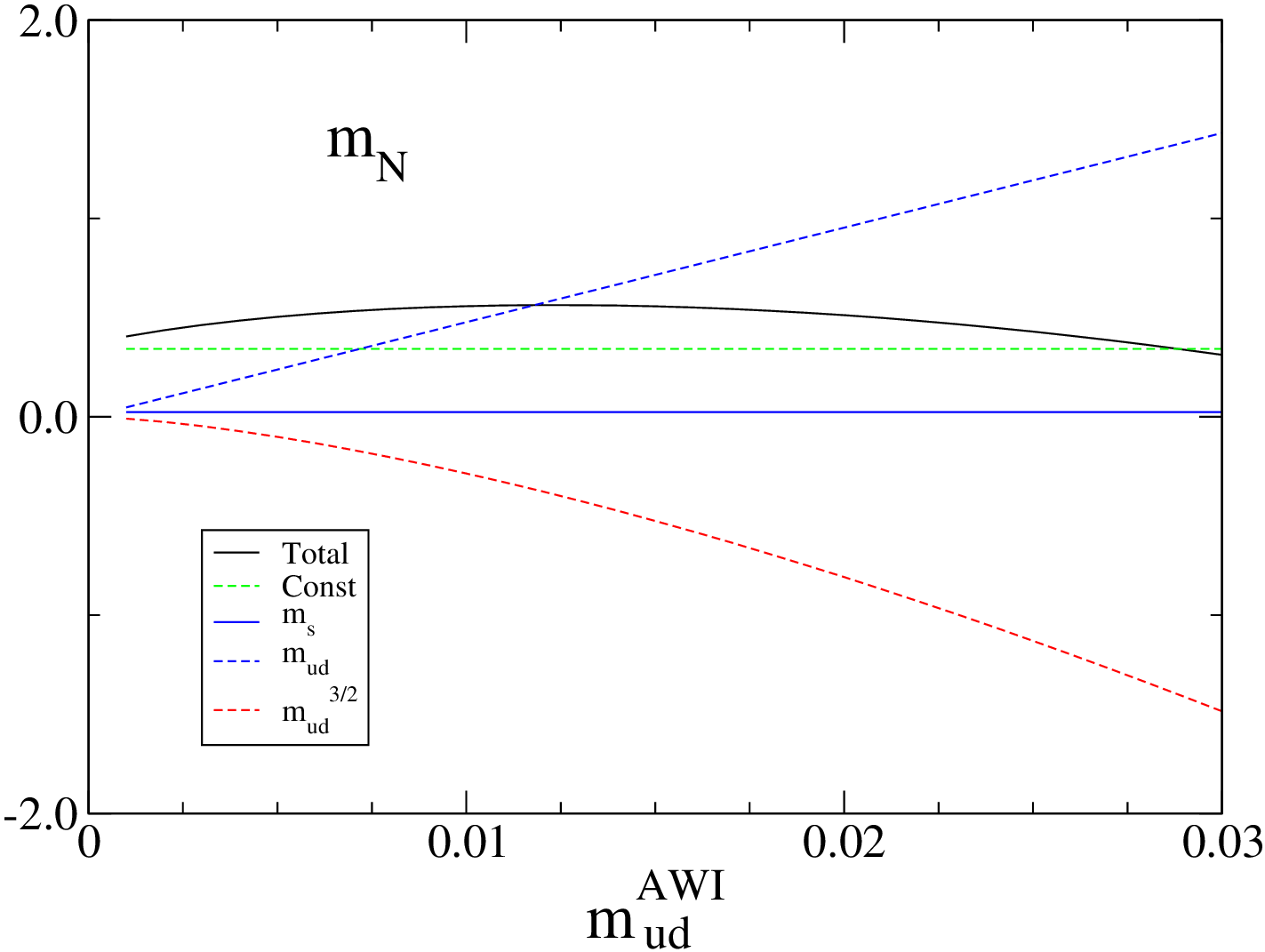}
\end{tabular}
\end{center}
\vspace{-.5cm}
\caption{\label{fig:hbchpt_su2_nlo_n}Fit results with the SU(2) HBChPT 
up to NLO for the nucleon mass (left)
and comparison of LO and NLO contributions (right).
Experimental values are given in lattice units with $a^{-1}=2.176$ GeV
in Ref.~\cite{pacscs_nf3}.} 
\end{figure*}

We apply the formula (\ref{eq:hbchpt_su2_n}) to the four data points 
with $a m_{\rm ud}\simlt 0.01$, employing the experimental axial coupling 
$g_A=1.267$,  and the value of $f$ already determined from 
the SU(2) ChPT fit for $m_\pi, f_\pi, f_K$ in Ref.~\cite{pacscs_nf3}.
The results are given in Fig.~\ref{fig:hbchpt_su2_nlo_n} 
and Table~\ref{tab:hbchpt_su2_n}.
The value of $\chi^2$/dof is sufficiently small. However,   
we should remark two points.
Firstly, Fig.~\ref{fig:hbchpt_su2_nlo_n} shows that the fit results
fail to predict the quark mass dependence of the data 
beyond $m_{\rm ud}^{\rm AWI}=0.01$. 
The extrapolated value at the physical point, for which 
we obtain $m_N=0.382(25)$, also undershoots
the experimental one sizably.
Secondly, the LO and NLO contributions quickly increase as the quark
mass increases so that the good convergence region is restricted 
near the chiral limit.

It is worthwhile to make a comparison of our fit results
with the 2 flavor twisted mass case given in Ref.~\cite{etm} where 
the authors apply the NLO formula (\ref{eq:hbchpt_su2_n}) to the nucleon mass
choosing $g_A=1.2695(29)$ and $f_\pi=0.092419(7)(25)$ GeV\footnote{The
definition of $f_\pi$ in Ref.~\cite{etm} is different from ours by a
factor of $\sqrt{2}$.}.
A particular interest exists in the parameter $c_1$ which is
responsible for the pion-nucleon sigma term: 
$\sigma(t=0)=-4c_1 m_\pi^2$ with $t$ the squared momentum transfer to
the nucleon. Our result is $c_1=-1.44(31)$ GeV$^{-1}$ 
with $a^{-1}=2.176$ GeV, which is consistent with $c_1=-1.19(1)$ GeV
in the continuum limit of the 2 flavor twisted mass case.
We obtain $\sigma(0)=75(15)$ MeV for the sigma term, 
which prefers 
$\sigma(0)=64\pm 7$ MeV given by a recent analysis with the new
experimental data\cite{sigma_ph_pasw} to the
conventional phenomenological 
estimate $\sigma(0)\simeq 45$ MeV\cite{sigma_ph_gls}.
A reliable way for the  precise determination of the sigma term, however, is
the direct calculation of the forward matrix element of the nucleon  
$\la N |{\bar q}{q}| N\ra$ which requires both the connected and the
disconnected diagrams\cite{sigma_tkb,sigma_ktk,sigma_can}.  

\section{Finite size effects}
\label{sec:fse}

The one-loop correction for the baryon mass is evaluated by the following function,
\ben
&&{\cal F}^{(\infty)}(m_\phi,\Delta) \nonumber \\
&&= -8\pi^2
\int_0^{\infty} \frac{{\rm d}^4p}{(2\pi)^4} \frac{{\vec p}^2}{(ip_4-\Delta)(p_4^2+{\vec p}^2 +m_\phi^2)},\nn\\
\een
where the integral is defined in the Euclidean space-time. 
This leads to Eq.~(\ref{eq:func_f}) after the dimensional regularization
 and the renormalization in the $\overline{\rm MS}$ scheme with the
 scale $\mu$.
In a finite spacial volume of linear size $L$, the integral over
the spatial components of the loop momentum ${\vec p}$ is replaced by a sum over discrete 
momenta ${\vec p}=(2\pi/L){\vec n}$,
\ben
&&{\cal F}^{(L)}(m_\phi,\Delta) \nonumber\\
&&= -8\pi^2 \int\, \frac{{\rm d} p_4}{2\pi} 
\sum_{\vec p} 
\frac{1}{L^3} \frac{{\vec p}^2}{(ip_4-\Delta)(p_4^2+{\vec p}^2 +m_\phi^2)},\nn\\
\een 
where we assume that the time direction is infinite. We define
the finite size correction for ${\cal F}$ as
\begin{widetext}
\ben
\delta_L {\cal F}(m_\phi,\Delta) &\equiv& 
{\cal F}^{(L)}(m_\phi,\Delta)- {\cal F}^{(\infty)}(m_\phi,\Delta)\\
& = & 4 \pi^2 \int_0^{\infty} {\rm d}\lambda \,
\left[ \delta_L ({\vec p}^2 + \beta_\Delta^2)^{-\frac{1}{2}}  
-\beta_\Delta^2 \delta_L ({\vec p}^2 +
 \beta_\Delta^2)^{-\frac{3}{2}}  \right],
\een
\end{widetext}
where $\beta_\Delta^2=\lambda^2+2\lambda \Delta +m_\phi^2$ and
\begin{widetext}
\ben
\delta_L ({\vec p}^2 + m^2)^{-r} &\equiv& \frac{1}{L^3}\sum_{\vec p}  ({\vec p}^2 + m^2)^{-r} 
-\int \frac{{\rm d}^3p}{(2\pi)^3} ({\vec p}^2 + m^2)^{-r}.
\een
\end{widetext}
With the use of the master formula\cite{AliKhan:2003rw,Beane:2004tw} 
\begin{widetext}
\ben
\delta_L ({\vec p}^2 + m^2)^{-r} &=&
\frac{1}{(4\pi)^{\frac{3}{2}} \Gamma(r)}\sum_{{\vec n} \neq 0} 
\left(\frac{L|{\vec n} |}{2 m}\right)^{r-\frac{3}{2}}K_{r-\frac{3}{2}}(mL|{\vec n}|),\nonumber\\
\een
\end{widetext}
where $K_n(z)$ is a modified Bessel function of the second kind,
one finds that the finite size corrections to the baryon masses at next-to-leading order are
\begin{widetext}
\ben
\delta_L m_{B_i}   &=& -\frac{2}{(4 \pi f_0)^2} \sum_{\phi=\pi,K,\eta} 
   \left[ 
           A_\phi^{B_i} \delta_L {\cal F}(m_\phi,0) 
          + {\cal C}^2B_\phi^{B_i} \delta_L {\cal F}(m_\phi,\Delta) 
   \right],\\
\delta_L m_{T_i} &= & -\frac{1}{(4\pi f_0)^2} \sum_{\phi=\pi,K,\eta} 
   \left[
           \frac{10}{9} {\cal H}^2A^{T_i}_\phi  \delta_L {\cal F}(m_\phi,0) 
            + {\cal C}^2 B_\phi^{T_i} \delta_L {\cal F}(m_\phi,-\Delta)
   \right ],
\een
\end{widetext}
where $\delta_L m = m(L)-m(\infty)$ and 
$\delta_L {\cal F}(m_\phi,\Delta)$ is given by
\begin{widetext}
\ben
\qquad \qquad \delta_L {\cal F}(m_\phi,\Delta) &=& 2 \int_0^{\infty}\,{\rm d}\lambda \,
 \beta_\Delta \sum_{{\vec n}\neq 0}
\left[\frac{1}{L |{\vec n}|}K_1(L\beta_\Delta |{\vec n}|)
-\beta_\Delta K_0(L\beta_\Delta |{\vec n}|)\right], 
  \label{first-eqn} \\
\delta_L {\cal F}(m_\phi,0) &=& -\pi m_\phi^2 
\sum_{{\vec n}\neq 0} \frac{1}{L|{\vec n}|} {\rm e}^{-m_\phi L
 |{\vec n}| }.
\label{second-eqn}
\een
\end{widetext}


In order to evaluate the magnitude of the finite size effects, 
let us consider the asymptotic form of $\delta_L {\cal F}$\cite{Beane:2004tw}:
\begin{widetext}
\ben
\delta_L {\cal F}(m_\phi,\Delta) &=& -6\sqrt{2\pi}m_\phi^{5/2}
\frac{1}{L^{3/2} \Delta}
{\rm e}^{-m_\phi L}+\cdots,\\ 
\delta_L  {\cal F}(m_\phi,0) &=& -6\pi m_\phi^2 
\frac{1}{L} {\rm e}^{-m_\phi L}.
\een
\end{widetext}
The finite size effect for the baryon mass 
$\delta_L m  = m(L)-m(\infty)$ is expressed as
\begin{widetext}
\ben
\delta_L m = A \left\{ \frac{3}{8\pi} 
\frac{m_\pi^3}{f_0^2}
\frac{1}{(m_\pi L)}\,{\rm e}^{-m_\pi L} \right\}+ 
B \left\{\frac{6\sqrt{2\pi}}{16\pi^2} 
\frac{m_\pi^4}{f_0^2 \Delta}
\frac{1}{(m_\pi L)^{3/2}}
{\rm e}^{-m_\pi L}\right\} \equiv A E_1 + BE_2,
\label{eq:fse}
\een
\end{widetext}
where we neglect the sub-leading contributions from $m_K$ and $m_\eta$.
An intriguing point is that $E_1$ and $E_2$ diminish as the pion mass
decreases if the product of $m_\pi L$ is kept fixed. 
The values of $E_1$ and $E_2$ with $aL=32$ at the physical point can be 
evaluated as
\ben
a E_1&=& 6.61\times 10^{-4},\\
a E_2&=& 1.43\times 10^{-4},
\een
where we employ the following results in Ref.~\cite{pacscs_nf3}: 
\ben
m_{\rm ud}^{\rm ph} B_0 &=& 0.00859(11)\;\;[{\rm GeV}^2],\\
af_0&=&0.0546(39),\\
a^{-1}&=&2.176(31)\;\;[{\rm GeV}].
\een
We also use $\Delta=m_T-m_B=0.16$ obtained by the LO SU(3) HBChPT fit
for the baryon masses.
Table~\ref{tab:fse} summarizes the coefficients $A,B$ and 
the relative finite size correction normalized by the baryon mass extrapolated
at the physical point with the formula (\ref{eq:linear_su2}).
To evaluate the numerical value of $A$ and $B$ we use the
phenomenological estimates $D=0.80$, $F=0.47$ and ${\cal C}=1.5$ and
assume ${\cal H}=1$ which is comparable to our fit results 
in Table~\ref{tab:hbchpt_su3_d}. 
We find that the HBChPT predicts fairly small finite size corrections
even at the physical point: The magnitude is less than 1\% 
for all the channels. 
We additionally remark that the $\Omega$ baryon 
mass is free from the leading contribution of the finite size effects.
This is another fascinating feature to choose $m_\Omega$ as one of the 
physical inputs.  

\section{Conclusion}
\label{sec:conclusion}

We have investigated the chiral behavior of the octet and the decuplet 
baryon masses based on the SU(3) HBChPT. At LO we find reasonable 
fit results both for the octet and the decuplet baryon masses, though
rather large strange quark contributions are observed.
This point is contrary to the SU(2) linear chiral expansion where
the LO strange quark contribution is well controlled around the physical
$m_{\rm s}$.
Inclusion of the NLO contributions makes the situation worse:
The fit results are incompatible with the
phenomenological estimates for the low energy constants $D,F,{\cal C}$ both for the
octet and the decuplet cases.
We have also applied the NLO SU(2) HBChPT to the nucleon mass.
The quark mass dependence is reasonably described 
below $m_{\rm ud}^{\rm AWI}\simlt 0.01$. The good convergence
property, however, is observed only near the chiral limit.
The finite size effects predicted by the SU(3) HBChPT turn out to be fairly small:
at most less than 1\%.
However, we should be aware of the possibility that the value $m_\pi L\approx 2$  
at the physical point for the $aL=32$ lattice may be beyond the applicability range 
of Eq.~(\ref{eq:fse}).  Comparisons with lattice data on larger lattices should 
settle the issue here.

In order to avoid the difficulties associated with the chiral extrapolation of the
baryonic quantities we are now carrying out a simulation directly on the physical point.
This simulation is being made on a larger lattice size so that a direct study of 
finite size effects is possible.

\begin{acknowledgments}
Numerical calculations for the present work have been carried out
on the PACS-CS computer 
under the ``Interdisciplinary Computational Science Program'' of 
Center for Computational Sciences, University of Tsukuba. 
A part of the code development has been carried out on Hitachi SR11000 
at Information Media Center of Hiroshima University. 
This work is supported in part by Grants-in-Aid for Scientific Research
from the Ministry of Education, Culture, Sports, Science and Technology
(Nos.
16740147,   
17340066,   
18104005,   
18540250,   
18740130,   
19740134,   
20340047,   
20540248,   
20740123,   
20740139    
).
\end{acknowledgments}


\begin{table*}[h]
\caption{\label{tab:m_o_ph}Octet baryon masses at the physical point.
NLO results are obtained without (case 1)
and with (case 2) fixing $D, F,{\cal C}$ at the phenomenological estimate.}
\begin{ruledtabular}
\setlength{\tabcolsep}{10pt}
\renewcommand{\arraystretch}{1.2}
\begin{tabular}{lllll}
 & \multicolumn{1}{c}{linear} & 
   \multicolumn{1}{c}{LO} & 
   \multicolumn{2}{c}{NLO}  \\ 
 &  & & \multicolumn{1}{c}{case 1} & 
        \multicolumn{1}{c}{case 2}   \\ \hline 
$N$ &0.438(20)& 0.4532(75) & 0.447(15) & 0.322(32) \\
$\Lambda$& 0.502(10) & 0.5199(72) & 0.517(11) & 0.387(82) \\
$\Sigma$ &0.531(11)& 0.5439(77) & 0.546(12) & 0.53(16) \\
$\Xi$ &0.5992(75)& 0.5987(80) & 0.600(14) & 0.62(18)  \\
\end{tabular}
\end{ruledtabular}
\end{table*}

\begin{table*}[h]
\caption{\label{tab:m_d_ph}Decuplet baryon masses at the physical point.
NLO results are obtained without  (case 1)
and with (case 2) fixing ${\cal C}$ at the phenomenological estimate.
}
\begin{ruledtabular}
\setlength{\tabcolsep}{10pt}
\renewcommand{\arraystretch}{1.2}
\begin{tabular}{lllll} 
 & \multicolumn{1}{c}{linear} & \multicolumn{1}{c}{LO} & \multicolumn{2}{c}{NLO} \\  
 & &   & case 1 & case 2    \\ \hline 
$\Delta$ & 0.586(19)& 0.612(9) & 0.604(19) & 0.545(20) \\
$\Sigma^*$ & 0.657(15)& 0.666(9) & 0.663(14) & 0.604(17) \\
$\Xi^*$ &0.718(12)&0.720(9) & 0.721(14) & 0.694(17) \\
$\Omega$ &0.769(11)& 0.774(11) & 0.777(17) & 0.815(21) \\
\end{tabular}
\end{ruledtabular}
\end{table*}

\begin{table*}[h]
\caption{\label{tab:linear_su2_o}Fit results with the linear 
formula Eq.~(\ref{eq:linear_su2}) for the octet baryon masses.}
\begin{ruledtabular}
\setlength{\tabcolsep}{10pt}
\renewcommand{\arraystretch}{1.2}
\begin{tabular}{c....}
  & \multicolumn{1}{c}{$N$} & 
    \multicolumn{1}{c}{$\Lambda$} & 
    \multicolumn{1}{c}{$\Sigma$} & 
    \multicolumn{1}{c}{$\Xi$} \\ \hline
 $\alpha$ & 0.371(50) & 0.375(28) & 0.381(33) & 0.408(21)\\
 $\beta$ & 11.6(2.4) & 9.6(1.0) & 8.0(1.0) & 4.24(46)\\
 $\gamma$& 1.8(1.3) & 3.90(76) & 4.7(9) & 6.24(58)\\ \hline 
 $\chi^2/{\rm dof}$ & 0.63(2.5) & 1.3(2.3) & 0.6(1.4) & 0.09(59)\\
\end{tabular}
\end{ruledtabular}
\end{table*}

\begin{table*}[h]
\caption{\label{tab:linear_su2_d}Same as Table~\ref{tab:linear_su2_o} 
for the decuplet baryon masses.}
\begin{ruledtabular}
\setlength{\tabcolsep}{10pt}
\renewcommand{\arraystretch}{1.2}
\begin{tabular}{c....}
  & \multicolumn{1}{c}{$\Delta$} & 
    \multicolumn{1}{c}{$\Sigma^*$} & 
    \multicolumn{1}{c}{$\Xi^*$} & 
    \multicolumn{1}{c}{$\Omega$} \\ \hline
 $\alpha$ &0.527(60) & 0.538(48) & 0.548(40) & 0.552(33) \\
 $\beta$ & 10.7(2.1) & 6.3(1.5) & 3.41(82) & 1.80(57) \\
 $\gamma$&1.6(1.6) & 3.8(1.3) & 5.5(1.1) & 7.2(9) \\ \hline 
 $\chi^2/{\rm dof}$ &0.4(1.4) & 0.2(1.3) & 0.00(21) & 0.6(1.7) \\
\end{tabular}
\end{ruledtabular}
\end{table*}

\begin{table*}
\caption{\label{tab:ac_octet} Coefficients for the octet baryons 
$A_\phi^{B_i}$ and $B_\phi^{B_i}$.}
\begin{ruledtabular}
\renewcommand{\arraystretch}{1.2}
\begin{tabular}{c|ccc|ccc}
 &  &{$A_\phi^{B_i}$}& & &{$B_\phi^{B_i}$}& \\
$\phi$& $\pi$ & $K$ & $\eta$ & $\pi$ & $K$ & $\eta$ \\ \hline
$N$ & $\frac{3}{2}(D+F)^2$ & $\frac{1}{3}(5D^2-6DF+9F^2)$ &
 $\frac{1}{6}(D-3F)^2$ & $\frac{4}{3}$ & $\frac{1}{3}$ & 0 \\
$\Lambda$ & $2D^2$ & $\frac{2}{3}(D^2+9F^2)$ & $\frac{2}{3}D^2$ & 1 & $\frac{2}{3}$ & 0\\
$\Sigma$ &  $\frac{2}{3}(D^2+6F^2)$ & $2(D^2+F^2)$ & $\frac{2}{3}D^2$ &$\frac{2}{9}$ & $\frac{10}{9}$ & $\frac{1}{3}$ \\
$\Xi$ & $\frac{3}{2}(D-F)^2$ & $\frac{1}{3}(5D^2+6DF+F^2)$ &
 $\frac{1}{6}(D+3F)^2$ & $\frac{1}{3}$ & 1 & $\frac{1}{3}$ \\
\end{tabular}
\end{ruledtabular}
\end{table*}

\begin{table*}[h]
\caption{\label{tab:hbchpt_su3_o}Fit results with the SU(3) HBChPT
for the octet baryon masses. NLO results are obtained with (case 2)
 and without (case 1) fixing $D, F, {\cal C}$ at the phenomenological estimate.}
\begin{ruledtabular}
\setlength{\tabcolsep}{10pt}
\renewcommand{\arraystretch}{1.2}
\begin{tabular}{ccccc}
 & \multicolumn{1}{c}{LO} & 
   \multicolumn{2}{c}{NLO} & 
   \multicolumn{1}{c}{phenom.} \\ 
 &    & \multicolumn{1}{c}{case 1} & 
        \multicolumn{1}{c}{case 2} &  \\ \hline 
$m_{B}$ & 0.410(14) & 0.391(39) & $-$0.15(9) &   \\
$\alpha_{M}$ & $-$2.262(62) & $-$2.62(62) & $-$15.3(2.0) &   \\
$\beta_{M}$ & $-$1.740(58) & $-$2.6(1.5) & $-$21.3(3.0) &  \\
$\sigma_{M}$ & $-$0.53(12) & $-$0.71(34) & $-$9.6(1.4) &    \\ \hline
$D$ &  & 0.000(16)$\times 10^{-8}$ & 0.80 fixed &    0.80 \\
$F$ &  & 0.000(9)$\times 10^{-8}$ & 0.47 fixed &   0.47 \\ 
${\cal C}$ &  & 0.36(30) & 1.5 fixed &    1.5 \\ \hline
$\chi^2$/dof & 1.10(63) & 1.39(77) & 153(82) &  \\
\end{tabular}
\end{ruledtabular}
\end{table*}

\begin{table*}[h!]
\caption{\label{tab:ac_decuplet} Coefficients for the decuplet baryons 
$A_\phi^{T_i}$ and $B_\phi^{T_i}$.}
\begin{ruledtabular}
\setlength{\tabcolsep}{10pt}
\renewcommand{\arraystretch}{1.2}
\begin{tabular}{c|ccc|ccc}
 &  &{$A_\phi^{T_i}$}& & &{$B_\phi^{T_i}$}& \\ 
$\phi$& $\pi$ & $K$ & $\eta$ & $\pi$ & $K$ & $\eta$ \\ \hline
$\Delta$ & $\frac{5}{6}$ & $\frac{1}{3}$ & $\frac{1}{6}$ & $\frac{2}{3}$ & $\frac{2}{3}$ & 0 \\
$\Sigma^*$ &$\frac{4}{9}$ & $\frac{8}{9}$ & 0 & $\frac{5}{9}$ & $\frac{4}{9}$ & $\frac{1}{3}$ \\
$\Xi^*$ &$\frac{1}{6}$ & $1$ & $\frac{1}{6}$ &$\frac{1}{3}$ & $\frac{2}{3}$ & $\frac{1}{3}$  \\
$\Omega$ & 0 & $\frac{2}{3}$ & $\frac{2}{3}$ & 0 & $\frac{4}{3}$ & 0 \\
\end{tabular}
\end{ruledtabular}
\end{table*}

\begin{table*}[h]
\caption{\label{tab:hbchpt_su3_d}Fit results with the SU(3) HBChPT
for the decuplet baryon masses. NLO results are obtained with (case 2)
 and without (case 1) fixing $D,F,{\cal C}$ at the phenomenological estimate.}
\begin{ruledtabular}
\setlength{\tabcolsep}{10pt}
\renewcommand{\arraystretch}{1.2}
\begin{tabular}{ccccc} 
 & \multicolumn{1}{c}{LO} & \multicolumn{2}{c}{NLO} & phenom. \\  
 &    & case 1 & case 2 &   \\ \hline 
$m_{T}$ & 0.570(16) & 0.550(43) & 0.359(43) &    \\
$\gamma_{M}$ & 2.745(80) & 3.4(1.1) & 3.88(25) &   \\
$\bar\sigma_{M}$ & $-$0.56(15) & $-$0.96(75) & -2.88(40) &     \\ \hline
${\cal C}$ &  & 0.000(21)$\times 10^{-8}$ & 1.5 fixed &       1.5 \\ 
${\cal H}$ &  &  0.50(49) & 0.000(4)$\times 10^{-8}$ &  \\ \hline
$\chi^2$/dof & 0.46(48) & 0.50(60) & 21.5(9.5) &  \\
\end{tabular}
\end{ruledtabular}
\end{table*}

\begin{table*}[h]
\caption{\label{tab:hbchpt_su2_n}Fit results with SU(2) HBChPT 
up to NLO for the nucleon mass.}
\begin{ruledtabular}
\renewcommand{\arraystretch}{1.2}
\centering
\begin{tabular}{lllllll}
  &    NLO             &  \\ \hline 
 ${\bar m}_0$ & 0.258(63)&\\ 
 $m_0^\prime$ & 2.8(1.4)&\\
 $c_1$& $-$3.14(68)&\\ \hline
 $\chi^2$/dof &0.2(9) & \\
\end{tabular}
\end{ruledtabular}
\end{table*}

\begin{table*}[h]
\caption{\label{tab:fse}Relative finite size correction 
normalized by the baryon mass extrapolated
at the physical point with the formula Eq.~(\ref{eq:linear_su2}).
LECs are chosen to be $D=0.80$, $F=0.47$, ${\cal C}=1.5$ and ${\cal H}=1$.}
\begin{ruledtabular}
\begin{tabular}{ccccc|c}
  &   $A$ &  $B$  & $R=\delta_L m/m[\%]$       \\ \hline
$m_N$                &  $3(D+F)^2 $             & $\frac{8}{3}{\cal C}^2$ 
& $0.89$\\
$m_\Lambda$    & $4D^2 $                 & $2{\cal C}^2$
& $0.45$ \\
$m_\Sigma$       & $\frac{4}{3}(D^2+6F^2) $ & $\frac{4}{9}{\cal C}^2$
& $0.34$ \\ 
$m_\Xi $           & $3(D-F)^2 $                 & $\frac{2}{3}{\cal C}^2$
& $0.07$ \\ 
\hline
$m_\Delta$          & $\frac{25}{27}{\cal H}^2 $     & $\frac{2}{3}{\cal C}^2$  
&  $0.13$ \\
$m_{\Sigma^*}$    &  $\frac{40}{81}{\cal H}^2$     & $\frac{5}{9}{\cal C}^2$  
& $0.08$ \\
$m_{\Xi^*}\ $       &  $\frac{5}{27}{\cal H}^2$       &
 $\frac{1}{3}{\cal C}^2$
& $0.03$  \\ 
$m_\Omega $      & $0$   & $0$
& $0$  \\ 
\end{tabular}
\end{ruledtabular}
\end{table*}

\clearpage

\clearpage


\begin{thebibliography}{99}
\bibitem{pacscs_nf3}
PACS-CS Collaboration, S.~Aoki {\it et al.}, 
Phys. Rev. {\bf D79}, 034503 (2009).

\bibitem{cppacs_nf3}
CP-PACS/JLQCD Collaborations, T.~Ishikawa {\it et al.}, 
Phys. Rev. {\bf D78}, 011502 (2008).

\bibitem{jenkins1}
E.~Jenkins and A.~V.~Manohar,
Phis. Lett. {\bf B255}, 558 (1991); {\it ibid.} {\bf B259}, 353 (1991)

\bibitem{jenkins2}
E.~Jenkins,
Nucl. Phys. {\bf B368}, 190 (1992).

\bibitem{walker-loud}
A.~Walker-Loud,
Nucl. Phys. {\bf A747}, 476 (2005);
B.~C.~Tiburzi and A.~Walker-Loud,
Nucl. Phys. {\bf A748}, 513 (2005).

\bibitem{hsueh}
S.~Y.~Hsueh {\it et al.},
Phys. Rev. {\bf D38}, 2056 (1988).

\bibitem{pdg}
Particle Data Group, C.~Amsler {\it et al.}, 
Phys. Lett. {\bf B667}, 1 (2008).

\bibitem{c_ph}
E.~Jenkins and A.~V.~Manohar, in
{\it Effective field theories of the standard model},
edited by U.-G. Meissner (World Scientific, Singapore, 1992).

\bibitem{qcdsf}
QCDSF-UKQCD Collaboration, A.~Ali~Khan {\it et al.}, 
Nucl. Phys. {\bf B689}, 175 (2004).

\bibitem{etm}
ETM Collaboration, C.~Alexandrou {\it et al.}, 
Phys. Rev. {\bf D78}, 014509 (2008).

\bibitem{sigma_ph_pasw}
M.~M.~Pavan, I.~I.~Strakovsky, R.~L.~Workman and R.~A.~Arndt,
PiN Newslett. {\bf 16}, 110 (2002). 

\bibitem{sigma_ph_gls}
J.~Gasser, H.~Leutwyler and M.~E.~Sainio,
Phys. Lett. {\bf B253}, 252 (1991).

\bibitem{sigma_tkb}
M.~Fukugita, Y.~Kuramashi, M.~Okawa and A.~Ukawa,
Phys. Rev. {\bf D51}, 5319 (1995).

\bibitem{sigma_ktk}
S.~J.~Dong, J.-F.~Laga{\"e} and K.~F.~Liu,
Phys. Rev. {\bf D54}, 5496 (1996).

\bibitem{sigma_can}
R.~Lewis, W.~Wilcox and R.~M.~Woloshyn,
Phys. Rev. {\bf D67}, 013003 (2003).


\bibitem{AliKhan:2003rw}
  QCDSF-UKQCD Collaboration, A.~Ali Khan {\it et al.},
  arXiv:hep-lat/0312029.

 
\bibitem{Beane:2004tw}
  S.~R.~Beane,
  Phys. Rev. {\bf D70}, 034507 (2004).




\end{thebibliography}
\end{document}